\documentclass[sn-nature,Numbered]{sn-jnl}%

\usepackage{subfiles}%
\usepackage{graphicx}%
\usepackage{multirow}%
\usepackage{amsmath,amssymb,amsfonts}%
\usepackage{amsthm}%
\usepackage{mathrsfs}%
\usepackage[title]{appendix}%
\usepackage{xcolor}%
\usepackage{textcomp}%
\usepackage{siunitx}%
\usepackage{manyfoot}%
\usepackage{booktabs}%
\usepackage{algorithm}%
\usepackage{algorithmicx}%
\usepackage{algpseudocode}%
\usepackage{listings}%
\usepackage{floatpag}%

\usepackage{etoc}

\usepackage{multibib}
\newcites{SI}{Supplementary References}

\newcommand{\pd}[2]{\frac{\partial#1}{\partial#2}}

\newcommand{\Exp}[1]{ \mathbb{E}\left[ #1 \right] }
\newcommand{\qbox}[1] { {\quad\mbox{#1}\quad}}

\begin{document}

\title[NeuralGCMs]{Neural General Circulation Models for Weather and Climate}

\author*[1]{\fnm{Dmitrii} \sur{Kochkov}}
\email{dkochkov@google.com}
\equalcont{These authors contributed equally to this work.}
\author*[1]{\fnm{Janni} \sur{Yuval}}
\email{janniyuval@google.com}
\equalcont{These authors contributed equally to this work.}
\author[1]{\fnm{Ian} \sur{Langmore}}
\equalcont{These authors contributed equally to this work.}
\author[1]{\fnm{Peter} \sur{Norgaard}}
\equalcont{These authors contributed equally to this work.}
\author[1]{\fnm{Jamie} \sur{Smith}}
\equalcont{These authors contributed equally to this work.}

\author[1]{\fnm{Griffin} \sur{Mooers}}
\author[4]{\fnm{Milan} \sur{Klöwer}}
\author[1]{\fnm{James} \sur{Lottes}}
\author[1]{\fnm{Stephan} \sur{Rasp}}

\author[3]{\fnm{Peter} \sur{Düben}}
\author[3]{\fnm{Sam} \sur{Hatfield}}

\author[2]{\fnm{Peter} \sur{Battaglia}}
\author[2]{\fnm{Alvaro} \sur{Sanchez-Gonzalez}}
\author[2]{\fnm{Matthew} \sur{Willson}}

\author[1,5]{\fnm{Michael} P. \sur{Brenner}}
\author*[1]{\fnm{Stephan} \sur{Hoyer}}
\email{shoyer@google.com}
\equalcont{These authors contributed equally to this work.}

\affil[1]{\orgname{Google Research, Mountain View, CA}}
\affil[2]{\orgname{Google DeepMind, London, UK}}
\affil[3]{\orgname{European Centre for Medium-Range Weather Forecasts, Reading, UK}}
\affil[4]{\orgname{Earth, Atmospheric and Planetary Sciences, Massachusetts Institute of Technology}}
\affil[5]{\orgname{School of Engineering and Applied Sciences, Harvard University}}

\abstract{
General circulation models (GCMs) are the foundation of weather and climate prediction.
GCMs are physics-based simulators which combine a numerical solver for large-scale dynamics with tuned representations for small-scale processes such as cloud formation.
Recently, machine learning (ML) models trained on reanalysis data achieved comparable or better skill than GCMs for deterministic weather forecasting.
However, these models have not demonstrated improved ensemble forecasts, or shown sufficient stability for long-term weather and climate simulations.
Here we present the first GCM that combines a differentiable solver for atmospheric dynamics with ML components, and show that it can generate forecasts of deterministic weather, ensemble weather and climate on par with the best ML and physics-based methods.
NeuralGCM is competitive with ML models for 1-10 day forecasts, and with the European Centre for Medium-Range Weather Forecasts ensemble prediction for 1-15 day forecasts.
With prescribed sea surface temperature, NeuralGCM can accurately track climate metrics such as global mean temperature for multiple decades, and climate forecasts with 140 km resolution exhibit emergent phenomena such as realistic frequency and trajectories of tropical cyclones.
For both weather and climate, our approach offers orders of magnitude computational savings over conventional GCMs.
Our results show that end-to-end deep learning is compatible with tasks performed by conventional GCMs, and can enhance the large-scale physical simulations that are essential for understanding and predicting the Earth system.
}

\maketitle

\section*{Introduction \label{sec:intro}}

Solving the equations of the Earth's atmosphere with general circulation models (GCMs) is the basis of weather and climate prediction \cite{Bauer2015, Balaji2022-kp}.
Over the past 70 years, GCMs have been steadily improved with better numerical methods and more detailed physical models, while exploiting faster computers to run at higher resolution.
Inside GCMs, the unresolved physical processes such as clouds, radiation and precipitation are represented by semi-empirical parameterizations.
Tuning GCMs to match historical data remains a manual process~\cite{hourdin2017_tuning}, and GCMs retain many persistent errors and biases~\cite{bony2005marine,webb2013origins,sherwood2014spread}.
The difficulty of reducing uncertainty in long-term climate projections~\cite{PalmerStevens2019} and estimating distributions of extreme weather events~\cite{fischer2013robust} presents major challenges for climate mitigation and adaptation \cite{field2012managing}.

Recent advances in machine learning (ML) have presented an alternative for weather forecasting~\cite{rasp2023weatherbench, keisler2022forecasting, bi2023accurate, lam2022graphcast}.
These models rely solely on ML techniques, using roughly 40 years of historical data from the European Center for Medium-Range Weather Forecasts (ECMWF) reanalysis v5 (ERA5)~\cite{hersbach2020era5} for model training and forecast initialization.
ML methods have been remarkably successful, demonstrating state-of-the-art deterministic forecasts for 1-10 day weather prediction at a fraction of the computational cost of traditional models~\cite{lam2022graphcast,bi2023accurate}.
ML atmospheric models also require considerably less code, for example GraphCast~\cite{lam2022graphcast} has \num{5417} lines vs \num{376578} lines for NOAA's FV3 atmospheric model \cite{Zhou2019-next-gen-GFS} (see Appendix~\ref{apx:sec:loc} for details).

Nevertheless, ML approaches have noteworthy limitations compared to GCMs.
Existing ML models have focused on deterministic prediction, and surpass deterministic numerical weather prediction in terms of the aggregate metrics for which they are trained~\cite{bi2023accurate, lam2022graphcast}.
However, they do not produce calibrated uncertainty estimates~\cite{bi2023accurate}, which is essential for useful weather forecasts~\cite{Bauer2015}.
Deterministic ML models using a mean-squared-error loss are rewarded for averaging over uncertainty, producing unrealistically blurry predictions when optimized for multi-day forecasts \cite{keisler2022forecasting, lam2022graphcast}.
Unlike physical models, ML models misrepresent derived (diagnostic) variables such as geostrophic wind~\cite{bonavita2023limitations}.
Furthermore, although there has been some success in using ML approaches on longer time scales \cite{weyn2020improving,watt2023ace}, these models have not demonstrated the ability to outperform existing GCMs. 

Hybrid models that combine GCMs with machine learning are appealing because they build on the interpretability, extensibility and successful track record of traditional atmospheric models~\cite{Bretherton2023-ym,Reichstein2019review}.
In the hybrid model approach, an ML component replaces or corrects the traditional physical parameterizations of a GCM.
Up until now, the ML component in such models has been trained  ``offline,'' by learning parameterizations independently of their interaction with dynamics.
These components are then inserted into an existing GCM. The lack of coupling between ML components and the governing equations during training potentially causes serious problems such as instability and climate drift \cite{brenowitz2019spatially}.
So far, hybrid models have mostly been limited  to idealized scenarios such as aquaplanets \cite{rasp2018deep,yuval2020stable}.
Under realistic conditions, ML corrections have reduced some biases of very coarse GCMs~\cite{kwa2023machine, arcomano2023hybrid,han2023ensemble}, but performance remains considerably worse than state-of-the-art models.

Here we present NeuralGCM, the first fully-differentiable hybrid general circulation model of the Earth's atmosphere.
NeuralGCM is trained on forecasting up to 5-day weather trajectories sampled from ERA5.
Differentiability enables end-to-end ``online training''~\cite{Gelbrecht2023differentiable}, with ML components optimized in the context of interactions with the governing equations for large-scale dynamics, which we find enables accurate and stable forecasts.
NeuralGCM produces physically consistent forecasts with accuracy comparable to best-in-class models across a range of time-scales, from 1-15 day weather to decadal climate prediction.

\subsection*{Neural general circulation models \label{sec:Differentiable model}}

\begin{figure*}
\begin{center}
\makebox[\textwidth]{\colorbox{white}{\includegraphics[width=0.8\paperwidth]{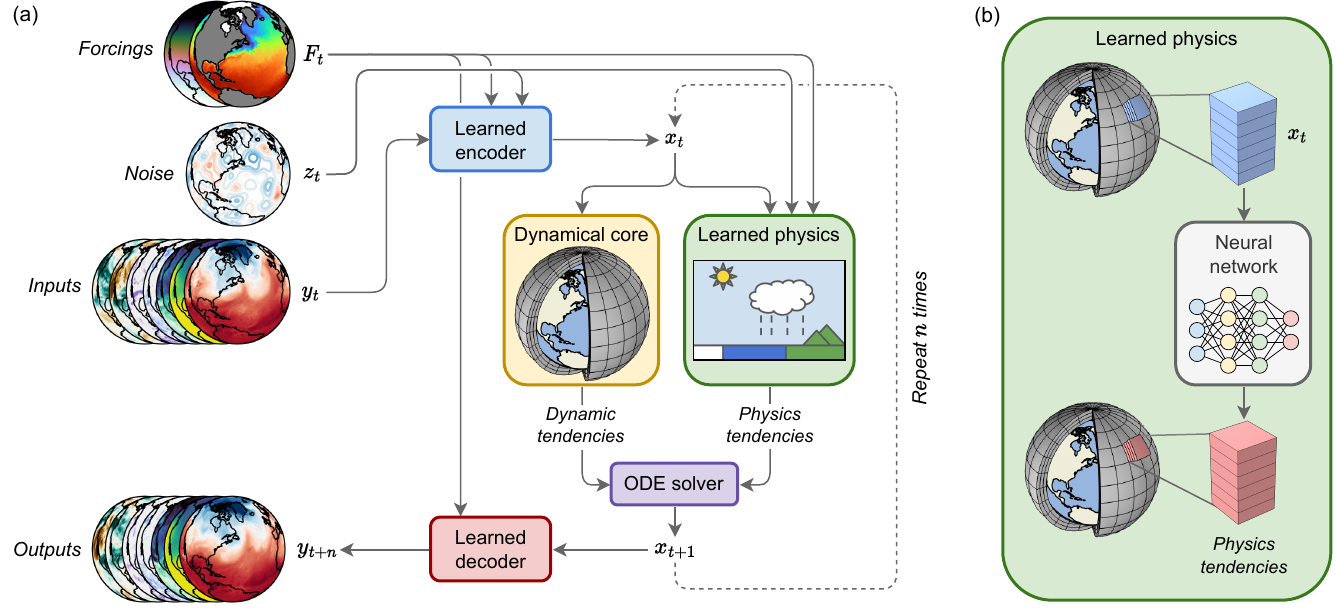}}}
\end{center}
\caption{
Structure of the NeuralGCM model.
(a) Overall model structure, showing how forcings $F_t$, noise $z_t$ (for stochastic models), and inputs $y_t$ are encoded into the model state $x_t$.
Model state is fed into the dynamical core, and alongside forcings and noise into the learned physics module. This produces tendencies (rates of change) used by an implicit-explicit ODE solver to advance the state in time.
The new model state $x_{t+1}$ can then be fed back into another time step, or decoded into model predictions.
(b) Inset of the learned physics module, which feeds data for individual columns of the atmosphere into a neural network used to produce physics tendencies in that vertical column.
}\label{fig:overview}
\end{figure*}

A schematic of NeuralGCM is shown in Fig.~\ref{fig:overview}.
The two key components of NeuralGCM are a differentiable dynamical core for solving the discretized governing dynamical equations, and a learned physics module that parameterizes physical processes with a neural network, described in full detail in Appendix \ref{apx:sec:dycore} and \ref{apx:sec:parameterization}.
The dynamical core simulates large-scale fluid motion and thermodynamics under the influence of gravity and the Coriolis force.
The learned physics module predicts the effect of unresolved processes such as cloud formation, radiative transport, precipitation and subgrid-scale dynamics, on the simulated fields using a neural network.

Our differentiable dynamical core is implemented in JAX, a library for high-performance code in Python that supports automatic differentiation~\cite{bradbury2018jax}. 
The dynamical core solves the hydrostatic primitive equations with moisture, using a horizontal pseudo-spectral discretization and vertical sigma coordinates~\cite{Bourke1974-spectral, durran2010numerical}.
We evolve seven prognostic variables:
vorticity and divergence of horizontal wind, temperature, surface pressure, and three water species (specific humidity, and specific ice and liquid cloud water content).

Our learned physics module uses the single-column approach of GCMs~\cite{Balaji2022-kp}, whereby information only from a single atmospheric column is used to predict the impact of unresolved processes occurring within that column.
These effects are predicted using a fully-connected neural network with residual connections, with weights shared across all atmospheric columns (\ref{apx:subsec:network_architecture}).
The inputs to the neural network include the prognostic variables in the atmospheric column, total incident solar radiation, sea ice concentration and sea surface temperature (\ref{apx:subsec:parameterization_features}).
We also provide horizontal gradients of the prognostic variables, which we found improves performance \cite{wang2022non}.
All inputs are standardized to have zero mean and unit variance using statistics precomputed during model initialization.
The outputs are the prognostic variable tendencies scaled by the fixed unconditional standard deviation of the target field~(\ref{apx:subsec:output_rescaling}).

To interface between ERA5~\cite{hersbach2020era5} data stored in pressure coordinates and the sigma coordinate system of our dynamical core, we introduce encoder and decoder components~(Appendix \ref{apx:sec:encoder_and_decoder}).
These components perform linear interpolation between pressure levels and sigma coordinates levels.
We additionally introduce learned corrections to both encoder and decoder steps, using the same column-based neural network architecture as the learned physics module.
Importantly, the encoder enables us to eliminate the gravity waves from initialization shock \cite{daley1981normal}, which otherwise contaminate forecasts.

Figure~\ref{fig:overview}(a) shows the sequence of steps that NeuralGCM takes to make a forecast. First, it encodes ERA5 data at $t=t_{0}$ on pressure levels to initial conditions on sigma coordinates.
To perform a time step, the dynamical core and learned physics [Fig.~\ref{fig:overview}(b)] then compute tendencies, which are integrated in time using an implicit-explicit ODE solver \cite{whitaker2013implicit} (Appendix~\ref{apx:sec:time_integration}).
This is repeated to advance the model from $t=t_{0}$ to $t=t_{final}$.
Finally, the decoder converts predictions back to pressure levels.

The time-step size of the ODE solver is limited by the CFL condition on dynamics, and can be small relative to the time-scale of atmospheric change.
Evaluating learned physics is approximately 1.5$\times$ as expensive as a time step of the dynamical core.
Accordingly, following the typical practice for GCMs, we hold learned physics tendencies constant for multiple ODE time-steps to reduce computational expense, typically corresponding to 30 minutes of simulation time.

\subsection*{Training \label{sec:Training}}

The differentiable dynamical core in NeuralGCM allows an end-to-end training approach, whereby we advance the model multiple time steps before employing stochastic gradient descent to minimize discrepancies between model predictions and conservatively regridded ERA5 data (\ref{apx:subsec:training_data_and_schedule}).
We gradually increase the rollout length from $6$ hours to $5$ days (Appendix \ref{apx:sec:training}), which we found to be critical because our models are not accurate for multi-day prediction early in training.
The extended backpropagation through hundreds of simulation steps enables our neural networks to take into account interactions between the learned physics and the dynamical core.
We train deterministic and stochastic NeuralGCM models, each of which uses a distinct training protocol, described in full detail in \ref{apx:subsec:deterministic_training}-\ref{apx:subsec:stochastic_training}.

We train deterministic NeuralGCM models using a combination of three loss functions (\ref{apx:subsec:deterministic_training}) to encourage accuracy and sharpness while penalizing bias.
During the main training phase, all losses are defined in a spherical harmonics basis. 
We use a standard MSE loss for prompting accuracy, modified to progressively filter out contributions from higher total wavenumbers at longer lead times.
This filtering approach tackles the ``double penalty problem''~\cite{gilleland2009intercomparison} as it prevents the model from being penalized for predicting high-wavenumber features in incorrect locations at later times, especially beyond the predictability horizon.
A second loss term encourages the spectrum to match the training data using squared loss on the total wavenumber spectrum of prognostic variables. These first two losses are evaluated on both sigma and pressure levels. Finally, a third loss term discourages bias by adding MSE on the batch-averaged mean amplitude of each spherical harmonic coefficient.
For analysis of the impact that various loss functions have, refer to~\ref{apx:subsec:ablation_loss}.
The combined action of the three training losses allow the resulting models trained on $3$-day rollouts to remain stable during years-to-decades long simulations~(Section \ref{sec:climate_simulations}.
Before final evaluations, we perform additional fine-tuning of just the decoder component on short rollouts of $24$ hours~(\ref{apx:subsubsec:decoder_finetuning}).

Stochastic NeuralGCM models incorporate inherent randomness in the form of additional random fields passed as inputs to neural network components.
Our stochastic loss is based on the the Continuous Ranked Probability Score (CRPS)~\cite{Gneiting2007ProperScoring,rasp2018neural,pacchiardi2021probabilistic}.
CRPS consists of mean absolute error that encourages accuracy, balanced by a similar term that encourages ensemble spread.
For each variable we use a sum of CRPS in grid-space and CRPS in the spherical harmonic basis below a maximum cutoff wavenumber (\ref{apx:subsec:stochastic_training}). We compute CRPS on rollout lengths from $6$ hours to $5$ days.
As illustrated in Fig.~\ref{fig:overview}, we inject noise to the learned encoder and leaned physics module by sampling from Gaussian random fields with learned spatial and temporal correlation~(\ref{apx:subsec:parameterization_features-stochastic}).
For training, we generate two ensemble members per forecast, which suffices for an unbiased estimate of CRPS.

\section*{Results}\label{sec:results}
We train a range of NeuralGCM models at horizontal resolutions with grid spacing of $2.8^{\circ}$, $1.4^{\circ}$, and $0.7^{\circ}$, each utilizing $32$ evenly spaced vertical levels on sigma coordinates. 
We evaluate the performance of NeuralGCM at a range of timescales appropriate for weather forecasting and climate simulation.
For weather, we compare against the best-in-class conventional physics-based weather models, ECMWF's high resolution model (ECMWF-HRES) and ensemble prediction system (ECMWF-ENS), and two of the recent ML-based approaches, GraphCast~\cite{lam2022graphcast} and Pangu~\cite{bi2023accurate}.
For climate, we compare against a Global Cloud Resolving Model and Atmospheric Model Inter-comparison Project (AMIP) runs.

\subsection*{Medium-range global weather forecasting \label{sec:weather_forecast}}

Our evaluation setup focuses on quantifying accuracy and physical consistency, following WeatherBench2~\cite{rasp2023weatherbench}.
We regrid all forecasts to a $1.5^{\circ}$ grid using conservative regridding, and average over all 732 forecasts made at noon and midnight UTC in the year $2020$, which was held-out from training data for all ML models.
NeuralGCM, GraphCast, and Pangu compare to ERA5 as ground-truth, whereas ECMWF ENS/HRES compare to the ECMWF operational analysis (i.e., HRES at 0-hour lead time), to avoid penalizing the operational forecasts for different biases than ERA5.

\begin{figure*}
\begin{center}
\makebox[\textwidth]{\colorbox{white}{\includegraphics[width=0.85\paperwidth]{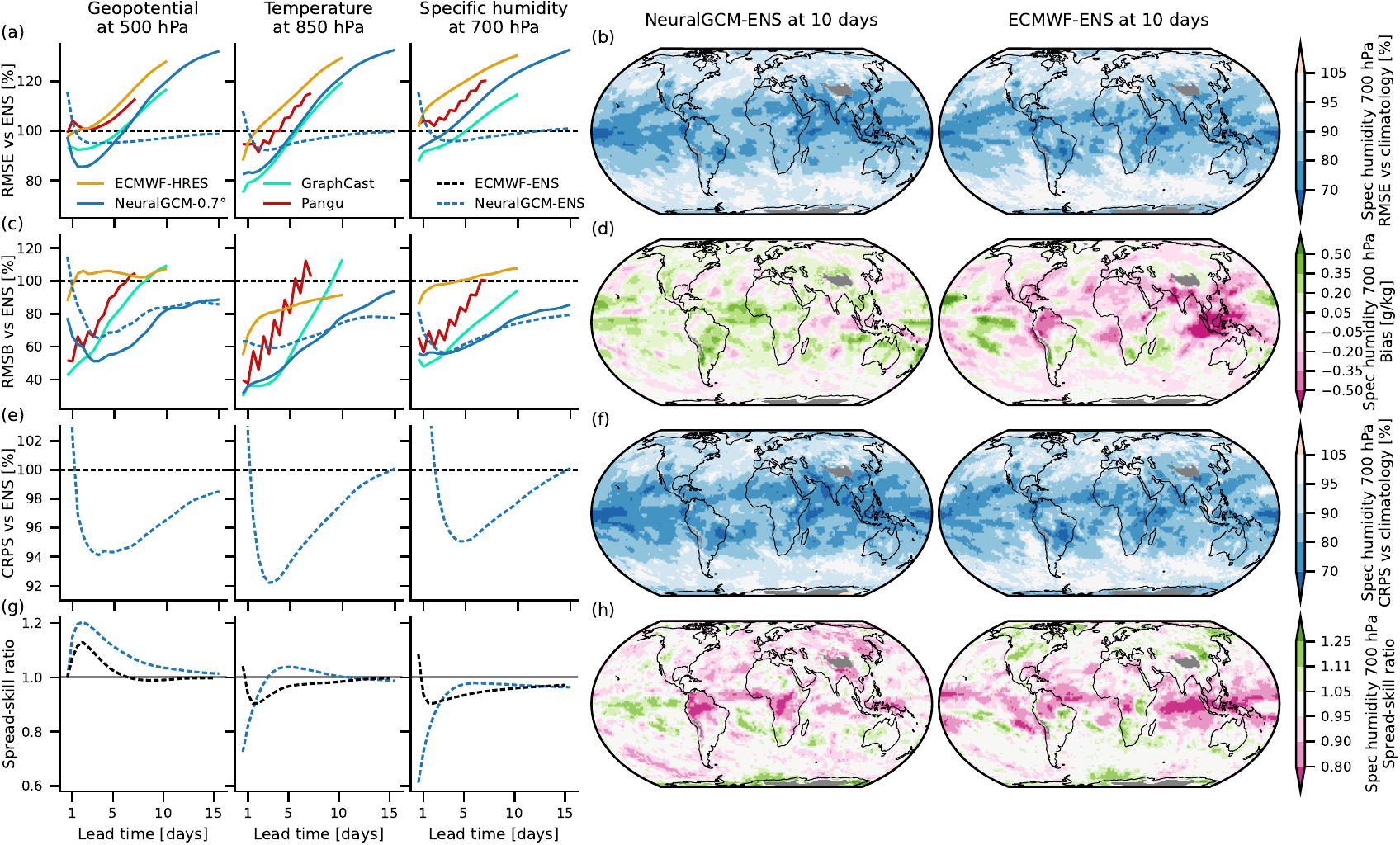}}}
\end{center}
\caption{Weather forecasting accuracy scores for deterministic and stochastic models. (a) RMSE and (c) RMSB for ECMWF-ENS, ECMWF-HRES, NeuralGCM-$0.7^{\circ}$, NeuralGCM-ENS, GraphCast~\cite{lam2022graphcast} and Pangu~\cite{bi2023accurate}  on the main WeatherBench variables, as a percent of the error of ECMWF-ENS. Deterministic and stochastic models are shown in solid and dashed lines respectively. (e) CRPS relative to ECMWF-ENS and (g) skill-spread ratio for ENS and NeuralGCM-ENS models. Spatial distributions of (b) RMSE, (d) Bias, (f) CRPS and (h) spread-skill ratio for NeuralGCM-ENS and ECMWF-ENS models for $10$-day forecasts of specific humidity at 700 hPa. Spatial plots of RMSE and CRPS are relative to a probabilistic climatology with an ensemble member for each of the years 1990-2019. Grey areas indicate regions where climatological surface pressure on average is below 700hPa. }\label{fig:accuracy_and_resolution}
\end{figure*}

\subsubsection*{Accuracy}

We use ECMWF's ensemble (ENS) model as a reference baseline as it achieves the best performance across the majority of lead times~\cite{rasp2023weatherbench}. We assess accuracy using (1) root mean squared error (RMSE), (2) bias, (3) continuous ranked probability score (CRPS) and (4) spread-skill ratio with results shown in Fig.~\ref{fig:accuracy_and_resolution}.
We provide more in-depth evaluations including scorecards, metrics for additional variables and levels, and maps in Appendix~\ref{apx:sec:additional_weather_evaluation}.

Deterministic models that produce a single weather forecast for given initial conditions can be compared effectively using RMSE skill at short lead times.
For the first $1$-$3$ days, depending on the atmospheric variable, RMSE is minimized by forecasts that accurately track the evolution of weather patterns.
At this timescale we find that NeuralCGM-$0.7^{\circ}$ and GraphCast achieve best results, with slight variations across different variables~(Fig.~\ref{fig:accuracy_and_resolution}a).
At longer lead times RMSE rapidly increases due to chaotic divergence of nearby weather trajectories, making RMSE less informative for deterministic models.
Root-mean-squared bias (RMSB) calculates persistent errors over time, which provides an indication of how models would perform at much longer lead times.
Here NeuralGCM models also compare favorably against previous approaches (Fig.~\ref{fig:accuracy_and_resolution}c), with notably much less bias for specific humidity in the tropics (Fig.~\ref{fig:accuracy_and_resolution}d).

Ensembles are essential for capturing intrinsic uncertainty of weather forecasts, especially at longer lead times.
Beyond about 7 days, the ensemble means of ECMWF-ENS and NeuralGCM-ENS forecasts have considerably lower RMSE than the deterministic models, indicating that these models better capture the average of possible weather.
A better metric for ensemble models is CRPS, which is a proper scoring rule that is sensitive to full marginal probability distributions~\cite{Gneiting2007ProperScoring}.
Our stochastic model (NeuralGCM-ENS) running at $1.4^{\circ}$ resolution has lower error compared to ECMWF-ENS across almost all variables, lead times and vertical levels for ensemble-mean RMSE, RSMB and CRPS (Fig.~\ref{fig:accuracy_and_resolution}a,c,e and Appendix~\ref{apx:sec:additional_weather_evaluation}), with similar spatial patterns of skill~(Fig.~\ref{fig:accuracy_and_resolution}b,f).
Like ECMWF-ENS, NeuralGCM-ENS has a spread-skill ratio of approximately one (Fig.~\ref{fig:accuracy_and_resolution}d), which is a necessary condition for calibrated forecasts~\cite{Fortin2014-spread-skill}.

\begin{figure*}
\begin{center}
\makebox[\textwidth]{\colorbox{white}{\includegraphics[width=0.85\paperwidth]{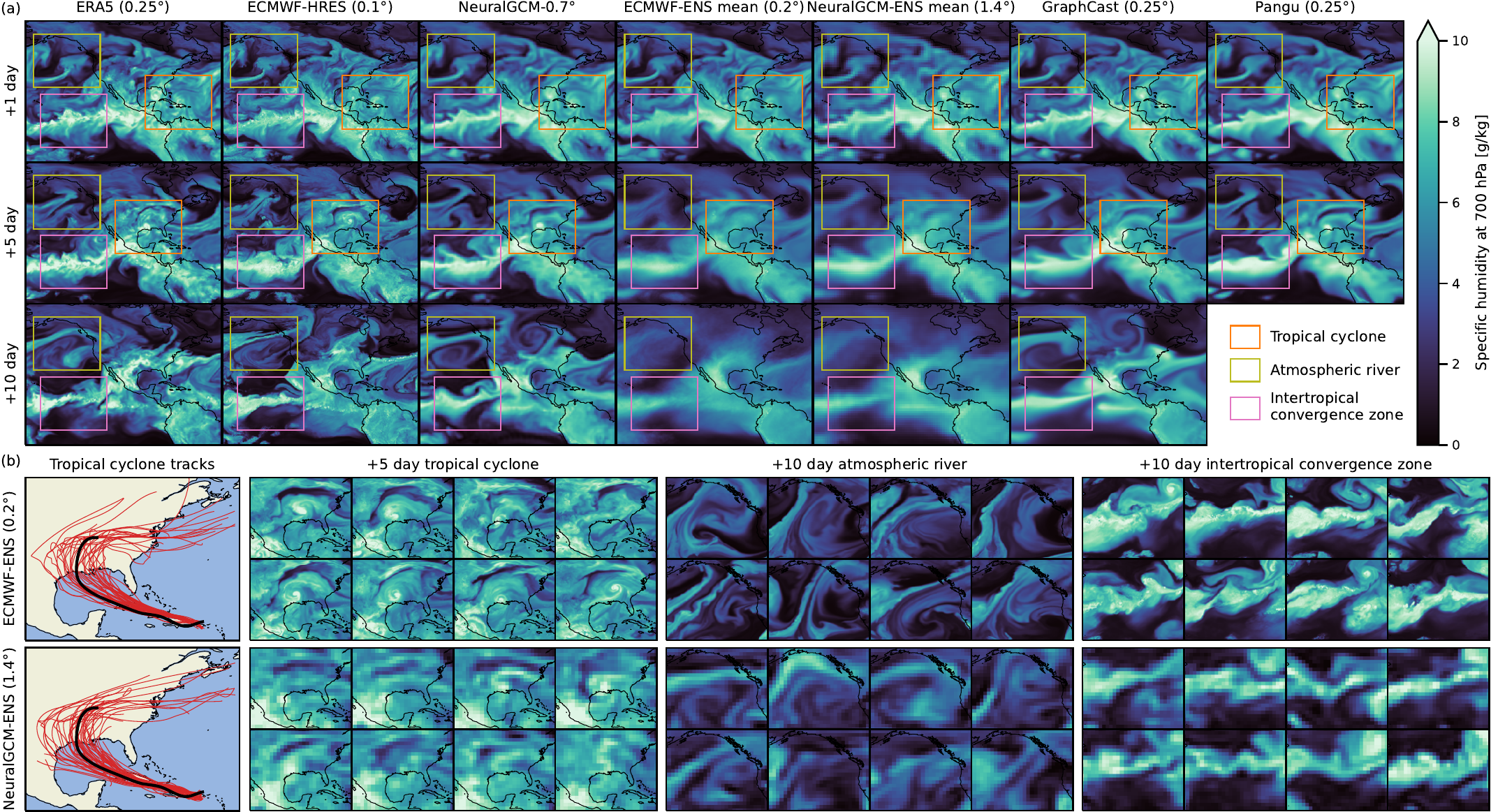}}}
\end{center}
\caption{Case study of a medium-range weather forecast. All forecasts are initialized at 2020-08-22T12z, chosen to highlight Hurricane Laura, the most damaging Atlantic hurricane of 2020.
(a) Specific humidity at 700 hPa for 1-day, 5-day and 10-day forecasts over North America and the North-East Pacific ocean from ERA5, ECMWF-HRES, NeuralGCM-$0.7^{\circ}$, ECMWF-ENS (mean), NeuralGCM-ENS (mean), GraphCast~\cite{lam2022graphcast} and Pangu~\cite{bi2023accurate}. 
(b) Forecasts from individual ensemble members from ECMWF-ENS and NeuralGCM-ENS over regions of interest, including predicted tracks of Hurricane Laura from each of the 50 ensemble members (Appendix \ref{apx:sec:TC_tracking}). The track from ERA5 is plotted in black.
}\label{fig:case-study}
\end{figure*}

\subsubsection*{Physical consistency}

\textbf{Case study}. 
An important characteristic of forecasts is their resemblance to realistic weather patterns. Figure~\ref{fig:case-study} shows a case study that illustrates the performance of NeuralGCM on three types of important weather phenomena: tropical cyclones, atmospheric rivers and the inter-tropical convergence zone.
The top panel shows that all the ML models make significantly blurrier forecasts than source data ERA5 and physics-based ECMWF-HRES forecast, but NeuralCGM-$0.7^{\circ}$ outperforms the pure ML models, despite its coarser resolution (0.7$^{\circ}$ vs 0.25$^{\circ}$ for GraphCast and Pangu).  
Blurry forecasts correspond to physically inconsistent atmospheric conditions, and misrepresent extreme weather.
Similar trends hold for other derived variables of meteorological interest~(Appendix~\ref{apx:subsec:derived_variables_and_consistency}).
Ensemble mean predictions, both from NeuralGCM and ECMWF, are closer to ERA5 in an average sense, and thus are inherently smooth at long lead times.
In contrast, as shown in the bottom panel and in Appendix~\ref{apx:subsec:ensemble_visualization}, individual realizations from the ECMWF and NeuralGCM ensembles remain sharp, even at long lead times.
Like ECMWF-ENS, NeuralGCM-ENS produces a statistically representative range of future weather scenarios for each weather phenomena, despite its $8\times$ coarser resolution.

\textbf{Spectra}.
We can quantify the blurriness of different forecast models via their power spectra. Appendix Fig.~\ref{fig:physical_consistency_thermo} and Fig.~\ref{fig:physical_consistency_wind} show that the power spectra of NeuralCGM-$0.7^{\circ}$ is consistently closer to ERA5 than the other ML forecast methods, but is still blurrier than ECMWF's physical forecasts.
The spectra of NeuralGCM forecasts is also roughly constant over the forecast period, in stark contrast to GraphCast, which worsens with lead time.
The spectrum of NeuralGCM becomes more accurate with increased resolution (\ref{sifig:multiple_resolutions_spectra}), which suggests the potential for further improvements of NeuralGCM models trained at higher resolutions.

\textbf{Water budget}.
In NeuralGCM, advection is handled by the dynamical core, while the ML parameterization models local processes within vertical columns of the atmosphere. Thus, unlike pure ML methods, local sources and sinks can be isolated from tendencies due to horizontal transport and other resolved dynamics.
This makes our results more interpretable and facilitates the diagnosis of the water budget.
Specifically, we diagnose precipitation minus evaporation (P-E, see \ref{apx:sec:diagnose_precipitation}) rather than directly predicting these as in ML-based approaches \cite{lam2022graphcast}.
For short weather forecasts, the mean of P-E has a realistic spatial distribution that is very close to ERA5 data (Fig.~\ref{sifig:P_minus_E_short_term}c-e).
The P-E rate distribution of NeuralGCM-$0.7^\circ$ closely matches ERA5 distribution in the extratropics (Fig.~\ref{sifig:P_minus_E_short_term}b), though it underestimates extreme events in the tropics (Fig.~\ref{sifig:P_minus_E_short_term}a).
Note that the current version of NeuralGCM directly predicts tendencies for an atmospheric column, and thus cannot distinguish between precipitation and evaporation.

\textbf{Geostrophic wind balance}.
We examined the extent to which NeuralGCM, GraphCast, and ECMWF-HRES capture the geostrophic wind balance, the near-equilibrium between the dominant forces that drive large-scale dynamics in the mid-latitudes \cite{holton2004introduction}.
A recent study~\cite{bonavita2023limitations} highlighted that Pangu misrepresents the vertical structure of the geostrophic and ageostrophic winds, and noted a deterioration at longer lead times.
Similarly, we observe that GraphCast exhibits an error that worsens with lead time.
In contrast, NeuralGCM more accurately depicts the vertical structure of the geostrophic and ageostrophic winds, as well as their ratio, compared to GraphCast across various rollouts, when compared against ERA5 data (Fig.~\ref{sifig:geostrophic_wind_vertical}).
However, ECMWF-HRES still exhibits a slightly closer alignment to ERA5 data than NeuralGCM does. Within NeuralGCM, the representation of the geostrophic wind's vertical structure only slightly degrades in the initial few days, showing no noticeable changes thereafter, particularly beyond day 5.

\textbf{Generalizing to unseen data}.
Physically consistent weather models should still perform well for weather conditions for which they were not trained.
To test generalization in the context of weather,
we compare versions of NeuralCGM-$0.7^{\circ}$ and GraphCast trained through 2017 on five years of weather forecasts beyond the training period (2018-2022) in Fig.~\ref{sifig:5years}.
Unlike GraphCast, NeuralGCM does not show a clear trend of increasing error when initialized further into the future from the training data.
To extend this test beyond five years, we trained  a NeuralGCM-$2.8^{\circ}$ model using data before 2000, and tested its skill for over 21 unseen years (Fig.~\ref{sifig:extrapolation_temperature_wind}).

\subsection*{Climate Simulations}\label{sec:climate_simulations}
Although our deterministic NeuralGCM models are trained to predict weather up to $3$ days ahead, they are generally capable of simulating the atmosphere far beyond medium-range weather timescales.
For extended climate simulations we prescribe historical sea surface temperature (SST) and sea ice concentration. These simulations feature many emergent phenomena of the atmosphere on timescales from months to decades.

For climate simulations with NeuralGCM, we use $2.8^{\circ}$ and $1.4^{\circ}$ deterministic models, which are relatively inexpensive to train (Appendix~\ref{apx:subsec:training_times}) and
allow us to explore a larger parameter space to find stable models.
Previous studies found that running extended simulations with hybrid models is challenging due to numerical instabilities and climate drift \cite{brenowitz2019spatially}.
To quantify stability in our selected models, we run multiple initial conditions and report how many of them finish without instability.

\subsubsection*{Seasonal cycle and emergent phenomena}
To assess the capability of NeuralGCM to simulate various aspects of the seasonal cycle, we run two-year simulations with NeuralGCM-$1.4^{\circ}$. for $37$ different initial conditions spaced every $10$ days for the year 2019.
Out of these 37 initial conditions, $35$ successfully complete the full two years without instability; for case studies of instability see Sec.~\ref{apx:sec:instability_case}.
We compare results from NeuralGCM-$1.4^\circ$ for 2020 to ERA5 data and to outputs from GFDL's X-SHiELD global cloud resolving model, which is coupled to an ocean model nudged towards reanalysis~\cite{cheng2022impact}.
This X-SHiELD run has been used as a target for training ML climate models~\cite{kwa2023machine}.
For comparison we evaluate models after regridding predictions to $1.4^{\circ}$ resolution.
This comparison slightly favors NeuralGCM because NeuralGCM was tuned to match ERA5, but the discrepancy between ERA5 and the actual atmosphere is small relative to model error.

\begin{figure*}
\centering
\makebox[\textwidth]{\colorbox{white}{\includegraphics[width=0.85\paperwidth]{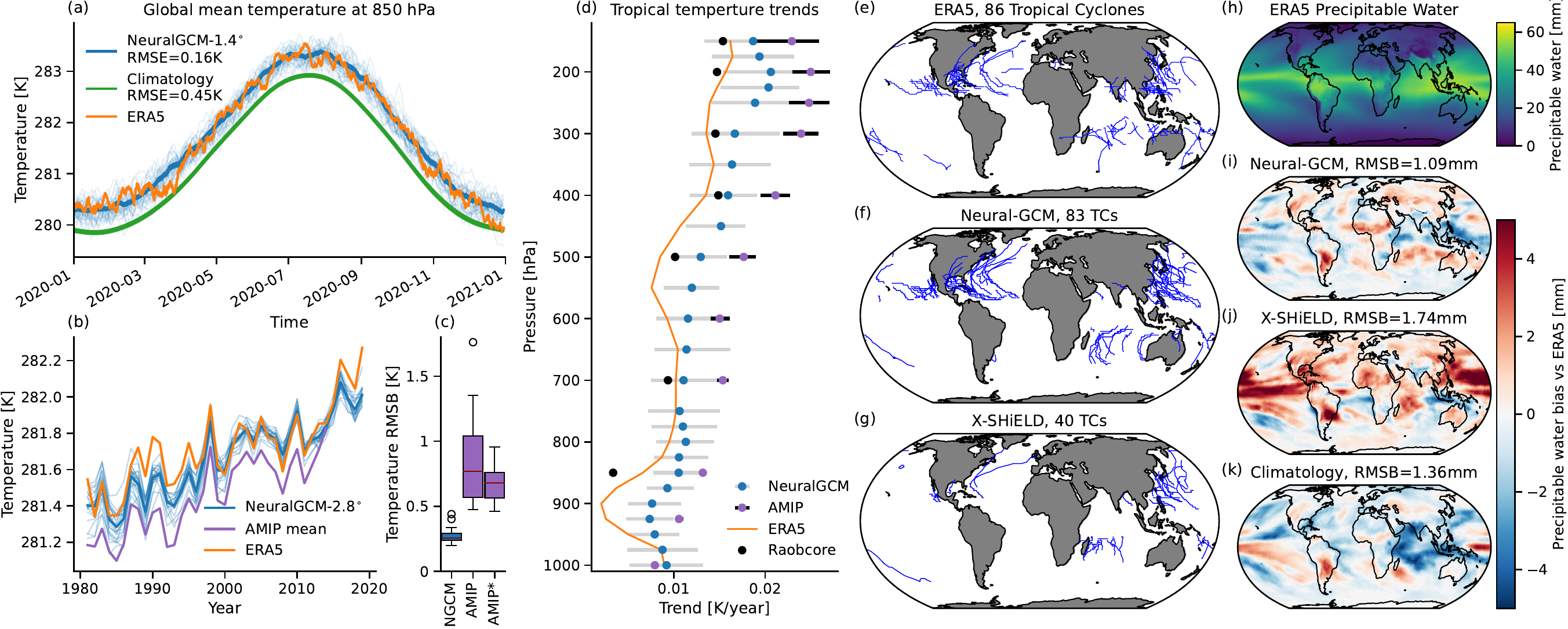}}}
\caption{Simulation of climate with NeuralGCM. (a) Global mean temperature for ERA5 for 2020 (orange), climatology (defined as the averaged temperature between 1990-2019; green), and for NeuralGCM-$1.4^{\circ}$ for 2020 for 35 simulations initialized every 10 days during 2019 (thick blue represents the ensemble mean; thin blue lines indicate different initial conditions).
(b) Yearly global mean temperature for ERA5 (orange), mean over 22 CMIP6 AMIP experiments for 1981-2014 (purple; model details are found in Appendix~\ref{apx:sec:AMIP_models_used}), and NeuralGCM-$2.8^{\circ}$ for 22 AMIP-like simulations with prescribed SST initialized every 10 days during 1980 (thick blue represents the ensemble mean, and thin blue lines indicate different initial conditions).
(c) The root mean square bias (RMSB) of the 850hPa temperature averaged between 1981-2014 for 22 NeuralGCM-$2.8^\circ$ AMIP runs (labeled NGCM), 22 CMIP6 AMIP experiments (labeled AMIP) and debiased 22 CMIP6 AMIP experiments (labeled as AMIP*; biased was removed by removing the 850hPa global temperature bias). In the box plots, the red line represents the median; the box delineates the first to third quartiles; the whiskers extend to 1.5 times the interquartile range (Q1 - 1.5IQR and Q3 + 1.5IQR), and outliers are shown as individual dots.
(d) Vertical profiles of tropical (20S-20N) temperature trends for the period 1981-2014. The orange line shows ERA5 reanalysis, and the black dots show the trends calculated from Radiosonde Observation Correction using Reanalyses~\cite{haimberger2008toward}. The blue dots shows the mean trends for NeuralGCM-$2.8^{\circ}$ 22 AMIP-like runs  and purple dots are the mean trends from CMIP6 AMIP runs (see appendix~\ref{apx:sec:AMIP_models_used} for full list of models used), where the grey (black) whiskers show the 25th and 75th percentiles for NeuralGCM-$2.8^{\circ}$ (CMIP6 AMIP runs).
(e,f,g) Tropical cyclone tracks for (d) ERA5, (e) NeuralGCM-$1.4^{\circ}$ and (f) X-SHiELD. 
(h) Mean precipitable water for ERA5 and the precipitable water bias in
(i) NeuralGCM-$1.4^{\circ}$; initialized 90 days before mid-January 2020 similarly to X-SHiELD, 
(j) X-SHiELD and (j) climatology (averaged between 1990-2019).
In panels d-i quantities are calculated between mid-January 2020 and mid-January 2021 (when X-SHiELD model data is available) and  all models were regridded to a 256x128 Gaussian grid before computation and tracking.}\label{fig:climatology}
\end{figure*}

Figure~\ref{fig:climatology}a displays the temporal variation of the global mean temperature throughout 2020, as captured by 35 simulations from NeuralGCM, in comparison to the ERA5 reanalysis and standard climatology benchmarks. The seasonality and variability of the global mean temperature from NeuralGCM are quantitatively similar to those observed in ERA5. The ensemble mean temperature RMSE for NeuralGCM stands at 0.16K when benchmarked against ERA5, which is a significant improvement over the climatology's RMSE of 0.45K. We find that NeuralGCM accurately simulates the seasonal cycle, as evidenced by metrics such as the annual cycle of the global precipitable water (Fig.~\ref{sifig:TKE_2020}a) and global total kinetic energy (Fig.~\ref{sifig:TKE_2020}b). Furthermore, the model captures essential atmospheric dynamics, including the Hadley circulation and the zonal-mean zonal wind (Fig.~\ref{sifig:Hadley_cell_u_wind}), as well as the spatial patterns of eddy kinetic energy in different seasons (Fig.~\ref{sifig:EKE_2020}), and the distinctive seasonal behaviors of monsoon circulation (Fig.~\ref{sifig:monsoon_2020}; additional details are provided in \ref{apx:sec:seasonal cycle}).

Next, we compare the annual biases of a single NeuralGCM realization with a single realization of X-SHiELD (the only one available), both initiated in mid-October 2019. We consider 19 January 2020 to 17 January 2021, the time frame for which X-SHiELD data is available. Global cloud-resolving models, such as X-SHiELD, are considered state-of-the-art, especially for simulating the hydrological cycle, due to their resolution being capable of resolving deep convection \cite{stevens2019dyamond}.
The annual bias in precipitable water for NeuralGCM (RMSE of 1.09 mm) is substantially smaller than the biases of both X-SHiELD (RMSE of 1.74 mm) and climatology (RMSE of 1.36 mm; Fig.~\ref{fig:climatology}i-k). Moreover, NeuralGCM shows a lower temperature bias in the upper and lower troposphere than X-SHiELD (Fig.~\ref{sifig:temperature_bias_200_850_2020}).
We also indirectly compare precipitation bias in X-SHiELD with precipitation minus evaporation bias in NeuralGCM-$1.4^{\circ}$, which shows slightly larger bias and grid-scale artifacts for NeuralGCM (Fig.~\ref{sifig:precip_climate_2020}).

Finally, to assess NeuralGCM capability of generating tropical cyclones (TCs) in an annual model integration, we use the TC tracker TempestExtremes \cite{ullrich2021tempestextremes}, as described in Appendix~\ref{apx:sec:TC_tracking}.
Figure~\ref{fig:climatology}e,f,g shows that NeuralGCM, even at a coarse resolution of $1.4^{\circ}$, produces realistic trajectories and counts of TCs (83 versus 86 in ERA5 for the corresponding period), while X-SHiELD, when regridded to $1.4^{\circ}$ resolution, substantially underestimates the TC count (40).
Additional statistical analyses of TCs can be found in Figs.~\ref{sifig:TCs_34_years} and \ref{sifig:TCs_wind_34_years}.

\subsubsection*{Decadal simulations - AMIP-like experiments}

To assess the capability of NeuralGCM to simulate historical temperature trends,
we conduct AMIP-like simulations over a duration of 40 years with NeuralGCM-$2.8^{\circ}$.
Out of 37 different runs with initial conditions spaced every 10 days during the year 1980, 22 simulations were stable for the entire 40-year period, and our analysis focuses on these results.
We compare to 22 simulations run with prescribed SST from the Coupled Model Intercomparison Project Phase 6 \cite{eyring2016overview}, listed in Appendix~\ref{apx:sec:AMIP_models_used}. 

We find that all 40-year simulations of NeuralGCM, as well as the mean of the 22 AMIP runs, accurately capture the global warming trends observed in ERA5 data (Fig.~\ref{fig:climatology}b). There is a strong correlation in the year-to-year temperature trends with ERA5 data, suggesting that NeuralGCM effectively captures the impact of SST forcing on climate.
When comparing spatial biases averaged over 1981-2014 we find that all 22 NeuralGCM-$2.8^{\circ}$ runs have smaller bias than CMIP6 AMIP runs, and this result remains even when removing the global temperature bias in CMIP6 AMIP runs (Figs.~\ref{fig:climatology}c, \ref{sifig:AMIP_1981_2014_bias_CESM_ens} and \ref{sifig:AMIP_1981_2014_bias_ngcm}).

Next, we investigated the vertical structure of tropical warming trends, which climate models tend to overestimate in the upper troposphere~\cite{mitchell2020vertical}.
As shown in Fig.~\ref{fig:climatology}d, the trends, calculated by linear regression, of NeuralGCM are closer to ERA5 than those of AMIP runs.
In particular, the bias in the upper troposphere is reduced.
However, NeuralGCM does show a wider spread in its predictions than the AMIP runs, even at levels near the surface where temperatures are typically more constrained by prescribed SST.

Lastly, we evaluated NeuralGCM's capability to generalize to unseen warmer climates by conducting AMIP simulations with increased SST (Section~\ref{apx:subsec:climate_extrapolation}). We find that NeuralGCM exhibits some of the robust features of climate warming response to modest SST increases (+1K and +2K), however, for more substantial SST increases (+4K), NeuralGCM's response diverges from expectations (Fig.~\ref{sifig:extrapolation_temperature_wind}).
Additionally, AMIP simulations with increased SST exhibit climate drift, underscoring NeuralGCM's limitations in this context (Fig.~\ref{sifig:extrapolation_temperature_treds_AMIP_SST_warming}).

\section*{Discussion \label{sec:discussion}}

NeuralGCM is a differentiable hybrid atmospheric model that combines the strengths of traditional GCMs with machine learning for weather forecasting and climate simulation. 
NeuralGCM is the first ML-based model to make accurate ensemble weather forecasts, with better CRPS than state-of-the-art physics-based models.
It is also the first hybrid model that achieves comparable spatial bias to global cloud resolving models, can simulate realistic tropical cyclone tracks, and that can run AMIP-like simulations with realistic historical temperature trends.
Overall, NeuralGCM demonstrates that incorporating machine learning is a viable alternative to building increasingly detailed physical models~\cite{stevens2019dyamond} for improving GCMs.

Compared to traditional GCMs with similar skill, NeuralGCM is compute efficient and low-complexity.
NeuralGCM runs at 8-40$\times$ coarser horizontal resolution than ECMWF IFS and global cloud resolving models, which enables 3-5 orders of magnitude savings in compute.
For example, NeuralGCM-$1.4^{\circ}$ simulates \num{70000} simulation days in $24$ hours using a single TPU vs 19 simulated days on \num{13824} CPU cores with X-SHiELD.
This can be leveraged for previously impractical tasks like large ensemble forecasting.
NeuralGCM's dynamical core uses global spectral methods~\cite{Bourke1974-spectral}, and learned physics is parameterized with fully-connected neural networks acting on single vertical columns.
Substantial headroom exists to pursue higher accuracy using advanced numerical methods and ML architectures.

Our results provide strong evidence for the disputed hypothesis \cite{ruiz2013estimating,schneider2017earth,Schneider2024opinion} that learning to predict short-term weather is an effective way to tune parameterizations for climate.
NeuralGCM models trained on 72-hour forecasts are capable of realistic multi-year simulation.
When provided with historical sea surface temperatures, they capture essential atmospheric dynamics such as seasonal circulation, monsoons, and tropical cyclones.
However, we will likely need alternative training strategies~\cite{schneider2017earth,Schneider2024opinion} to learn important processes for climate with subtle impact on weather timescales, such as a cloud feedback.

The NeuralGCM approach is compatible with incorporating either more physics or more ML, as required for operational weather forecasts and climate simulations.
For weather forecasting, we expect that end-to-end learning~\cite{sutskever2014sequence} with observational data will allow for better and more relevant predictions, including key variables such as precipitation.
Such models could include neural networks acting as corrections to traditional data assimilation and model diagnostics.
For climate projection, NeuralGCM will need to be reformulated to enable coupling with other Earth system components (e.g., ocean, land), and integrating data on the atmospheric chemical composition (e.g., greenhouse gasses, aerosols).
There are also research challenges common to current ML-based climate models \cite{Bretherton2023-ym}, including the capability to simulate unprecedented climates (i.e., generalization), adhering to physical constraints, and resolving numerical instabilities and climate drift.
NeuralGCM’s flexibility to incorporate physics-based models (e.g., radiation) offers a promising avenue to address these challenges.

Models based on physical laws and empirical relationships are ubiquitous in science.
We believe the differentiable hybrid modeling approach of NeuralGCM has the potential to transform simulation for a wide range of applications, such as materials discovery, protein folding, and multiphysics engineering design.

\textbf{Acknowledgments}.
We thank Anna Kwa, Alex Merose, and Kunal Shah for assistance with data acquisition and handling,
Leonardo Zepeda-Núñez for feedback on the manuscript,
and
John Anderson, Christopher Van Arsdale, Rei Chemke, Gideon Dresdner, Justin Gilmer, Jason Hickey, Nicholas Lutsko, Grey Nearing, Adam Paszke, John Platt, Sameera Ponda, Mike Pritchard, Daniel Rothenberg, Fei Sha, Tapio Schneider, and Octavian Voicu for productive discussions.

\section*{Data availability}
For training and evaluating NeuralGCM models we used publicly available ERA5 dataset~\cite{hersbach2020era5}, originally downloaded from \url{https://cds.climate.copernicus.eu/} and available via Google Cloud Storage in Zarr format at \path{gs://gcp-public-data-arco-era5/ar/full_37-1h-0p25deg-chunk-1.zarr-v3}.
To compare NeuralGCM to operational and data-driven weather models we used forecast datasets distributed as part Weatherbench2~\cite{rasp2023weatherbench} at \url{https://weatherbench2.readthedocs.io/en/latest/data-guide.html}, to which we have added NeuralGCM forecasts for 2020.
To compare NeuralGCM to atmospheric models in climate settings we used CMIP6 data available at \url{https://catalog.pangeo.io/browse/master/climate/}, as well as X-SHiELD \cite{kwa2023machine} outputs available on Google Cloud storage in a ``requester pays'' bucket at  \path{gs://ai2cm-public-requester-pays/C3072-to-C384-res-diagnostics}.
The Radiosonde Observation Correction using Reanalyses (RAOBCORE) V1.9 that was used as reference tropical temperature trends was downloaded from
\url{https://webdata.wolke.img.univie.ac.at/haimberger/v1.9/}.

\section*{Code availability}
NeuralGCM code base is separated into two open source projects: Dinosaur and NeuralGCM, both publicly available on GitHub at \url{https://github.com/google-research/dinosaur} and \url{https://github.com/google-research/neuralgcm}. The Dinosaur package implements a differentiable dynamical core used by NeuralGCM, while the NeuralGCM package provides ML models and checkpoints of trained models.
Evaluation code for NeuralGCM weather forecasts is included in WeatherBench2~\cite{rasp2023weatherbench}, available at \url{https://github.com/google-research/weatherbench2}.

\newpage

\begin{appendices}
\renewcommand{\contentsname}{Appendices}
\localtableofcontents
\setcounter{figure}{4} %

\clearpage
\section{Lines of code in atmospheric models}\label{apx:sec:loc}

To measure of the complexity of different code-bases, we counted the number of ``core model'' lines of code, excluding files devoted to tests, examples, input/output, coupling between different frameworks and running models.

\subsection{fv3atm}

We counted a total of \num{376578} lines of core model code for fv3atm, the atmospheric component of the NOAA's Unified Forecast System (UFS) weather model, which we downloaded from \url{https://github.com/NOAA-EMC/fv3atm} on 7 November, 2023.

Excluding files with ``test'' or ``example'' in their paths, we counted the following number of lines in Fortran files in the two major modules of fv3atm (ending in .f or .f90):

\begin{enumerate}
    \item \num{42387} lines in the FV3 dynamical core (atmos\_cubed\_sphere/model)
    \item \num{334191} lines in the Common Community Physics Package (CCPP) Physics (ccpp/physics/physics)
\end{enumerate}

\subsection{GraphCast}

We counted a total of \num{5417} lines of core model code for GraphCast~\cite{lam2022graphcast}, which we downloaded from \url{https://github.com/deepmind/graphcast} on 7 November, 2023.

\subsection{NeuralGCM}

We counted a total of \num{20136} lines of core model code for NeuralGCM, split between two major modules:

\begin{enumerate}
    \item \num{8609} lines in the spectral dynamical core
    \item \num{11527} lines in the machine learning code
\end{enumerate}

\clearpage
\section{Dynamical core of NeuralGCM}\label{apx:sec:dycore}
The dynamical core provides NeuralGCM with strong physics priors based on well understood and easy to simulate phenomena. In section \ref{apx:subsec:dycore_discretization} we provide more details on spatial discretization of the atmospheric state in NeuralGCM. In section \ref{apx:subsec:dycore_equations} we summarize the governing equations of the dynamical core. In section \ref{apx:subsec:dycore_numerics} we provide references to numerical implementations and rationale for our choices.

\subsection{Discretization of the dynamical core}\label{apx:subsec:dycore_discretization}
Our dynamical core uses a Gaussian grid and sigma coordinates \cite{Bourke1974-spectral} to discretize the computational domain. Gaussian grids enable fast and accurate transformations between the grid space representation and spherical harmonics basis. They result in equiangular longitude lines and unequal spacing latitudes defined by the Gaussian quadrature. Terrain-following sigma coordinates discretize the vertical direction by the fraction of the surface pressure, and thus correspond to non-stationary vertical height since surface pressure changes with time. Cell boundaries in sigma coordinates take values $\sigma \in \left[0,1\right]$, with $\sigma=0$ corresponding to the top of the atmosphere ($p=0$ pressure boundary) and $\sigma=1$ representing the earth's surface.

In this work we have trained a lineup of models that make forecasts at varying horizontal resolutions: $2.8^{\circ}$, $1.4^{\circ}$, and $0.7^{\circ}$, corresponding to truncated linear Gaussian grids TL$63$, TL$127$, TL$255$. The number in the grid name corresponds to the maximum total wavenumber of spherical harmonic that the grid can represent. These grids provide a framework for transforming data from grid space (nodal) to spherical harmonic representations with minimal loss of information. When solving model equations we use cubic truncation Gaussian grids T$62$, T$125$ and T$253$, that capture a similar number of spherical harmonics, while avoiding aliasing errors and minimizing the need to increase array dimensions above a multiple of 128, which is expensive on the Google TPU. See Table \ref{apx:table:gaussian_grids} for resolution details. All models use $32$ equidistant sigma levels for vertical discretization. We suspect that using higher vertical resolution with assimilation data from more levels could further improve the performance.

\begin{table}[ht]
\begin{tabular}{|c|c|c|c|}
\hline
\textbf{Grid name} & \textbf{Longitude nodes} & \textbf{Latitude nodes} & \textbf{Max total wavenumber} \\ \hline
TL63               & $128$                      & $64$                      & $63$                            \\ \hline
TL127              & $256$                      & $128$                     & $127$                           \\ \hline
TL255              & $512$                      & $256$                     & $255$                           \\ \hline
T62                & $190$                      & $95$                      & $62$                            \\ \hline
T125               & $379$                      & $190$                     & $125$                           \\ \hline
T254               & $766$                      & $383$                     & $254$                           \\ \hline
\end{tabular} \label{apx:table:gaussian_grids}
\caption{Spatial and spectral resolutions of horizontal grids used by NeuralGCM.}
\end{table}

\subsection{Primitive equations}\label{apx:subsec:dycore_equations}
The dynamical core of NeuralGCM solves the primitive equations, which represent a combination of (1) momentum equations, (2) the second law of thermodynamics, (3) a thermodynamic equation of state (ideal gas), (4) continuity equation and (5) hydrostatic approximation. For solving the equations we use a divergence-vorticity representation of the horizontal winds, resulting in equations for the following seven prognostic variables: divergence $\delta$, vorticity $\zeta$, temperature $T$, logarithm of the surface pressure $\log p_{s}$, as well as $3$ moisture species (specific humidity $q$, specific cloud ice $q_{c_i}$ and specific liquid cloud water content $q_{c_l}$). To facilitate efficient time integration of our models we split temperature $T$ into a uniform reference temperature on each sigma level $\bar{T}_{\sigma}$ and temperature deviations per level $T^{\prime}_{\sigma} = T_\sigma - \bar{T}_{\sigma}$. The resulting equations are:
\begin{align}
\begin{split}
&\pd{\zeta}{t} =-\nabla\times\left((\zeta + f)\mathbf k\times\mathbf u
    +\dot\sigma\frac{\partial\mathbf u}{\partial\sigma} + RT^\prime\nabla\log p_s \right) \\
&\pd{\delta}{t} =-\nabla\cdot\left((\zeta + f)\mathbf k\times\mathbf u
    +\dot\sigma\frac{\partial\mathbf u}{\partial\sigma} + RT^\prime\nabla\log p_s \right)
    -\nabla^2\left(\frac{||\mathbf u||^2}{2} + \Phi + R \bar{T}\log p_s \right) \\
&\pd{T}{t} =-\mathbf u\cdot\nabla T -\dot\sigma \frac{\partial T}{\partial \sigma}
    +\frac{\kappa T\omega}{p} =-\nabla\cdot \mathbf u T^\prime + T^\prime\delta
    -\dot\sigma \frac{\partial T}{\partial \sigma} +\frac{\kappa T\omega}{p} \\
&\pd{q_i}{t} =-\nabla\cdot \mathbf{u}q_{i} + q_{i}\delta -\dot{\sigma}\pd{q_{i}}{\sigma}  \\
&\pd{\log p_s}{t} =-\frac{1}{p_s}\int_0^1\nabla\cdot(\mathbf up_s)\,d\sigma
    =-\int_0^1\left(\delta +\mathbf u\cdot\nabla\log p_s\right)\,d\sigma
\label{eq:primitive_equations}
\end{split}
\end{align}
with horizontal velocity vector $\mathbf{u}=\nabla(\Delta^{-1}\delta) + \mathbf{k} \times\nabla(\Delta^{-1}\zeta)$, Coriolis parameter $f$, upward-directed unit vector parallel to the z-axis $\mathbf{k}$, ideal gas constant $R$, heat capacity at constant pressure $C_{p}$,  $\kappa= \frac{R}{C_{p}}$, diagnosed vertical velocity in sigma coordinates $\dot{\sigma}$, diagnosed change in pressure of a fluid parcel $\omega \equiv \frac{dp}{dt}$, diagnosed geopotential $\Phi$, diagnosed virtual temperature $T_{\nu}$ and each moisture species denoted as $q_{i}$.

Diagnostic quantities are computed as follows:
\begin{align}
    \dot\sigma_{k + \frac{1}{2}} &= -\sigma_{k + \frac{1}{2}}\frac{\partial\log p_s}{\partial t} -\frac{1}{p_s}\int_0^{\sigma_{k + \frac{1}{2}}} \nabla\cdot(p_s\mathbf u)\, d\sigma \\
\frac{\omega_k}{p_s\sigma_k}
&= \mathbf u_k\cdot\nabla \log p_s
-\frac{1}{\sigma_k}\int_0^{\sigma_k}\left(\delta + \mathbf u\cdot\nabla\log p_s \right)\,d\sigma \\
    \Phi_k &= \Phi_{s} + R\int_{\log \sigma_k}^{0} T_{\nu}\,d\log\sigma \label{apx:eq:diagnostic_variables} \\
    T_{\nu} &= T(1 + \left(\frac{R_{vap}}{R} - 1 \right)q - q_{c_{i}} - q_{c_{l}})
\end{align}
where $\Phi_{s}=gz_{s}$ is the geopotential at the surface.

\subsection{Numerics}\label{apx:subsec:dycore_numerics}
Our choice of the numerical schemes for interpolation, integrals and diagnostics exactly follows Durran's book \cite{durran2010numerical} $\S8.6$, with the addition of moisture species (which are advected by the wind and only affect the dynamics through through their effect on the virtual temperature). We use semi-implicit time-integration scheme, where all right hand side terms are separated into groups that are treated either explicitly or implicitly. This avoids severe time step limitations due to fast moving gravity waves.

Our choice of dynamical core was also informed by our desire to run efficiently on machine learning accelerators, in particular Google TPUs~\citeSI{Jouppi2023-rw}.
TPUs have dedicated hardware for low-precision matrix-matrix multiplication, which conveniently is well suited for the bottleneck in spectral methods, which are forward and inverse spherical harmonic transformations. Accordingly, we use single-precision arithmetic throughout.
We found that full single precision for spherical harmonic transformations was not required to obtain accurate results even on our largest grid sizes, and according use only three passes of bfloat16 matrix-multiplication rather than the six passes that would required for full single precision~\citeSI{Henry2019-it}.
Our implementation supports parallelism across spatial dimensions (x, y, and z) for running on multiple accelerator cores, using XLA SPMD~\citeSI{Xu2021-fe}, with JAX's \texttt{shard\_map} for parallelizing key bottlenecks including matrix-multiplications in spherical harmonic transforms~\citeSI{shard_map}.

\clearpage
\section{Learned physics of NeuralGCM}\label{apx:sec:parameterization}

\begin{figure*}[b]
\centering
\includegraphics[width=1.0\textwidth]{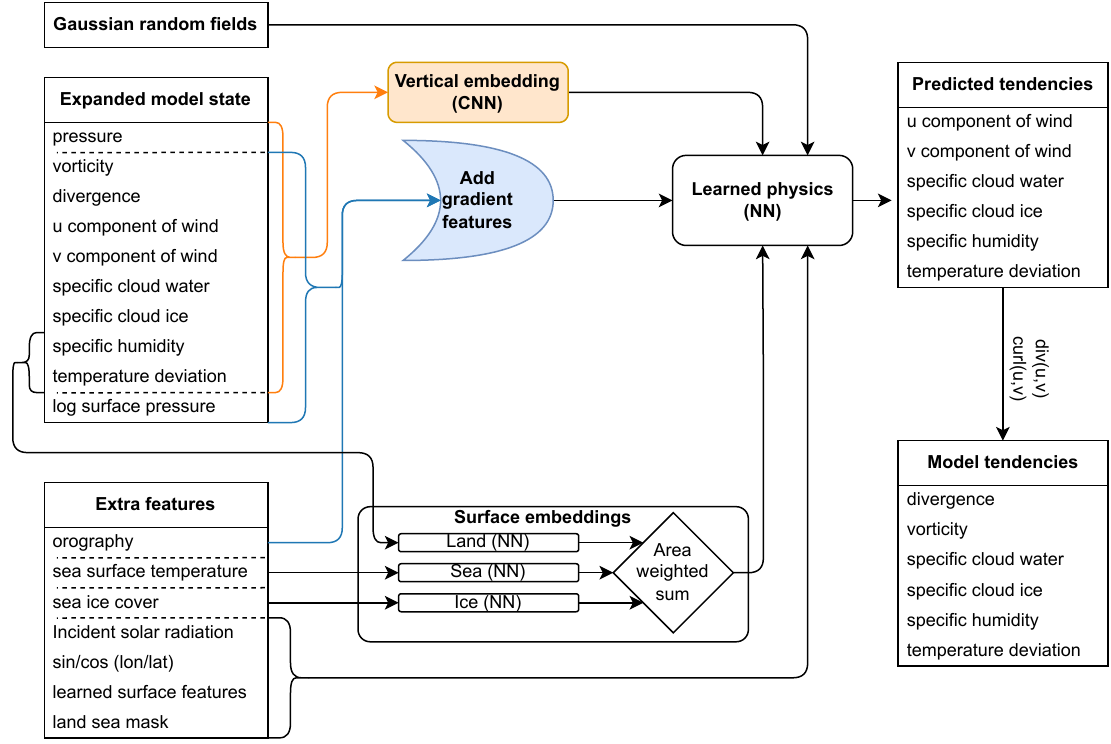}
\caption{
Visualization of the data flow in the learned physics module of NeuralGCM.}\label{apx:fig:learned_physics}
\end{figure*}
In NeuralGCM, effects of physical processes not accounted for by the dynamical core, as well as computational errors within the dynamical core,  are approximated by neural networks. Associated tendencies of the atmospheric state are computed in two steps: (1) features extraction and normalization (sections \ref{apx:subsec:parameterization_features}, \ref{apx:subsec:feature_normalization}) (2) neural network forward pass and rescaling (sections \ref{apx:subsec:network_architecture}, \ref{apx:subsec:output_rescaling}).

The overall data flow diagram, with the exclusion of feature normalization steps, is shown in Fig. \ref{apx:fig:learned_physics}. To describe the initial value problem solved by each NeuralGCM forecast, we let $Y(t)$ be the ERA5 state at time $t$, $X(t)$ the (decoded) NeuralGCM state, and $\tilde{X}(t)$ the encoded state. The network tendencies $\Psi(\tilde{X})$ are added to $\Phi(\tilde{X})$, the right hand side of the primitive equations \eqref{eq:primitive_equations} in encoded space. In other words, NeuralGCM implements
\begin{align*}
    \frac{\partial \tilde{X}}{\partial t} &= \Phi(\tilde{X}) + \Psi(\tilde{X}),
    \quad t_0 < t < t_0 + \tau,
    \\
    \tilde{X}(t_0) &= \mbox{Encode}(Y(t_0)).
\end{align*}
Then finally, $X = \mbox{Decode}(\tilde{X})$ is evaluated against $Y$.

\subsection{Input features for all models}\label{apx:subsec:parameterization_features}
The core input features to the neural network include the vertical structure of the divergence, vorticity, wind vector, temperature deviation, specific humidity, specific cloud ice water content, specific cloud liquid water content and (log) surface pressure.

Additionally, our model incorporates various supplementary features to capture critical information relevant to the atmospheric conditions. These features include spatial derivatives of the core features ($\partial_\phi$, $\partial_\theta$ and $\nabla^2$), a land-sea mask, incoming solar radiation, orography (along with spatial gradients), cosine, and sine of the latitude, pressure levels (i.e., sigma level times the pressure surface), and an $8$-dimensional (for NeuralGCM-$2.8^\circ$ and NeuralGCM-ENS) or $32$-dimensionanl (for NeuralGCM-$0.7^\circ$ and NeuralGCM-$1.4^\circ$) location-specific embedding vector for each horizontal grid-point.
This embedding vector aims to represent static location-specific information. It is initialized to random values and optimized during training.

We use two embedding modules in NeuralGCM that aim to extract helpful representations that are used as input features by the main network described in \ref{apx:subsec:network_architecture}. Both are computed by small neural networks that use the same weights shared across all spatial locations to promote feature learning. Each embedding module takes a restricted set of inputs. Vertical embedding network~\ref{apx:subsubsection:vertical_embedding} aims to extract common features across the vertical structure of the atmosphere. Surface embedding network ~\ref{apx:subsubsection:surface_embedding} aims to estimate the state of the atmosphere's surface boundary. To accomplish this, it receives additional surface-related inputs, in particular sea surface temperature (SST) and sea ice concentration.

When running weather forecast evaluation the provided SST and sea ice concentration remain constant throughout each forecast, with values taken from the day before the initial time of the forecast launch. This approach ensures that the model relies only on data available at the forecast initialization.

When running seasonal and climate evaluation forecasts, we do prescribe the SST and sea ice concentration from ERA5 data and update them at $6$-hour intervals for the $2.8^{\circ}$ resolution model and $12$-hour intervals for the $1.4^{\circ}$ resolution model (details on how land surfaces were used can be found in \ref{apx:subsubsection:surface_embedding}). This simulates a one way coupling to an ocean model that follows historical evolution.

In some NeuralGCM variations we experimented with adding core features from the previous time step as additional inputs. We did not find this modification to have any significant effect on model performance.

\subsection{Additional input features for stochastic models}\label{apx:subsec:parameterization_features-stochastic}
In stochastic models, the encoder and forward step (but not decoder) each make use of ten additional space-time correlated Gaussian Random Fields (GRF), for a total of twenty fields. All fields are independent of each other. The encoder's fields are used only once in each simulation, and the forward step's fields evolve independently of the model state. Every forecast uses a different seed to form new, independent GRFs. This provides sources of randomness needed to generate an ensemble of statistically independent forecasts.

\begin{figure}
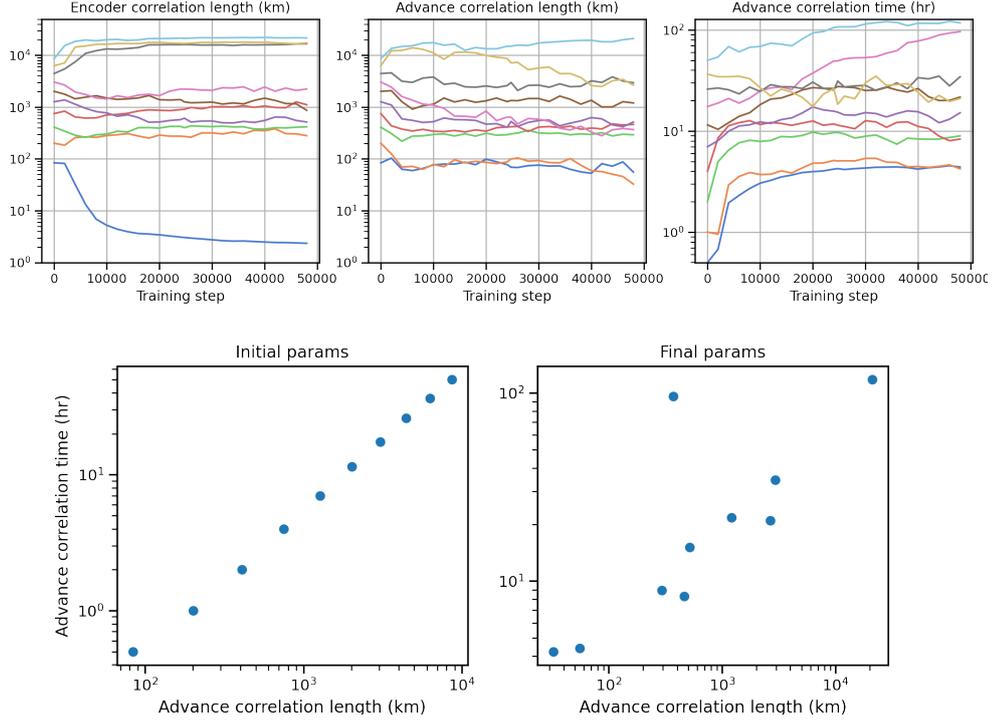

\centering
\includegraphics[width=1.0\textwidth]{images/GRF-params.pdf}\\
\includegraphics[width=1.0\textwidth]{images/GRF-scatter.pdf}
\caption{Random field parameters evolution during training. Top row shows correlation lengths and times as they evolve. Bottom row shows initial and final (advance) correlation time vs. length. The two outlier fields with correlation time $\approx 100$ hours are of interest. Since lead time during training is at most 120 hours, the model preferred these fields to be fixed in time.}\label{apx:fig:grf-params}
\end{figure}

Each GRF is constructed in a spherical harmonic basis, with adjustable length scale $\lambda$ and time scale $\gamma$.
\begin{align*}
    Z(t) &= \sum_{l,m}Z_{l,m}(t) Y_{l,m},
\end{align*}
where $Y_{l,m}$ are the spherical harmonics, and $Z_{l,m}$ are scalar Gaussian processes satisfying (with $R_e$ the earth's radius)
\begin{align*}
    \Exp{Z_{l,m}(t)} &\equiv 0, \\
    \Exp{Z_{l,m}(t)Z_{l',m'}(t')} &= 
    \left\{
    \begin{matrix}
    F_0\exp\left\{-\left(\frac{\lambda}{R_e}\right)^2 \frac{l (l+1)}{2}\right\} \exp\{-|t - t'|/\gamma\},&\quad l=l', m=m'\\
    0&\qbox{otherwise.}
    \end{matrix}
    \right.
\end{align*}
The normalization constant $F_0$ is chosen so that the global mean variance is 1.
This is the same construction used by ECMWF in \citeSI{Palmer2009SPPT}. In training NeuralGCM-ENS, the fields are initialized with length scales ranging from $85$km to $10,000$km and time scales from $30$ minutes to $60$ hours. These values were fixed for the first $1000$ training iterations and optimized afterwards. Figure \ref{apx:fig:grf-params} shows parameter values during training. Most changed very little. Some parameters became constant over space/time (still with variance = 1). We interpret them as representing random model parameters, but more investigation is needed.

\subsection{Normalization of input features }\label{apx:subsec:feature_normalization}
We normalize all input features to be approximately distributed with zero mean and unit variance to improve training dynamics. We do so by estimating the mean and the standard deviation of all of the features at initialization and adding appropriate shift and rescale transformations before feeding features into the neural network.
All features were normalized uniformly across all atmospheric levels, except for specific humidity $q$ which is normalized per-level. Features corresponding to trainable parameters (termed learned features in section \ref{apx:subsec:parameterization_features}), Gaussian random field features (section \ref{apx:subsec:parameterization_features-stochastic}) and embedding features were skipped during normalization as they are normally distributed by construction.

\subsection{Network architecture}\label{apx:subsec:network_architecture}

For our fully-connected neural networks with residual connections,
we use a Encode-Process-Decode (EPD) architecture~\citeSI{battaglia2018relational} with $5$ fully connected MLP blocks in the ``Process'' component. All input features are concatenated together before being passed to the ``Encode'' layer, which is a linear layer that maps input features to a latent vector of size $384$. All ``Process'' blocks use $3$-layer MLP block with $384$ hidden units to compute updates to the latent vector. Finally, linear ``Decode'' layer maps the vector of size $384$ to the output vector, which is is then split into per-level values for different variables. See table \ref{table:training_times} for a count of learnable parameters by model.
NeuralGCM models differ in parameter count primarily due to the location-specific embedding vector assigned to each horizontal grid point. Higher-resolution models require more parameters. Additionally, NeuralGCM-$1.4^\circ$ and NeuralGCM-$0.7^\circ$ models use 32-length embeddings, while NeuralGCM-ENS and NeuralGCM-$2.8^\circ$ models use 8-length embeddings.

In all models we predict tendencies of the wind vector, temperature and moisture species at all levels.

\subsubsection{Vertical embedding network}\label{apx:subsubsection:vertical_embedding}
We use $1$D convolutional network to compute vertical embeddings. We use $5$ layers with $64$ hidden and $32$ output channels. Input channels correspond to the $8$ atmospheric variables (divergence, vorticity, two components of the wind vector, temperature and three moisture species), aligned by their vertical location in the atmosphere.

\subsubsection{Surface embedding network}\label{apx:subsubsection:surface_embedding}
The surface embeddings are calculated by aggregating outputs from three different embeddings corresponding to distinct surface types: land, sea, and sea ice. The ``sea embedding'' is a neural network that receives the sea surface temperature (SST) as input. The settings for the ``land embedding'' and ``sea ice embedding'' vary slightly across different models. For the deterministic models with resolutions of 1.4$^{\circ}$ and 0.7$^{\circ}$, we employ a constant learned embedding for both land embedding and sea ice embedding. Conversely, for the 2.8$^{\circ}$ resolution deterministic model and the 1.4$^{\circ}$ ensemble model, we use neural networks that take the lowest-level atmospheric temperature and moisture as inputs. The output from each network is an embedding of size $8$. Each of these smaller networks is a fully connected network comprising three hidden layers with $16$ hidden neurons, followed by a readout linear layer that produces an embedding of size $8$. The aggregation step is determined based on the relative fraction of each surface type at each grid point.

\subsection{Network output scaling}\label{apx:subsec:output_rescaling}
When predicting learned physics tendencies, the outputs of the neural network are scaled by $0.01$ standard deviation of the tendencies for each variable, which are estimated from ERA5 data using finite-difference method at $1$-hour intervals. The standard deviation is estimated using $10$ snapshots averaged over the globe.

\subsection{Interpretability of learned physics tendencies}
\begin{figure*}
\centering
\includegraphics[width=1.0\textwidth]{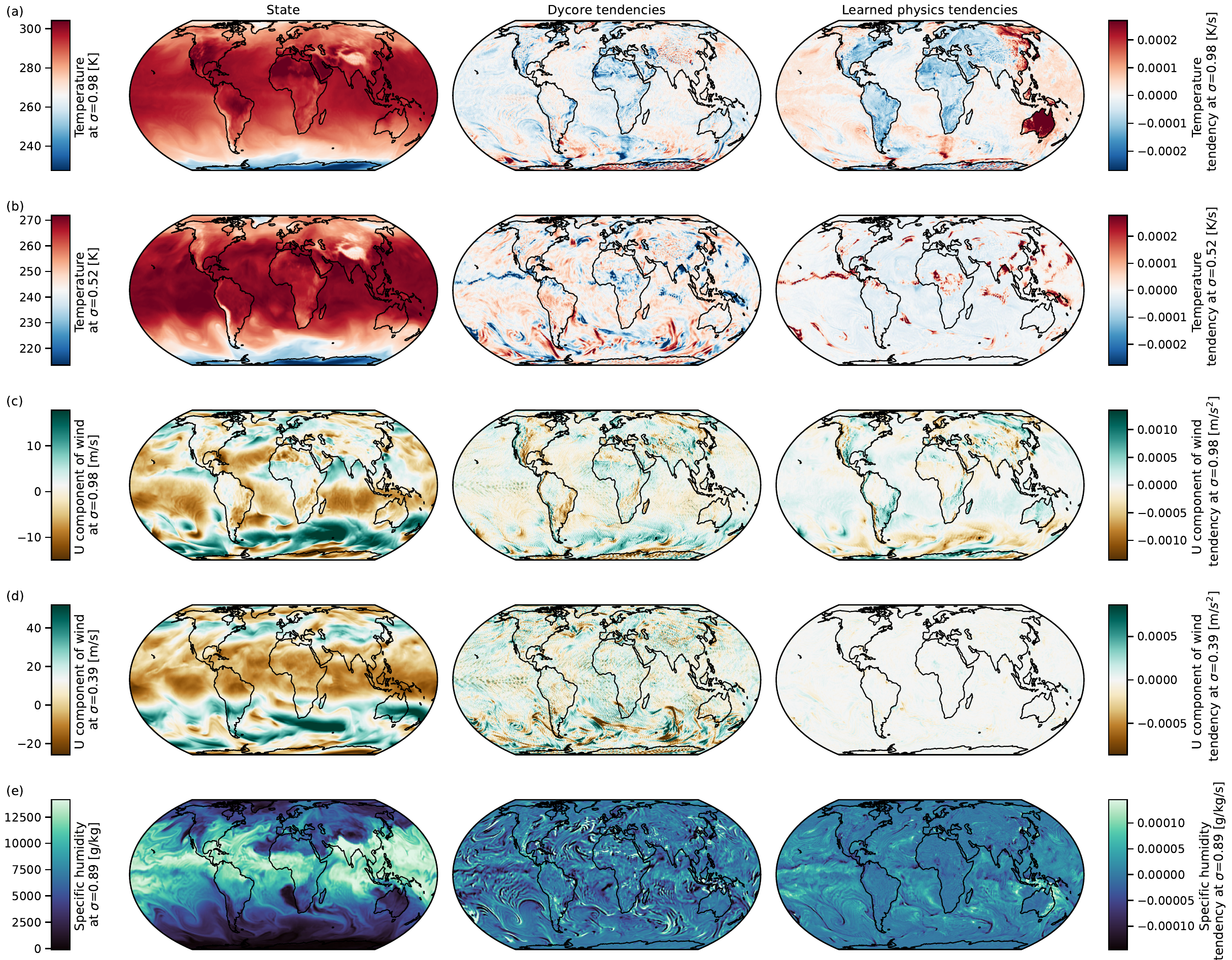}
\caption{
Comparison on tendencies produced by the dynamical core and the learned physics component of NeuralGCM-$0.7^\circ$. All tendencies are computed on 2020-08-
24T00z. The plot shows $\sigma$ level slices of temperature, zonal component of wind and specific humidity, as well as tendencies associated with the dynamical core and learned physics.}\label{apx:fig:tendencies_comparison}
\end{figure*}

In Fig.~\ref{apx:fig:tendencies_comparison} we show a comparison of tendencies associated with the dynamical core and trained learned physics component of NeuralGCM. While current instantiation of NeuralGCM does not allow for separation of the tendencies associated with different physical processes, we find several physically plausible signatures. In particular, near surface temperature tendencies shown in Fig. \ref{apx:fig:tendencies_comparison} (a) resemble surface heating region (where land is cooled during night time and is heated during daytime). In Fig. \ref{apx:fig:tendencies_comparison} (b) we find significant warming signal in the upper atmosphere, potentially associated with convective processes. Fig. \ref{apx:fig:tendencies_comparison} (c) show that learned physics generally counteracts the wind tendencies that of the dynamical core, potentially representing drag effects, which are not found further away from the surface \ref{apx:fig:tendencies_comparison} (d).

\clearpage
\section{Encoder and Decoder of NeuralGCM}\label{apx:sec:encoder_and_decoder}

\begin{figure*}[b]
\centering
\includegraphics[width=0.85\textwidth]{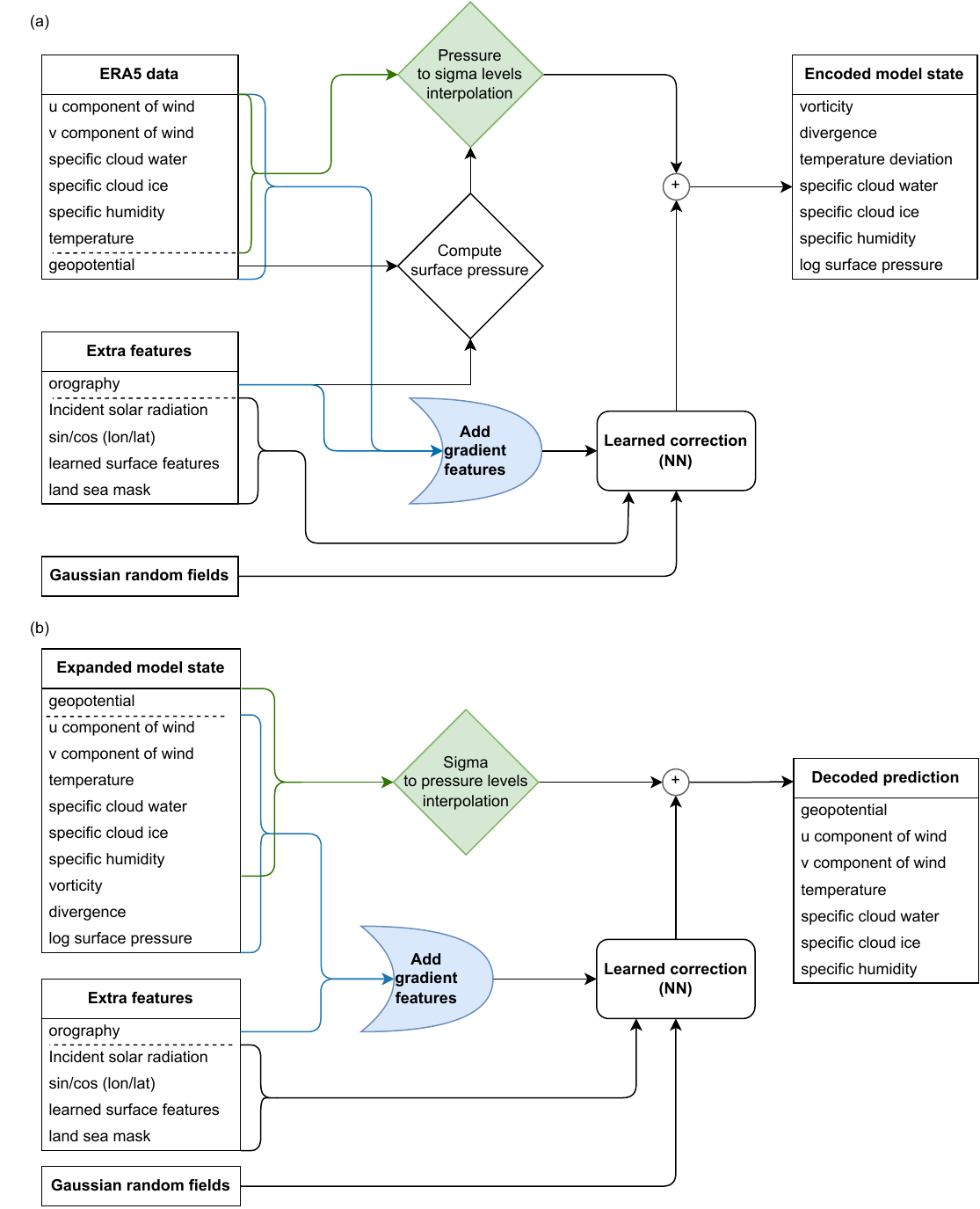}
\caption{
Visualization of the data flow in encoder and decoder modules of NeuralGCM.}\label{apx:fig:encoder_decoder_viz}
\end{figure*}
To interface NeuralGCM that uses sigma coordinate representation of the atmosphere with ERA5 data we use Encoder and Decoder modules. Each component is based on regridding between sigma and pressure levels combined with learned correction. The overall data flow for encoder and decoder components are shown in Fig. \ref{apx:fig:encoder_decoder_viz} (a) and (b) respectively.

\subsection{Encoder}
We use Encoder to obtain a model state in sigma coordinates from an ERA5 snapshot on pressure coordinates in three steps. First, we compute surface pressure for each (longitude, latitude), by calculating the pressure levels at which geopotential field matches that of the surface. Next, we linearly interpolate all relevant atmospheric variables to NeuralGCM's sigma coordinates. Finally, we add a correction computed using a neural network of the same structure as in learned physics module (section \ref{apx:subsec:network_architecture}) to the interpolation results. The last step significantly alleviates the initialization shock, which otherwise contaminates forecasts with rapid oscillations.

\begin{figure*}
\centering
\includegraphics[width=1.0\textwidth]{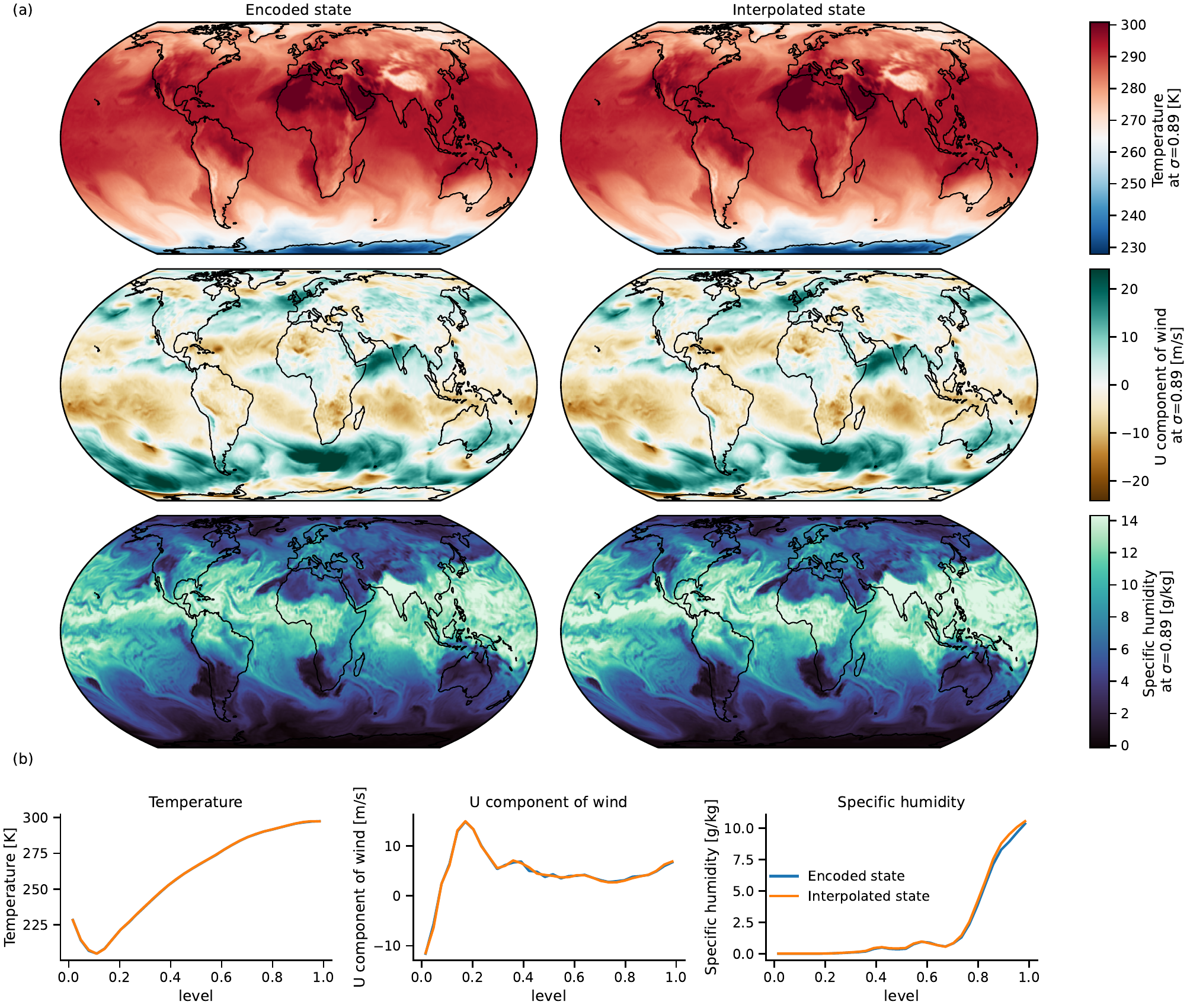}
\caption{Comparison of NeuralGCM-$0.7^{\circ}$ model states produced by the encoder and that of the pressure to sigma level interpolation scheme. (a) Visualization of temperature, u component of wind and specific humidity with full encoder and that with learned components disabled. (b) Vertical profiles for the same variables evaluated at (longitude$\approx238$ and latitude$\approx37$).}\label{apx:fig:encoder_vs_interpolation}
\end{figure*}

When computing the correction, input features to the network are extracted from input data on pressure levels. We omit embedding features from the Encoder network inputs. Network outputs include corrections to divergence, vorticity, temperature, logarithm of the surface pressure and all moisture species. Before being combined with linear interpolation, network outputs are scaled by $0.02$ standard deviation of corresponding variables. Fig.\ref{apx:fig:encoder_vs_interpolation} compares NeuralGCM-$0.7^{\circ}$ model states (on vertical sigma levels) computed with encoder against using pressure to sigma level interpolation. As shown in Fig.\ref{apx:fig:encoder_vs_interpolation} (a), the overall structure of the atmospheric state is hardly affected.

\subsection{Decoder}
We use Decoder to map model state on sigma coordinates back to ERA5 snapshot on pressure coordinates in three steps. First, we diagnose geopotential from temperature and moisture using Eq.~\ref{apx:eq:diagnostic_variables}. Next, we interpolate the results to pressure levels of ERA5. For pressure levels that correspond to values above the Earth's surface, we uses linear interpolation. To extrapolate data from sigma coordinates (which are terrain-following) to pressure coordinates (that extend below the Earth's surface) we use linear extrapolation for geopotential and temperature. For all other variables we uses boundary values for extrapolation. These choices aim to approximate a more complicated procedure used by ECMWF for pressure extrapolation. Finally, we add a correction computed using a neural network of the same structure as in learned physics module (section \ref{apx:subsec:network_architecture}) to the interpolation results.

\begin{figure*}
\centering
\includegraphics[width=1.0\textwidth]{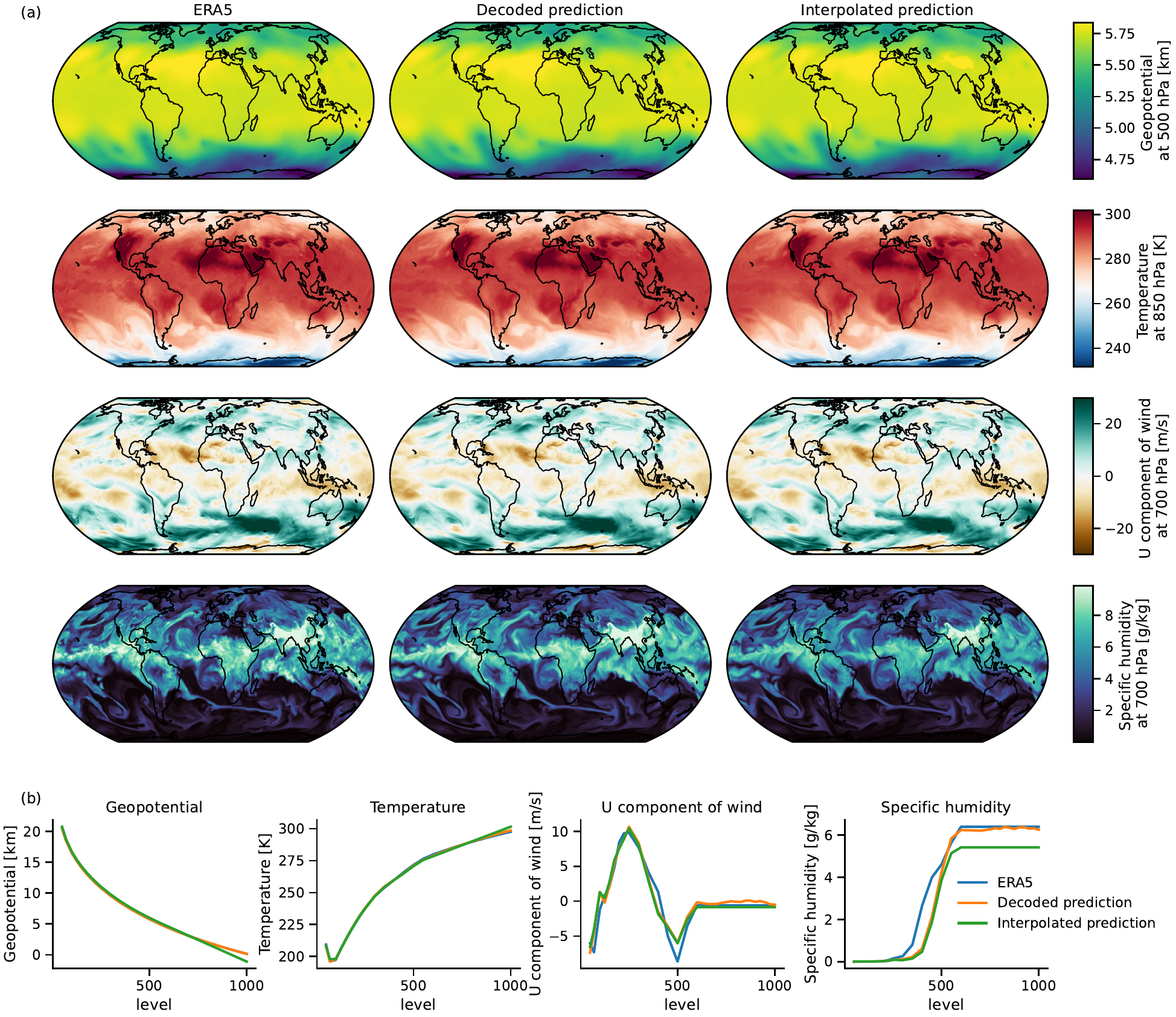}
\caption{Comparison of NeuralGCM decoder and underlying sigma level to pressure level interpolation scheme. (a) Visualization of geopotential, temperature, u component of wind and specific humidity from ERA5, $2$ day NeuralGCM-$0.7^{\circ}$ prediction processed with NeuralGCM decoder and $2$ day NeuralGCM-$0.7^{\circ}$ prediction interpolated from sigma to pressure levels without learned components of the decoder. (b) Vertical profiles for the same variables evaluated around Tibetan plateau (longitude$\approx86$ and latitude$\approx32$).}\label{apx:fig:decoder_vs_interpolation}
\end{figure*}

Similarly to the Encoder, embedding features are omitted. Network outputs include corrections to the horizontal wind vector, temperature, geopotential and all moisture species. Before being combined with the result of interpolation, network outputs are scaled by $0.02$ standard deviation of corresponding variables. Fig.~\ref{apx:fig:decoder_vs_interpolation} compares NeuralGCM-$0.7^{\circ}$ predictions processed with decoder and those  simply interpolated from sigma to pressure levels. Fig.~\ref{apx:fig:decoder_vs_interpolation} (a) shows that the overall structure of the atmospheric state is hardly affected, for the exception of a few minor artifacts around mountainous areas. Fig.~\ref{apx:fig:decoder_vs_interpolation} (b) shows that corrections introduced by the neural network mostly correspond to high pressure values that fall close to or below topographic features.

\clearpage
\section{Time integration}\label{apx:sec:time_integration}
In NeuralGCM, the state of the atmosphere is advanced in time by integrating model equations that combine effects from the dynamical core and learned physics parameterizations. This is done iteratively using Implicit-Explicit integration scheme~\cite{whitaker2013implicit} described in \ref{apx:subsec:sil3}. Integration time step varies with resolution, as shown in Table \ref{apx:table:timestep_filters}. This results in iterative updates to the model state every $4$-$30$ minutes, depending on model resolution. In contrast, data-driven methods commonly make predictions at $6$-hour jumps \cite{lam2022graphcast,keisler2022forecasting}.
Throughout time integration, dynamical core tendencies are computed at every time step, while learned physics tendencies are only recomputed once every $60$ minutes for our lowest resolution ($2.8^{\circ}$) model and every $30$ minutes for all others. This is done to avoid excessive backpropagation through the neural networks in learned physics. At higher resolutions it might be advantageous to include more frequent updates to learned physics tendencies to be able to account for short-time processes (rather than statistical effect that varies smoothly in time). 
Similar to traditional spectral GCMs we introduce spectral filters to improve numerical stability~\citeSI{jablonowski2011pros}, which are described in \ref{apx:subsec:timestep_filtering}.

\subsection{Time integration scheme}\label{apx:subsec:sil3}

As is typical for atmospheric models, in NeuralGCM we use semi-implicit ODE solvers to the solve the primitive equations, by partitioning dynamical tendencies into ``implicit'' and ``explicit'' terms. ``Implicit'' tendencies include linear terms of Eq.~\ref{eq:primitive_equations} that give rise to the low amplitude, fast moving gravity waves. These terms are treated implicitly, allowing for longer stable time steps, while the rest of the terms are computed explicitly.

Rather than the traditional semi-implicit leapfrog method, we use implicit-explicit Runge-Kutta methods to avoid the complexity of keeping track of multiple time-steps and time-filtering required by the traditional semi-implicit leapfrog method.
Specifically, we use the semi-implicit Lorenz three cycle scheme (SIL3), which was developed specifically for global spectral weather models~\cite{whitaker2013implicit}.

\begin{table}[h]
\begin{tabular}{ cc }
    \begin{tabular}{c|cccc}
    $1 / 3$ & $1 / 3$ & & & \\
    $2 / 3$ & $1 / 6$ & $1 / 2$ & & \\
    1 & $1 / 2$ & $-1 / 2$ & 1 & \\
    \hline & $1 / 2$ & $-1 / 2$ & 1 & 0 
    \end{tabular}
    \begin{tabular}{c|cccc}
    $1 / 3$ & $1 / 6$ & $1 / 6$ & & \\
    $1 / 3$ & $0$ & $1 / 3$ & & \\
    1 & $3 / 8$ & $0$ & $3 / 8$ & $1 / 4$ \\
    \hline & $3 / 8$ & $0$ & $3 / 8$ & $1 / 4$
    \end{tabular}
\end{tabular}
\caption{Butcher tableau for the IMEX SIL3 scheme.}\label{apx:table:butcher_tableau}
\end{table}

\subsection{Filtering}\label{apx:subsec:timestep_filtering}
During time integration we use two exponential filters of different strengths (``hard'' and ``soft'').
These filters correspond to hyper-diffusion, a standard component of spectral atmospheric models used to stabilize dynamics~\citeSI{Jablonowski2011-qf}. 
Each transform a scalar field $x_{hml}$ in spherical harmonic representation as:
\begin{equation}
    x_{hml} \rightarrow x_{hml} * e^{-a\left(\frac{k-c}{1-c}\right)^{2p}}
\end{equation}
with filter attenuation $a$, filter cutoff $c$, filter order $p$, and normalized total wavenumber $k\equiv\frac{l}{l_{max}}$.

Filter parameters used by different NeuralGCM models are summarized in Table \ref{apx:table:timestep_filters}, where filter attenuation is specified via attenuation time $\alpha$ and time step $dt$ via $a=\frac{\alpha}{dt}$. Both hard and soft filters are applied to the model state at the end of each integration step. We additionally apply hard filter to the outputs of learned physics parameterizations to avoid injection of high frequency noise in each model step. The filtering strength sets the true length scale of the simulation, which is generally slightly larger than the grid spacing.

\begin{table}[h]
\begin{tabular}{|c|c|c|c|c|c|}
\hline
Model resolution & Time step [minutes] & Filter & Attenuation time [minutes] & Order & Cutoff \\ \hline
\multirow{2}{*}{$2.8^{\circ}$} & \multirow{2}{*}{12} & hard & 4 & 10 & 0.4 \\ 
 &  & soft & 120 & 3 & 0.0 \\ \hline
\multirow{2}{*}{$1.4^{\circ}$} & \multirow{2}{*}{6} & hard & 8 & 6 & 0.4 \\ 
 &  & soft & 120 & 3 & 0.0 \\ \hline
\multirow{2}{*}{$0.7^{\circ}$} & \multirow{2}{*}{3.75} & hard & 4 & 6 & 0.4 \\ 
 &  & soft & 120 & 3 & 0.0 \\ \hline
\end{tabular}
\caption{Time step and filtering parameters of NeuralGCM models.}\label{apx:table:timestep_filters}
\end{table}

\clearpage
\section{Evaluation metrics}\label{apx:subsec:error-metrics-and-losses}
Evaluation metrics compare forecasts $X$ with ground truth $Y$. In most cases, $Y$ is ERA5 reanalysis. However, for evaluating ECMWF-HRES and ECMWF-ENS forecasts, we use the lead time = $0$ (or ``analysis'') from ECMWF-HRES. This prevents data-driven approaches from having an unfair advantage. All evaluations reported in the paper were done after regridding $X$ and $Y$ to $1.5^{\circ}$. We use the WeatherBench2 library \cite{rasp2023weatherbench} to standardize model evaluation. Our primary evaluation metrics include: root mean square error (RMSE), root mean squared bias (RSMB), continuous ranked probability score (CRPS) and skill-spread ratio.

\subsection{Root mean square error (RMSE)}\label{apx:subsec:root-mean-square-error}
RMSE compares $Y(t+\tau)$, the ground truth at time $t+\tau$, with $X(t\to t+\tau)$, a forecast initialized at time $t$ (to $Y(t)$), at lead time $\tau$ hours into the future. This is reported separately for each variable $v\in\mathcal{V}$ and pressure level $p$, but averaged over initial times.
\begin{align*}
    \mathcal{E}_{RMSE}(\tau, v, p) 
    &:= \sqrt{\frac{1}{|\mathcal{T}|}\sum_{t \in \mathcal{T}}\|X(t\to t+\tau, v, p) - Y(t+\tau, v, p)\|_{nodal}^2},
\end{align*}
where $\mathcal{T}$ are the set of $12$-hour spaced times in 2020. The \emph{nodal} norm above is area weighted, to avoid polar regions (where points are more dense) from having an undue influence:
\begin{align}
  \label{align:RMSE-norm}
    &\|X(t\to t+\tau, v, p) - Y(t+\tau, v, p)\|_{RMSE}^2 \\
    &\qquad := \frac{1}{IJ} \sum_{i=1}^{121}\sum_{j=1}^{240}  w(i) |X(t\to t+\tau, v, p, i, j) - Y(t+\tau, v, p, i, j)|^2.
\end{align}
where $(i, j)$ are the (latitude, longitude) indices and $w(i)\propto (\sin{\phi_{i+0.5}} - \sin{\phi_{i-0.5}})$.

For models that produce a single deterministic forecast $X$, MSE $\equiv \Exp{\|X(t+\tau) - Y(t+\tau)\|^2}$ (and hence RMSE) is minimized by $X \equiv \Exp{Y}$. So the best RMSE scores will be had by a forecast that predicts the (conditional) mean $\Exp{Y(t+\tau)\,|\,Y(t)}$. We primarily use this metric for assessing accuracy of deterministic forecasts at shorter lead times, when the distribution of $Y(t+\tau)$ is sharply peaked. As lead time $\tau$ increases, the distribution $Y(t+\tau)$ spreads out, as true dynamics is depends on information not fully contained in the initial state $Y(t)$. When this happens, $\Exp{Y(t+\tau)\,|\,Y(t)}$ becomes blurry and unphysical.

For models that generate ensemble forecasts we compute RMSE scores using ensemble mean $\mathbb{E}_{i}\left[X_{i}\right]$. In this setting RMSE remains informative even at later times, as the ensemble mean is taken explicitly. It's important to note that RMSE of the ensemble mean alone is not sufficient to assess the skill of the forecasting system. An ensemble of identical blurry realizations could still achieve good RMSE skill, but fail to capture extreme events and provide probabilistic insight.

\subsection{Root mean squared bias (RMSB)}\label{apx:subsec:bias_rmse}
Biases estimate persistent differences between forecasts $X(t+\tau)$ and ground truth $Y(t+\tau)$ averaged over time. This is reported separately for each variable $v\in\mathcal{V}$ and pressure level $p$, lead time $\tau$, and (latitude, longitude) coordinates $(i, j)$.
\begin{align*}
    \mathcal{E}_{bias}(\tau, v, p, i, j)
    &:= \frac{1}{|\mathcal{T}|}\sum_{t \in \mathcal{T}}\left[X(t\to t+\tau, v, p, i, j) - Y(t+\tau, v, p, i, j)\right],
\end{align*}

RMSB is computed by taking RMSE of the bias and aggregating it over spatial dimensions.
\begin{align*}
    \mathcal{E}_{RMSB}(\tau, v, p)
    &:= \sqrt{\frac{1}{IJ} \sum_{i=1}^{121}\sum_{j=1}^{240}  w(i) \|\mathcal{E}_{bias}\|^{2}}
\end{align*}

\subsection{Continuous Ranked Probability Score (CRPS)}\label{apx:subsec:crps}
Ideally, evaluation metrics should be minimized when $X(t)\sim Y(t)$, rather than $\mathbb{E}Y(t)$ as is the case for MSE. One such metric is CRPS.

To understand CRPS, consider ground truth scalar random variable $V$, and two independent forecasts $U$ and $U'$. CRPS takes the form
\begin{align}
  \label{align:crps-ideal}
    \mathbb{E}|U - V| - \frac{1}{2} \mathbb{E}|U - U'|.
\end{align}
The skill term $\mathbb{E}|U - V|$ penalizes forecasts that deviate from ground truth, while the spread term $(1/2) \mathbb{E}|U - U'|$ encourages well dispersed forecasts. As it turns out, CRPS is minimized just when $U$ has the same distribution as $V$ \cite{Gneiting2007ProperScoring}.

Following \cite{rasp2023weatherbench}, we extend CRPS to multiple dimensions by summing over components. We estimate this using $M$ ensemble members $(X^{(1)}, \ldots, X^{(M)})$ as
\begin{align}
\begin{split}
  \label{align:crps-nodal}
\mathcal{E}_{CRPS}&(\tau, v, p) := \\
&\frac{1}{|\mathcal{T}|}\sum_{t \in \mathcal{T}} \frac{1}{IJ} \sum_{i=1}^{121}\sum_{j=1}^{240} w(i)\,\bigg[ \\
  &\frac{1}{M}\sum_{m=1}^M \big|X^{(m)}(t\to t+\tau, v, p, i, j) - Y(t+\tau, v, p, i, j)\big| \\
  &-\frac{1}{2M(M-1)}\sum_{m, k=1}^M \big|X^{(m)}(t\to t+\tau, v, p, i, j) - X^{(k)}(t\to t+\tau, v, p, i, j) \big|
\bigg]
 \end{split}
\end{align}
The resultant ``CRPS'' is minimized by any distribution $X$ such that the marginals $X_n$, have the same distribution as $Y_n$. CRPS will therefore not require correct forecasts, but will not penalize them either.

\subsection{Spread-Skill ratio}\label{apx:subsec:spread_skill_ratio}
Spread-skill ratio represents the ratio of the ensemble spread (standard deviation) to skill (RMSE) of the ensemble mean. For generic scalar ground truth $Y_t$ and ensemble of forecasts $\{X_t^{(n)}\}_{n=1}^N$, each given at times $t=1,\ldots,T$, define
\begin{align*}
    \mu_t :&= \frac{1}{N}\sum_{n=1}^N X_t^{(n)}, \\
    \mbox{S}^2 :&= \frac{1}{T}\sum_{t=1}^T\frac{1}{N-1}\sum_{n=1}^{N}(X_t^{(n)} - \mu_t)^2, \\
    \epsilon^2 :&= \frac{1}{T}\sum_{t=1}^T\frac{1}{N}\sum_{n=1}^{N}(Y_t - \mu_t)^2.
\end{align*}
Above, $S^2$ is an unbiased estimate of ensemble variance.
However, the mean square error estimate $\epsilon^2$ is biased. Taking expectations, we see it is too large by a term equal to the variance divided by $N$. To un-bias it, we therefore subtract $S^2/N$. The resultant spread skill ratio is
\begin{align*}
    SSR &= \sqrt{\frac{S^2}{\epsilon^2 - S^2/N}}.
\end{align*}
This de-biasing is only used for spread-skill-ratio, and we report biased RMSE for all ensemble models.

If ensemble members $X$ are distributed identically to the ground truth $Y$, then the spread-skill ratio should be equal to $1$.
Similar to other evaluate metrics we compute spread-skill ratio for ECMWF-ENS and NeuralGCM-ENS models for all lead times by averaging over initial time. When reporting global spread-skill ratio we perform spatial aggregation prior to computing the ratio. For spatial visualizations spread-skill ratio is computed for each latitude, longitude and level independently.

\clearpage
\section{Training}\label{apx:sec:training}
We train NeuralGCM using Adam~\citeSI{kingma2014adam}, optimizing model parameters to minimize the loss function between predictions and coarsened ERA5 trajectories. Optimizer parameters are summarized in \ref{apx:subsec:training_optimizer}. For all models we progressively extend the lead time horizon over which the loss is computed as the model trains. The schedule and the data are discussed in \ref{apx:subsec:training_data_and_schedule}. Before computing the loss we rescale errors between trajectories to address the differences in magnitudes of the target distributions (see \ref{apx:subsec:error_rescaling} for details). Finally, the two loss functions used to train deterministic and stochastic NeuralGCM are discussed in \ref{apx:subsec:deterministic_training} and \ref{apx:subsec:stochastic_training} respectively. Computational resources and training times are reported in \ref{apx:subsec:training_times}.
      
\subsection{Optimizer settings}\label{apx:subsec:training_optimizer}
All NeuralGCM models were trained using Adam~\citeSI{kingma2014adam}. We used values $\beta_{1}=0.9$, $\beta_{2}=0.95$ and $\epsilon=10^{-6}$ for all experiments based on empirical evidence from initial results at coarsest resolution. Learning rate was adjusted as a function of training iterations, linearly increasing to its maximum value over the first $2000$ training steps, then remaining constant until training step $15000$ when exponential decay with rate $0.5$ was initiated with decay time set to $10000$ steps. The only exception to these parameters was the decoder fine-tuning stage described in \ref{apx:subsubsec:decoder_finetuning}, where warm-up is completed by step $1000$ and the learning rate decays with rate $0.5$ every $1000$ steps starting from step $2000$. Peak learning rates and total number of steps for different model configurations are reported in Table \ref{table:learning_rates}.

\begin{table}[h]
\begin{tabular}{|c|c|c|c|c|}
\hline
 & NeuralGCM-$2.8^{\circ}$    & NeuralGCM-$1.4^{\circ}$   & NeuralGCM-$0.7^{\circ}$ & NeuralGCM-ENS($1.4^{\circ})$ \\ \hline
\begin{tabular}[c]{@{}c@{}}Peak learning\\ rate\end{tabular}       & $0.002$ & $0.002$ & $0.001$ & $0.001$  \\ \hline
\begin{tabular}[c]{@{}c@{}}Number of\\ training steps\end{tabular} & 38000 & 26000   & 25000          &     43000     \\ \hline
\end{tabular}
\caption{Learning rate and number training steps used for NeuralGCM at different resolutions.}\label{table:learning_rates}
\end{table}

\begin{table}[h]
\begin{tabular}{|l|l|l|}
\hline
Model & Unroll lengths & Transition boundaries \\ \hline
NeuralGCM-$2.8^{\circ}$ & $(12, 24, 36, 48, 60, 72)$ & $(2000, 5656, 10392, 16000, 22360)$ \\ \hline
NeuralGCM-$1.4^{\circ}$ & $(12, 24, 36, 48, 60, 72)$ & $(2000, 5656, 10392, 16000, 22360)$ \\ \hline
NeuralGCM-$0.7^{\circ}$ & $(6, 12, 18, 24, 36, 48, 60)$ & $(500, 2000, 4500, 8000, 12500, 18000)$ \\ \hline
NeuralGCM-ENS ($1.4^{\circ})$ & \begin{tabular}[c]{@{}c@{}}$(6, 12, 18, 24, 36, 48, 60,$ \\ $ 72, 96, 120)$\end{tabular} & \begin{tabular}[c]{@{}c@{}}$(500, 2000, 4500, 8000, 12500, 18000, $ \\ $24500, 32000, 40500)$\end{tabular} \\ \hline
\end{tabular}
\caption{Training curriculum used for NeuralGCM at different resolutions.}\label{apx:table:unroll_lengths}
\end{table}

\begin{figure*}
\centering
\includegraphics[width=1.0\textwidth]{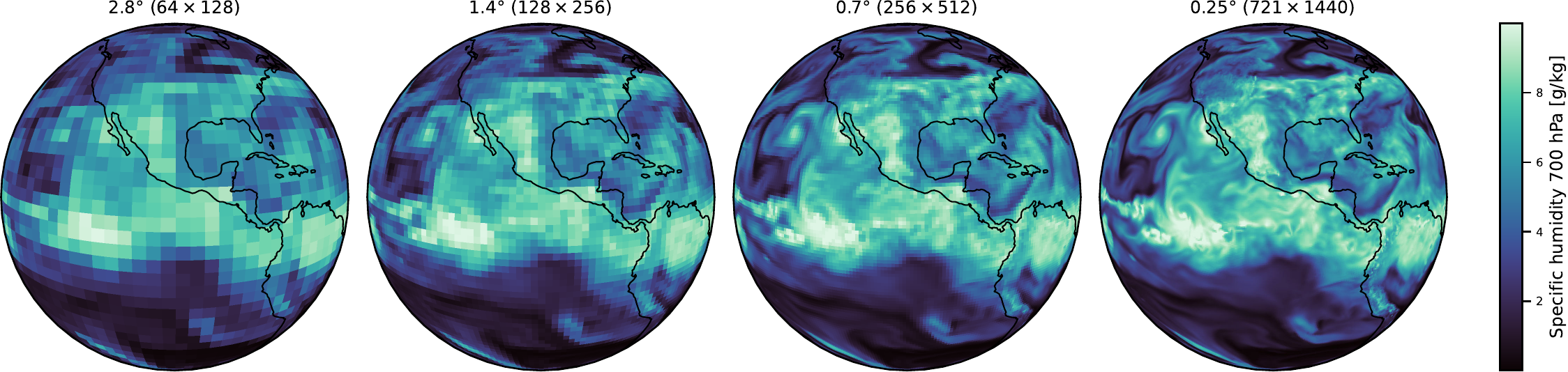}
\caption{Visualization of specific humidity at $700$ hPa from ERA5 conservatively regridded to Gaussian grids $2.8^{\circ}$, $1.4^{\circ}$, and $0.7^{\circ}$, and the original ERA5 data.}\label{apx:fig:training_data}
\end{figure*}

\subsection{Training data and unroll schedules}\label{apx:subsec:training_data_and_schedule}
To train NeuralGCM models at $2.8^{\circ}$, $1.4^{\circ}$ and $0.7^{\circ}$ resolutions we regridded ERA5 data to the corresponding Gaussian grids. We use a conservative regridding scheme, which linearly aggregates contributions weighted by the relative area overlap. Final $2.8^{\circ}$ and $1.4^{\circ}$ models were trained on data from years $1979$ through $2017$ and evaluated on $2018$ during the development cycle. $0.7^{\circ}$ model and stochastic model variations were trained with data from $1979$ through $2019$. None of the models had any exposure to data from $2020$ prior to running final model evaluations.
During training we also change lead time unroll length as a function of training iterations, increasing the prediction horizon as the model improves over the course of training. The values are reported in Table \ref{apx:table:unroll_lengths}.

\subsection{Variable rescaling for losses}\label{apx:subsec:error_rescaling}
Before computing losses, trajectories of all variables are rescaled to address: (1) differences in natural scales of atmospheric variables, and (2) growth of their deviations from the ground truth as a function of lead time. The former is accomplished by dividing each atmospheric variable by standard deviation of the temporal difference between snapshots that are $24$ hours apart, similar to \cite{keisler2022forecasting}. We further adjusted scaling factors across variables to better balance out the initial skill of the model by the following additional scaling factors: geopotential $2$, specific humidity $0.66$, logarithm of the surface pressure (internal model representation) $5$ and cloud moisture species $0.05$. Our rationale for additional balancing is to prevent the loss landscape being dominated by just a few atmospheric variables. Standard deviations were estimated based on $10$-$60$ snapshots of $24$-hour differences in ERA5 data averaged over longitude, latitude, level and sample. The only trajectory that was normalized per-level was specific humidity as its values vary significantly with level.  When estimating scales for variables in internal representation such as divergence,  vorticity,  log surface pressure, corresponding quantities were compute from ERA5 data using linear interpolation to sigma coordinates and appropriate transformations relating them to the data variables.

Each trajectory is also rescaled to account for a near linear increase in expected error variance with lead time $\tau$. The scaling factor is $(1 + (\tau/24))^{-1/2}$ for all losses except the spectral losses, which use $(1 + (\tau/40)^4)^{-1/2}$.

Finally, in the case of filtered mean squared loss an additional time-dependent rescaling is performed which described in detail in \ref{apx:subsubsec:filtered_mse_loss}.

\subsection{Loss for deterministic models}\label{apx:subsec:deterministic_training}
Our deterministic models are trained in two stages that differ in loss objectives and parameters being optimized: (1) primary training and (2) decoder fine tuning. During both stages, the loss terms take the general form of a mean squared error (MSE) on rescaled variables (Section \ref{apx:subsec:error_rescaling}) with varying definitions of distances $\rho_\ast$
\begin{align}
\label{align:mse-loss}
  \mathcal{M}_\ast(\tau)
    &:= \frac{1}{|\mathcal{T}|}\sum_{t\in\mathcal{T}}
    \sum_{\sigma, v}
    \sum_{\substack{\tau'\in\mathcal{D}\\ \tau'\leq \tau}} \rho_\ast(X(t\to t+\tau', v, \sigma), Y(t+\tau', v, \sigma))^2,
\end{align}
where the lead time $\tau$, in hours, is selected from
\begin{align*}
    \mathcal{D} := \{6, 12, 18, 24, 36, 48, 60, 72, 96, 120\}.
\end{align*}

During the primary stage we used a combination of three loss types to account for different types of discrepancies between NeuralGCM predictions and ERA5 data: accuracy (\ref{apx:subsubsec:filtered_mse_loss}), sharpness (\ref{apx:subsubsec:spectral_loss}) and biases (\ref{apx:subsubsec:spectral_loss}). All losses, except for the bias loss, are imposed on both ``data''(i.e., pressure levels) and ``model''(i.e., sigma levels) representations. This results in a training objective that is a sum of five loss terms:
\begin{align*}
    \mathcal{L}_{Deterministic}
    =& \lambda_{data} \mathcal{M}_{Data}
    + \lambda_{spec} \mathcal{M}_{DataSpec}
    + \lambda_{model} \mathcal{M}_{Model} \\
    &+ \lambda_{spec} \mathcal{M}_{ModelSpec}
    + \lambda_{bias} \mathcal{M}_{MSBias},
\end{align*}
where loss scales $\lambda_{data}=20$, $\lambda_{spec}=0.1$, $\lambda_{model}=1$ and $\lambda_{bias}=2$ were selected empirically. We found that even though the contributions corresponding to ``spec'' and ``bias'' terms were small (compared to the total loss), they had a positive effect on sharpness of NeuralGCM predictions. This is possibly related to the fact that the dynamical core simulates an energy cascade, maintaining the spectral energy of the simulated fields.

After the main training phase is complete, we run a short decoder fine-tuning optimization discussed in \ref{apx:subsubsec:decoder_finetuning} to remove spectral artifacts that arise from truncation errors in transformations from spherical harmonic to grid-point space. During this training phase we only update parameters of the decoder component of the model.

\subsubsection{Accuracy loss: filtered MSE}\label{apx:subsubsec:filtered_mse_loss}
Traditional mean squared error loss between forecast $X$ and targets $Y$ can be computed on pressure levels $p$ and in spherical harmonic basis indexed by total $(l)$, and zonal $(m)$, wavenumbers using $\rho_{Data}(X, Y) = \|X - Y\|_{spectral}$, where
\begin{align}
    \|\epsilon\|_{spectral}^2
    &:= \sum_l\sum_m |\epsilon(l, m)|^2. \label{apx:eq:modal_mse}
\end{align}
Up to discretization error, this is equivalent to $\|\cdot\|_{nodal}^2$, the area-weighted MSE in grid-point space representation.

When used to evaluate forecasts at longer lead time, this loss suffers from the ``double penalty problem''~\cite{gilleland2009intercomparison} as it penalizes the model for predicting sharp features that are slightly misplaced. We address this issue by using ``filtered MSE'', which applies filtering to $X$ and $Y$ prior to computing the norm \ref{apx:eq:modal_mse}.

\begin{figure*}
\centering
\includegraphics[width=1.0\textwidth]{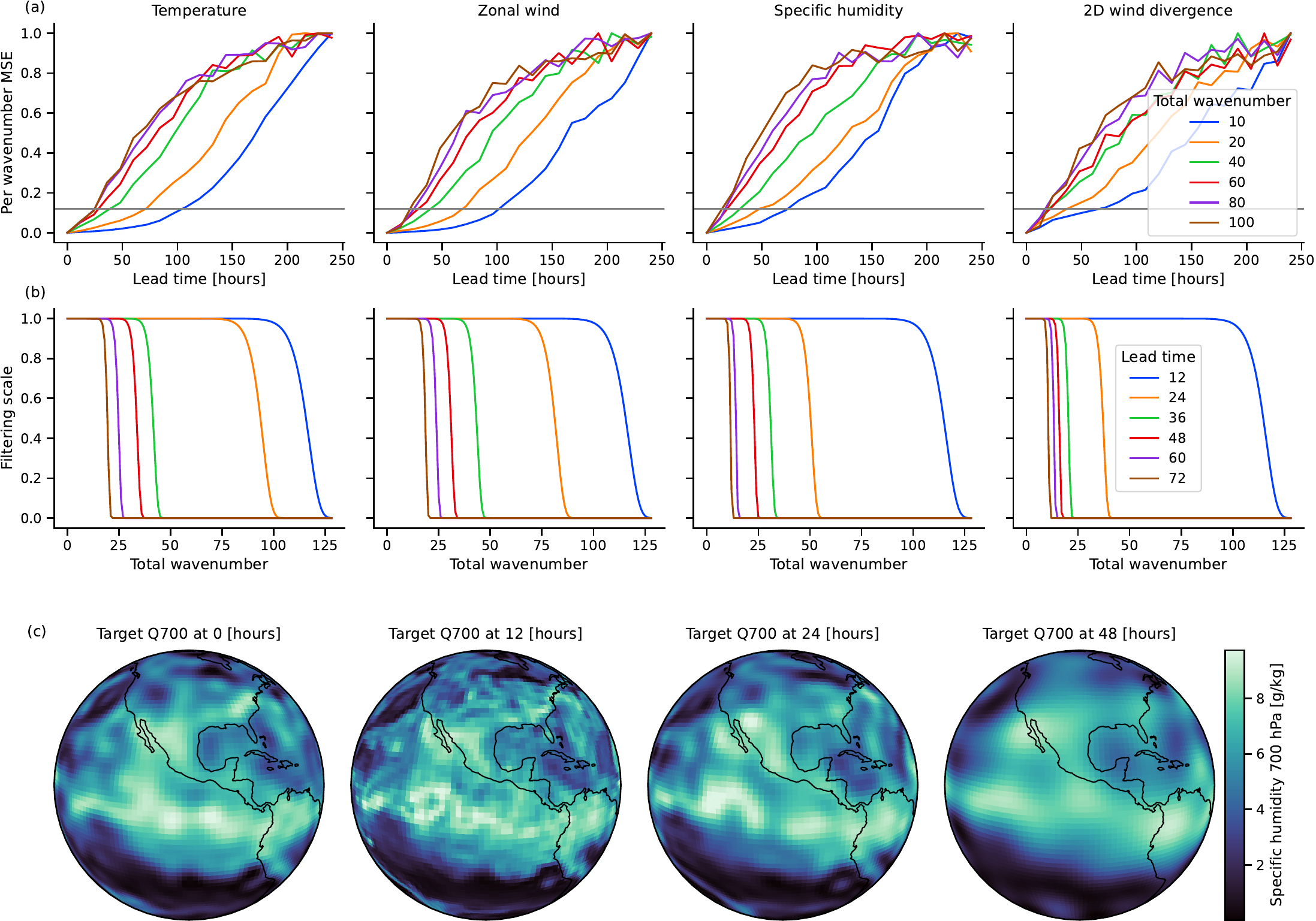}
\caption{Estimation of predictability based on ECMWF-HRES errors.
(a) Normalized mean squared error between ECMWF-HRES and ERA5 for each total wavenumber as a function of time for temperature, zonal wind, specific humidity and horizontal wind divergence. Grey line corresponds to threshold that was used to estimate the predictability horizon. (b) Resulting filtering profiles for varying lead times and different variables. (c) Visualization of effective resolution at which errors are estimated at varying lead times. Zero hour forecasts are additionally blurred in the loss so the model does not need to match exact initial conditions.
}\label{apx:fig:mse_filtering}
\end{figure*}

The filtering step aims to retain components of $X$ and $Y$ that we expect to be able to predict and drop components that cross the predictability horizon due to chaotic dynamics. We estimate such filtering parameters by analyzing the normalized MSE growth between ECMWF-HRES and ERA5 as a function of time for each variable and total wavenumber $l$ separately Fig.~\ref{apx:fig:mse_filtering}(a). We chose a predictability threshold of relative error of $0.12$, and determined the largest value of $l^{*}$ for each lead time that should be included in the loss. Then we determined attenuation parameters for the exponential filters of order $12$ that would remove total wavenumbers higher than $l^{*}$ for each group, shown in Fig.~\ref{apx:fig:mse_filtering}(b). The effect of applying filters is shown in Fig.~\ref{apx:fig:mse_filtering}(c).

The \emph{Model} terms are defined similarly to their Data counterparts, except they are computed in the encoded ``model'' space. This encourages the encoder/decoder round-trip to be the identity. It was found to enhance stability.

\subsubsection{Sharpness loss: spectrum MSE}\label{apx:subsubsec:spectral_loss}
Spectrum MSE is defined by the distance
\begin{align*}
    &\rho_{DataSpec}(X(t\to t+\tau, v, \sigma, l), Y(t+\tau, v, \sigma, l))^2 \\
    &= \sum\limits_{l}^{\tilde{l}} \left[ \mathcal{S}(X)(t\to t+\tau, v, \sigma) - \mathcal{S}(Y)(t+\tau, v, \sigma) \right]^2,
\end{align*}
where $S(X)^2(..., l) = \sum_m X(..., l, m)^2$ is the spectral energy at total wavenumber $l$ and $\tilde{l}$ corresponds to a spectral cutoff for the loss. For NeuralGCM models at $2.8^{\circ}$, $1.4^{\circ}$, $0.7^{\circ}$ we used $\tilde{l}$ of $42$, $80$ and $120$ respectively.

\subsubsection{Bias loss: spectral bias MSE}\label{apx:subsubsec:bias_loss}
Spectral bias MSE is slightly different from the previous two terms, as it averages over the batch and lead time dimensions prior to computing the distance norm:
\begin{align*}
    \mathcal{M}_{MSBias}(\tau)
    &= \sum_{\sigma,v}  \left\|
    \frac{1}{|\mathcal{T}|}\sum_{t\in\mathcal{T}} \sum_{\substack{\tau'\in\mathcal{D}\\\tau'\leq \tau}}
    \left[
    X(t\to t+\tau, v, \sigma) - Y(t+\tau, v, \sigma)
    \right]
    \right\|_{spectral}^2
\end{align*}
This loss term discourages accumulation of bias.
 
\subsection{Decoder fine-tuning}\label{apx:subsubsec:decoder_finetuning}
The final stage of training deterministic NeuralGCMs is focused on optimizing only the ``Decoder'' component. After the primary training phase, ``Decoder'' outputs contain high-frequency artifacts that are not captured by the losses imposed in spherical harmonics representation (because numerical spherical harmonics transform has non-zero kernel). To address that we freeze all of the ``Encoder'' and ``learned physics'' components of the model and optimize the ``Decoder'' parameters using traditional MSE computed in grid-space representation on predictions at $6$, $12$, $18$ and $24$ hours ($12$ and $24$ hours for NeuralGCM-$1.4^{\circ}$). We use $4000$ gradient descent steps, but generally find that loss and evaluation metrics plateau after the first $1000$ training steps.

\subsection{Loss for stochastic models}\label{apx:subsec:stochastic_training}

The CRPS loss function is defined as the sum of spectral and nodal terms.

For any variable $v$, and initial/lead times $t$ and $\tau$, the \emph{spectral} CRPS term on pressure level $p$ is
\begin{align*}
    \mathcal{C}_{spectral}(t, p, v, \tau)
    = \frac{1}{2}\sum_{l,m}
      & \bigg[
      |X(t\to t+\tau, \ldots, l, m) - Y(t+\tau, \ldots, l, m)| \\
      & + |X'(t\to t+\tau, \ldots, l, m) - Y(t+\tau, \ldots, l, m)| \\
      & - |X(t\to t+\tau, \ldots, l, m) - X'(t\to t+\tau, \ldots, l, m)|
    \bigg],
\end{align*}
where $X$ and $X'$ denote two independent ensemble members.
$\mathcal{C}_{spectral}$ is minimized just when the forecast has the correct spherical harmonic components.
CRPS contributions from lower wavenumbers $l$ encourage forecasts with correct long range correlations, which we found to be critical for training models with good performance.
We excluded wavenumbers $l$ and $m$ from spectral CRPS loss above a maximum wavenumber of 80, because we filter out higher wavenumbers for stability in our dynamical core.

The \emph{nodal} CRPS loss is defined similarly, but uses an area-weighted sum over latitude/longitude points, as in \eqref{align:crps-nodal}.
These two terms are combined to give
\begin{align*}
    \mathcal{L}_{CRPS}(\tau)
    :&= \sum_{t\in\mathcal{T}} \sum_{p,v} \sum_{\substack{\tau'\in\mathcal{D}\\\tau'\leq \tau}}
    \left[
    \mathcal{C}_{spectral}(t, \tau', v) + \mathcal{C}_{nodal}(t, \tau', v)
    \right].
\end{align*}

During training, for each initial time $t$, and forecast time $t+\tau$, we use one observation $Y(t+\tau)$ and exactly two forecasts $X(t+\tau)$ and $X'(t+\tau)$. This is the minimum needed for an unbiased CRPS estimate \eqref{align:crps-ideal}.
It is possible to construct a CRPS loss with more than two forecast samples for every observation. As it turns out, the compute budget (for each minibatch) is better spent on greater variety of initial times. That is, rather than increasing ensemble size, we select a new observation at a new initial time, $Y(s+\tau)$ for $s\neq t$, and once again create the minimum number of forecasts $X(s+\tau)$, $X’(s+\tau)$. This is due to the fact that while adding more ensemble members reduces the variance contribution of forecasts, the variance due to having one single observation is fixed. The result of adding more forecasts would be a sub-linear variance reduction. On the other hand, increasing the number of distinct observations in a minibatch gives a linear reduction in variance.

We note that the loss functions used in NeuralGCM, which promoted a more realistic spectrum and the ability to produce ensemble forecasts, could also be applied in ``pure'' ML approaches.

\subsection{Training resources}\label{apx:subsec:training_times}
Training resources are described in Table~\ref{table:training_times}.
In all cases, we use data parallelism, running our model with one single example on each TPU device, with a batch of examples distributed across multiple TPU cores.
For NeuralGCM-$0.7^{\circ}$ and NeuralGCM-ENS, we additionally use model parallelism across spatial dimensions.

\begin{table}[ht]
\begin{tabular}{|c|c|c|c|c|}
\hline
 & NeuralGCM-$2.8^{\circ}$ & NeuralGCM-$1.4^{\circ}$   & NeuralGCM-$0.7^{\circ}$ & NeuralGCM-ENS ($1.4^{\circ})$ \\ \hline
\begin{tabular}[c]{@{}c@{}}Training time\end{tabular}       & 1 day & 1 week & 3 weeks & 10 days  \\ \hline
\begin{tabular}[c]{@{}c@{}}Device\end{tabular} & 16 TPU v4   & 16 TPU v4         & 256 TPU v4 & 128 TPU v5e     \\ \hline
\begin{tabular}[c]{@{}c@{}}Parallelism\end{tabular} & batch=16   & batch=16        & \begin{tabular}[c]{@{}c@{}} batch=16 \\x=2 \\ y=2 \\ z=4\end{tabular} & \begin{tabular}[c]{@{}c@{}}batch=16 \\x=2 \\ z=2 \\ ensemble=2 \end{tabular}    \\ \hline
\begin{tabular}[c]{@{}c@{}}Parameter count\end{tabular} & 14.5 M   & 18.3 M         & 31.1 M & 11.5 M     \\\hline
\begin{tabular}[c]{@{}c@{}}Inference time\\ on one TPU v4 core \\($10$ day forecast) \end{tabular} & 2.5s  & 12.6s  & 119s & 12.4s \\\hline
\begin{tabular}[c]{@{}c@{}}Inference speed\\ on one TPU v4 core \\(sim days/day) \end{tabular} & \num{350000} & \num{69000}  & \num{7300} & \num{70000} \\\hline
\begin{tabular}[c]{@{}c@{}}Peak memory usage\\ during inference \end{tabular} & 255MB   & 1100MB  & 4177MB & 1011MB \\\hline
\end{tabular}
\caption{Resource requirements for different NeuralGCM models trained on Google Cloud TPUs}.\label{table:training_times}
\end{table}

\clearpage
\section{Additional weather evaluations}\label{apx:sec:additional_weather_evaluation}

\subsection{Accuracy}\label{apx:subsec:additional_evaluation_accuracy}
\begin{figure*}
\thisfloatpagestyle{empty}
\begin{center}
\makebox[\textwidth]{\colorbox{white}{\includegraphics[width=0.7\paperwidth]{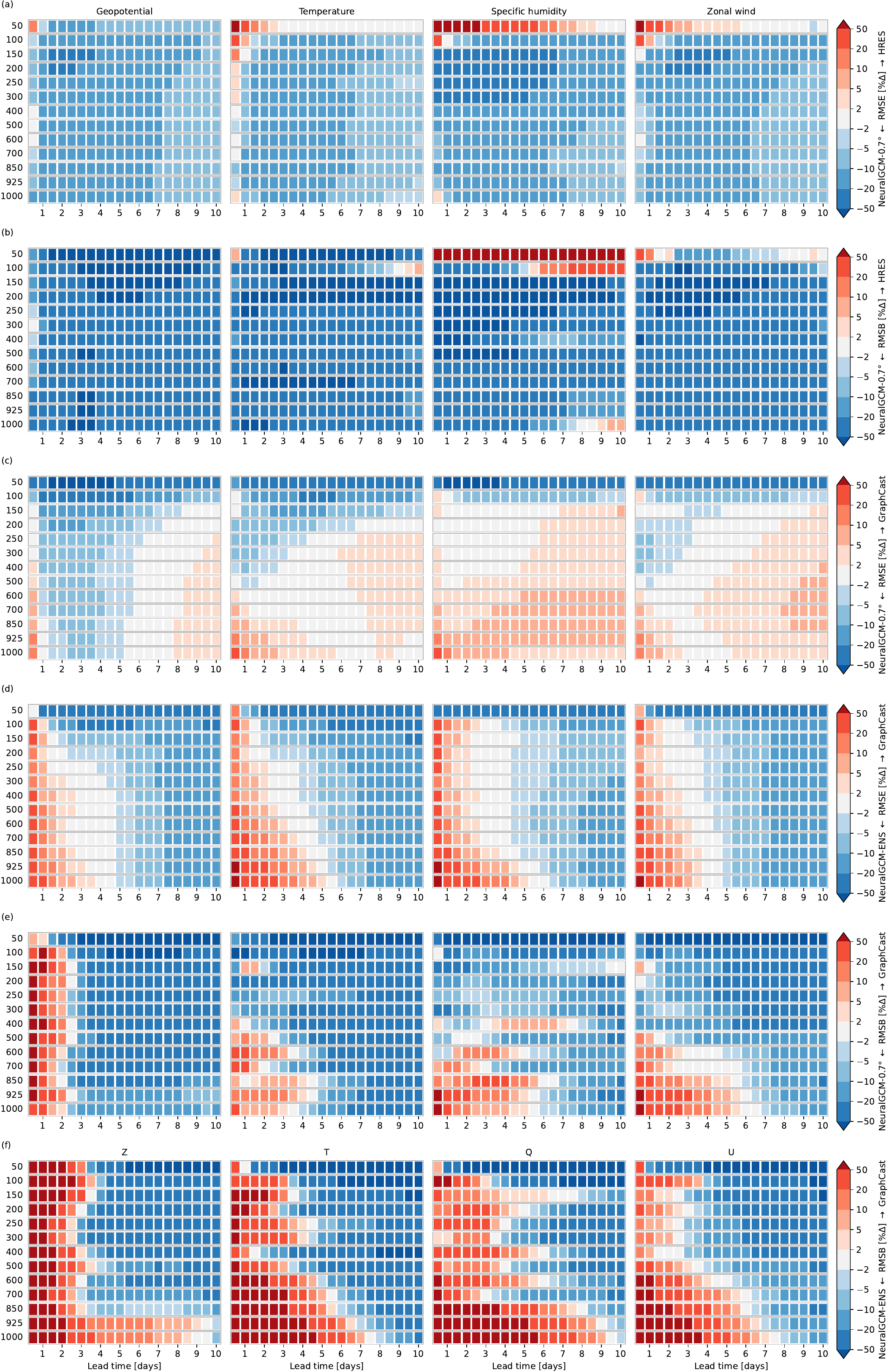}}}
\end{center}
\caption{Scorecards comparing NeuralGCM-$0.7^\circ$ and NeuralGCM-ENS models against ECMWF-HRES and GraphCast on RMSE and RSMB scores across atmospheric variables and pressure levels. (a) RMSE scorecard for NeuralGCM-$0.7^{\circ}$ vs ECMWF-HRES, (b) RSMB scorecard for NeuralGCM-$0.7^{\circ}$ vs ECMWF-HRES, (c) RMSE scorecard for NeuralGCM-$0.7^{\circ}$ vs GraphCast, (d) RMSE scorecard for NeuralGCM-ENS vs GraphCast, (e) bias RMSE scorecard for NeuralGCM-$0.7^{\circ}$ vs GraphCast, (f) bias RMSE scorecard for NeuralGCM-ENS vs GraphCast.
}\label{apx:fig:scorecards_10day}
\end{figure*}

We assess the accuracy of deterministic NeuralGCM, along with other ML models and ECMWF-HRES, by quantifying their skill using the root mean square error (RMSE) and root-mean-square-bias (RMSB), both computed against the relevant ground truth data. As NeuralGCM was trained to predict ERA5 data, we evaluate its errors against ERA5 as the ground truth. In contrast, the ECMWF-HRES model utilizes ECMWF-HRES analysis as input, leading us to compare ECMWF-HRES against ECMWF-HRES analysis, thereby reducing the reported ECMWF-HRES errors. We additionally include scores for the NeuralGCM-ENS mean. This model is not as accurate as NeuralGCM-$0.7^{\circ}$ at short lead times, but indicates the time when the RMSE skill starts to be dominated by forecast uncertainty.

\begin{figure*}
\begin{center}
\makebox[\textwidth]{\colorbox{white}{\includegraphics[width=0.7\paperwidth]{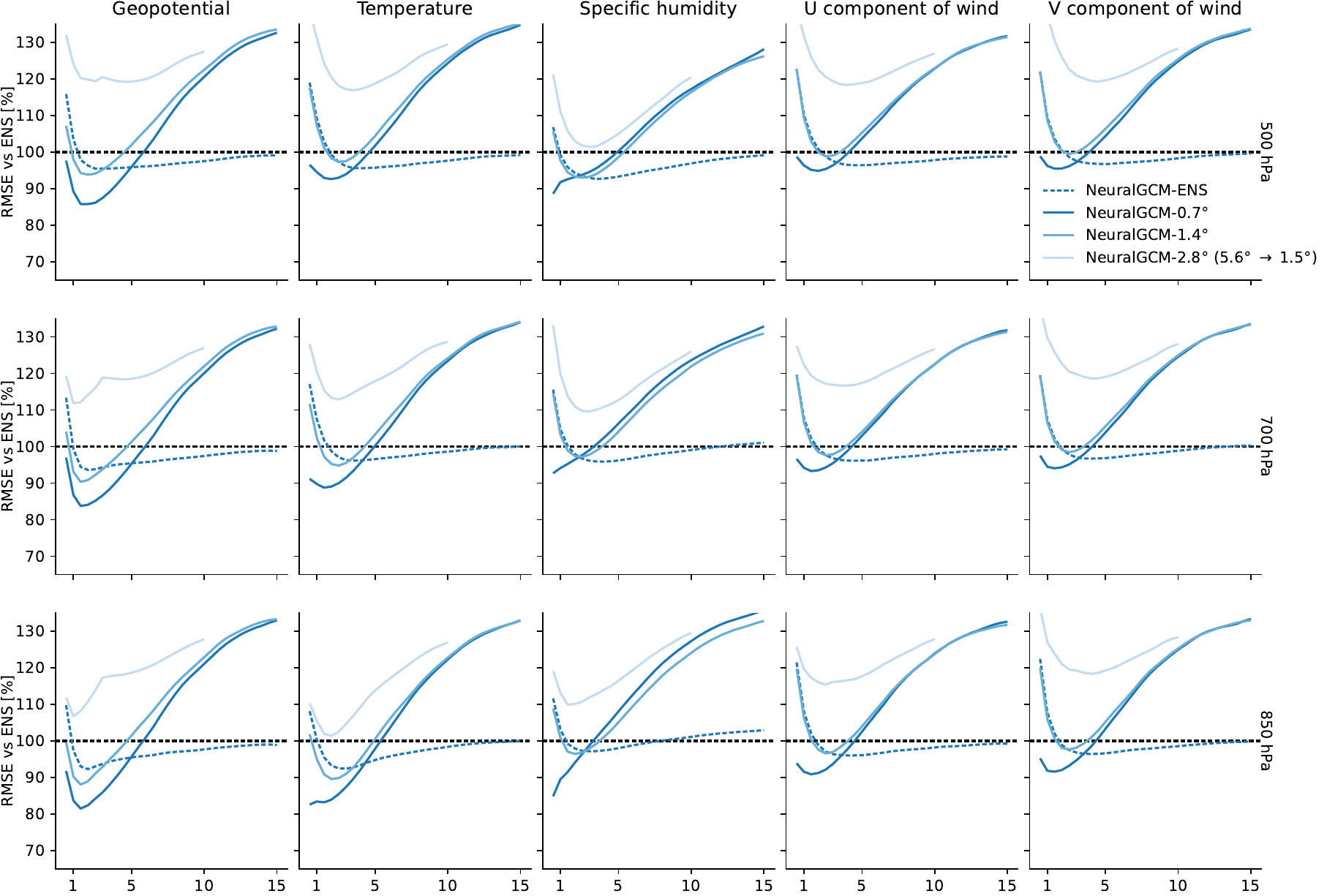}}}
\end{center}
\caption{Root-mean-squared-error (RMSE) relative to ECMWF-ENS for all NeuralGCM forecasts in 2020. 
}
\label{apx:fig:rmse-all-levels}
\end{figure*}

In Fig.~\ref{apx:fig:scorecards_10day} we plot RMSE and RMSB scorecard comparisons of NeuralGCM-$0.7^{\circ}$ and NeuralGCM-ENS against ECMWF-HRES and GraphCast across all core atmospheric variables (geopotential, temperature, specific humidity and u-component of wind) and $13$ pressure levels. Similar to previous ML approaches we find that NeuralGCM-$0.7^{\circ}$ generally outperforms ECMWF-HRES across all scores (Fig.~\ref{apx:fig:scorecards_10day}(a)), except for the top level of the atmosphere. When compared to GraphCast, we find that on short lead times NeuralGCM-$0.7^{\circ}$ performs similarly on RMSE, with some variables better modeled by GraphCast (specific humidity) and some by NeuralGCM-$0.7^{\circ}$ (geopotential) (Fig.~\ref{apx:fig:scorecards_10day}(c)).
We note that NeuralGCM-$0.7^{\circ}$is almost 3 times coarser than GraphCast, and it is likely that higher-resolution NeuralGCM models would improve their short lead time predictions, a trend already observed in NeuralGCM models [Fig.~\ref{apx:fig:rmse-all-levels}].
At longer lead times ($5$-$6$ days) NeuralGCM-ENS achieves better RMSE scores than GraphCast across the board, underlining the utility of stochastic models at such timescales. The general trend of NeuralGCM-$0.7^{\circ}$ to outperform GraphCast at higher levels and being slightly worse at lower levels of the atmosphere is potentially caused by an additional loss weight factor introduced in GraphCast that discounts levels at lower pressure.

\begin{figure*}
\thisfloatpagestyle{empty}
\begin{center}
\makebox[\textwidth]{\colorbox{white}{\includegraphics[width=0.7\paperwidth]{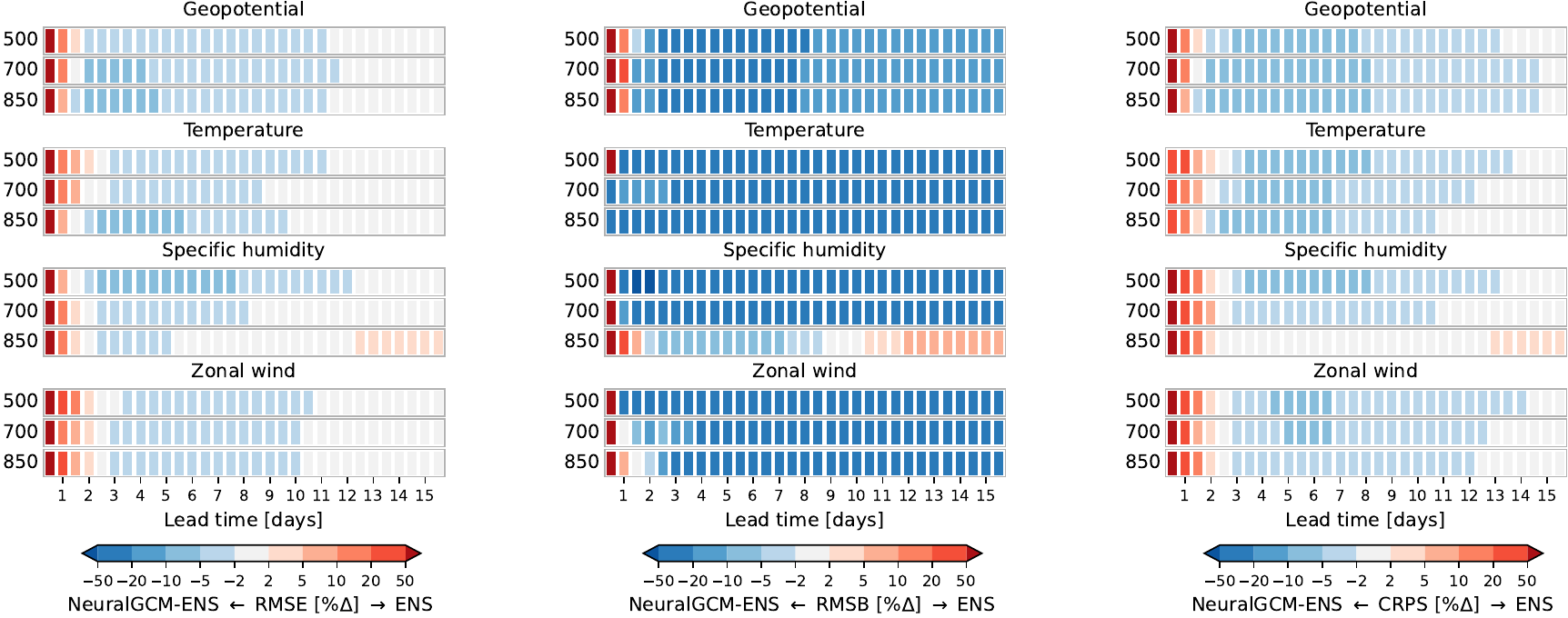}}}
\end{center}
\caption{Scorecards comparing NeuralGCM-ENS against ECMWF-ENS. Scorecards are shown for (a) RMSE, (b) RMSB and (c)CRPS for different atmospheric variables and pressure levels.}\label{apx:fig:scorecards_15day}
\end{figure*}

Contrary to RMSE, RMSB is not a priori expected to rapidly increase with lead time. NeuralGCM-$0.7^{\circ}$ model maintains consistently lower bias than ECMWF-HRES for the $10$ days (Fig.~\ref{apx:fig:scorecards_10day}(b)). GraphCast achieves lowest bias at very early lead times, but quickly degrades afterwards (Fig.~\ref{apx:fig:scorecards_10day}(e)). NeuralGCM-ENS model has slightly higher biases compared to NeuralGCM-$0.7^{\circ}$ model, likely due to coarser spatial resolution. In Fig.~\ref{apx:fig:scorecards_15day} we provide similar RMSE and RMSB scorecards comparing NeuralGCM-ENS to ECMWF-ENS on 15 day forecasts.

\begin{figure*}
\begin{center}
\makebox[\textwidth]{\colorbox{white}{\includegraphics[width=0.7\paperwidth]{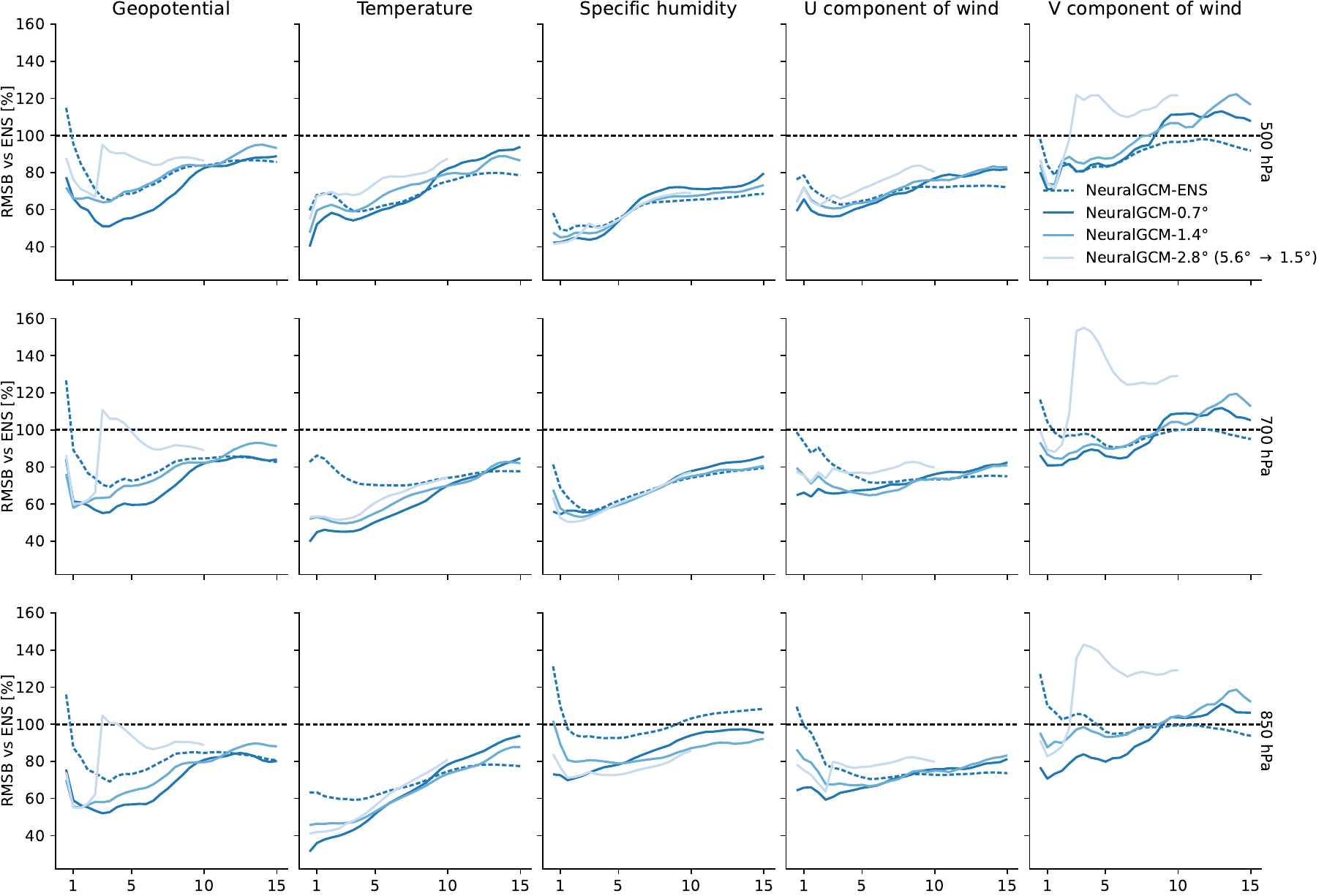}}}
\end{center}
\caption{Root-mean-squared-bias (RMSB) relative to ECMWF-ENS for all NeuralGCM forecasts in 2020. 
}
\label{apx:fig:rmsb-all-levels}
\end{figure*}

\begin{figure*}
\begin{center}
\makebox[\textwidth]{\colorbox{white}{\includegraphics[width=0.7\paperwidth]{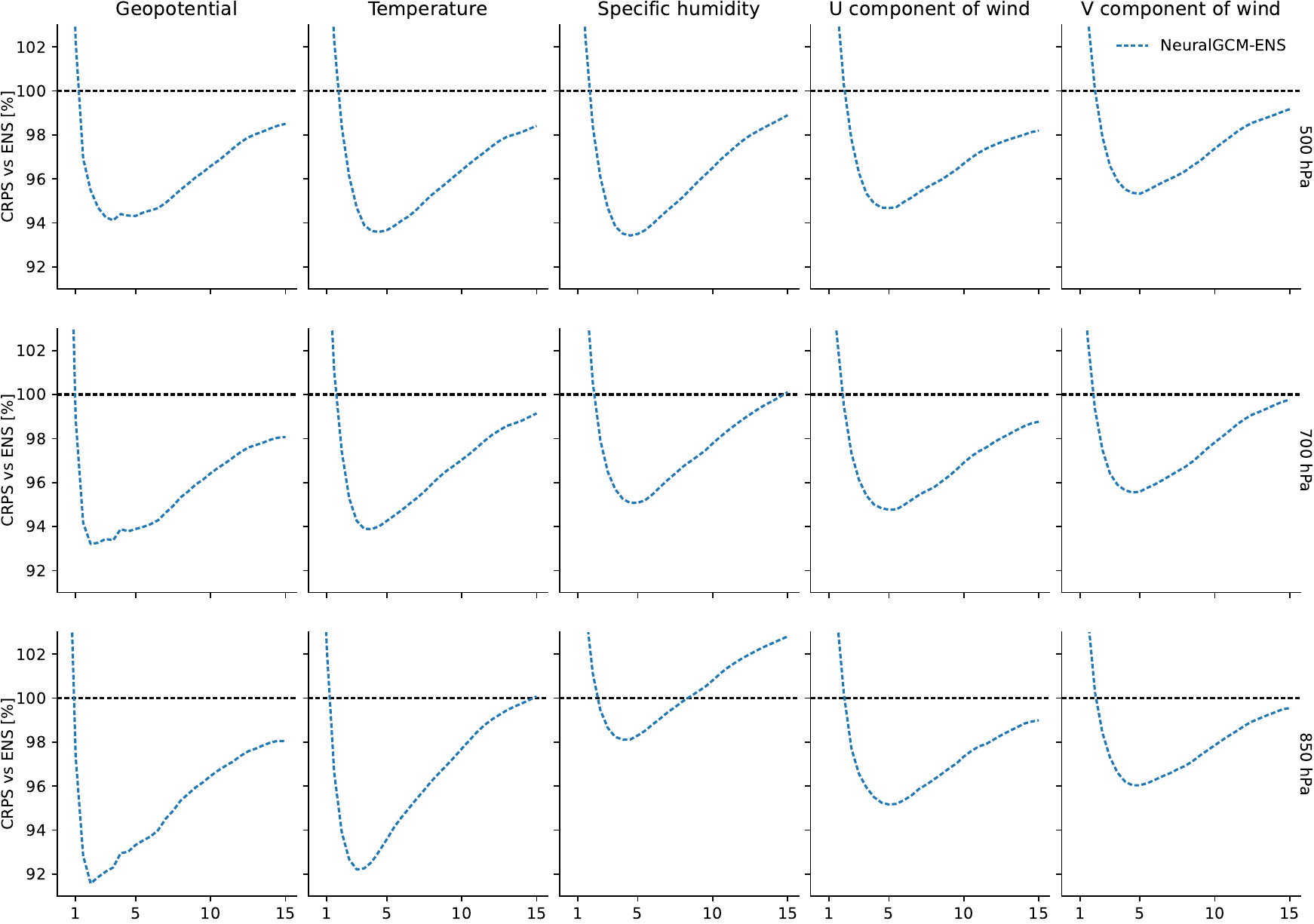}}}
\end{center}
\caption{Continuous ranked probability score (CRPS) relative to ECMWF-ENS for NeuralGCM-ENS forecasts in 2020.}
\label{apx:fig:crps-all-levels}
\end{figure*}

\begin{figure*}
\begin{center}
\makebox[\textwidth]{\colorbox{white}{\includegraphics[width=0.7\paperwidth]{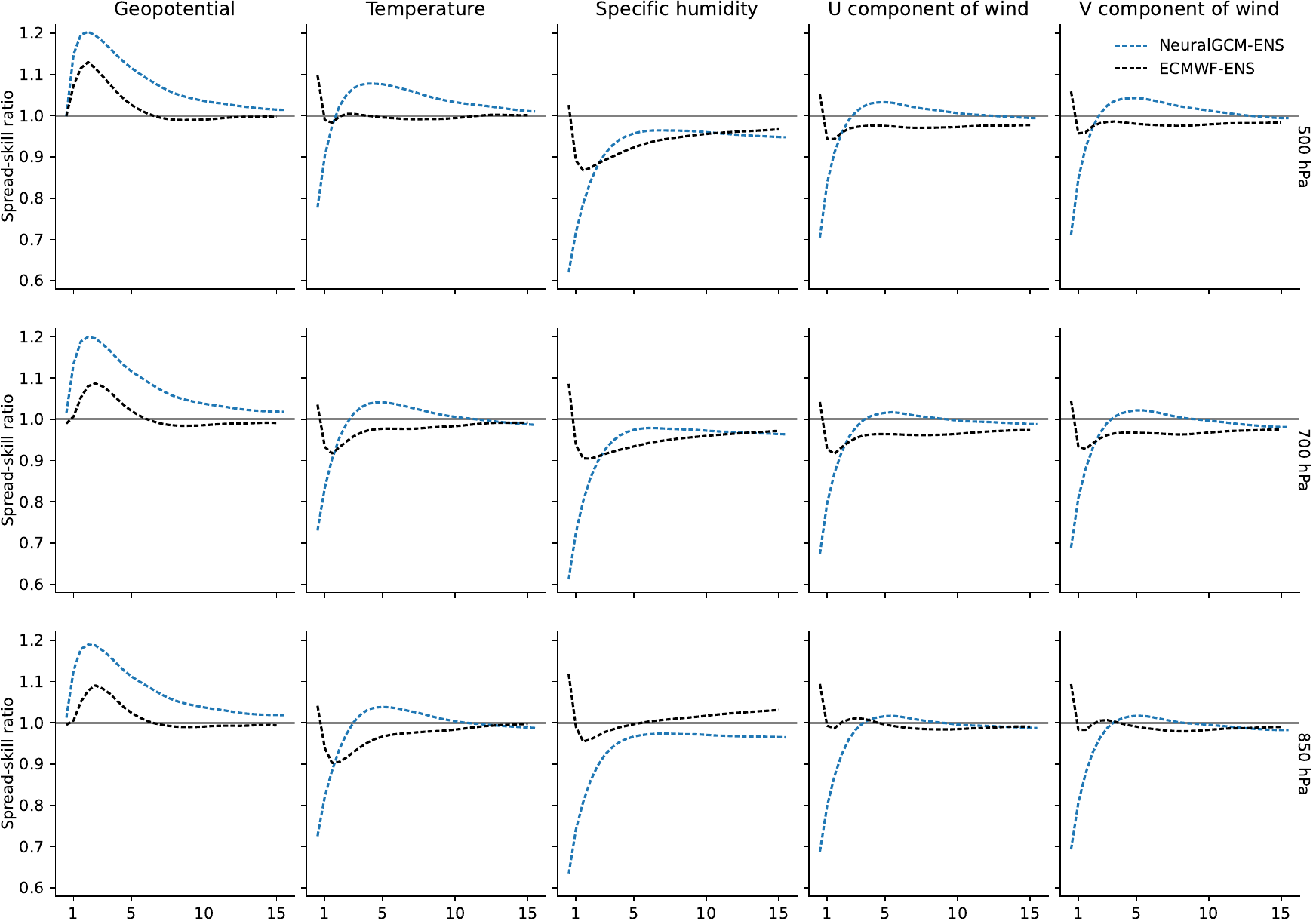}}}
\end{center}
\caption{Spread-skill-ratio for NeuralGCM-ENS and ECMWF-ENS forecasts in 2020.}
\label{apx:fig:ssr-all-levels}
\end{figure*}

Figures~\ref{apx:fig:rmse-all-levels}, \ref{apx:fig:rmsb-all-levels}, \ref{apx:fig:crps-all-levels} and  \ref{apx:fig:ssr-all-levels} compare NeuralGCM models against ECMWF-ENS across all core atmospheric variables and $3$ pressure levels ($500$, $700$ and $850$ hPa), for which we had ECMWF-ENS data to compare against.
Beyond two days, NeuralGCM-ENS has lower or similar error (within 1\%) compared to  ECMWF-ENS across all metrics (RMSE, RSMB and CRPS), with the exception of specific humidity at 850 hPa at longer lead times. 
At very early lead times ECMWF-ENS has better scores than NeuralGCM-ENS. However, both RMSE and CRPS values at these early times are small, indicating a negligible absolute difference between the models. ECMWF-ENS's advantage possibly stems from two factors:
a) NeuralGCM-ENS was optimized for 5-day rollouts, and initial condition perturbations likely improve later-stage predictions while introducing minor errors (in absolute terms) at earlier stages. Overall, these perturbations likely reduce loss.
b) NeuralGCM's lower resolution ($1.4^{\circ}$) compared to ECMWF-ENS ($0.2^{\circ}$) limits its ability to incorporate fine-scale initial perturbations. Namely, ECMWF-ENS can introduce these perturbations while still matching IFS-analysis at a coarser scale, whereas perturbations in NeuralGCM would always increase RMSE.

\begin{figure*}
\begin{center}
\makebox[\textwidth]{\colorbox{white}{\includegraphics[width=0.7\paperwidth]{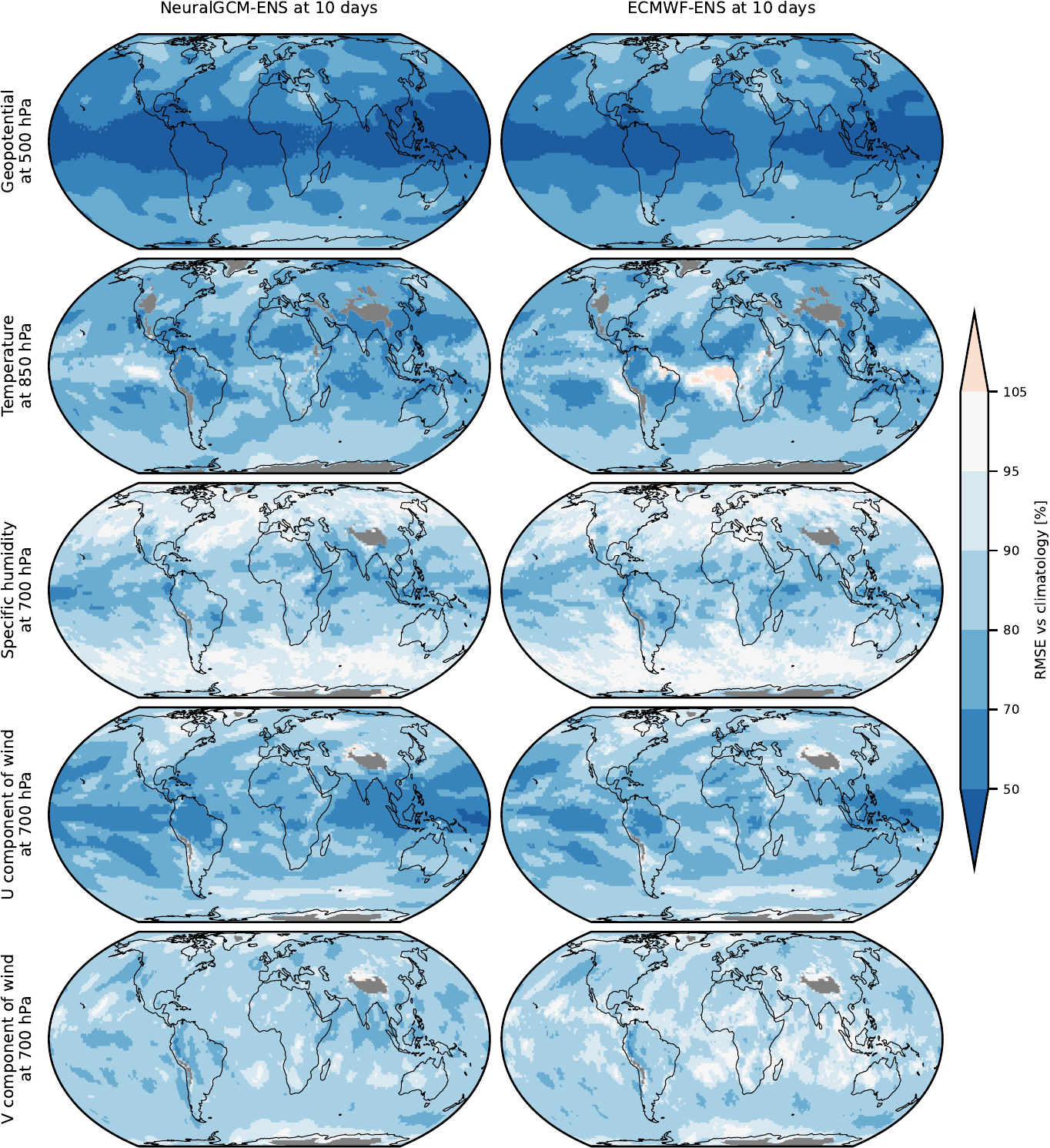}}}
\end{center}
\caption{Maps of root-mean-squared-error for NeuralGCM-ENS and ECMWF-ENS relative to 1990-2019 climatology for all forecasts in 2020.}
\label{apx:fig:rmse-map}
\end{figure*}

\begin{figure*}
\begin{center}
\makebox[\textwidth]{\colorbox{white}{\includegraphics[width=0.7\paperwidth]{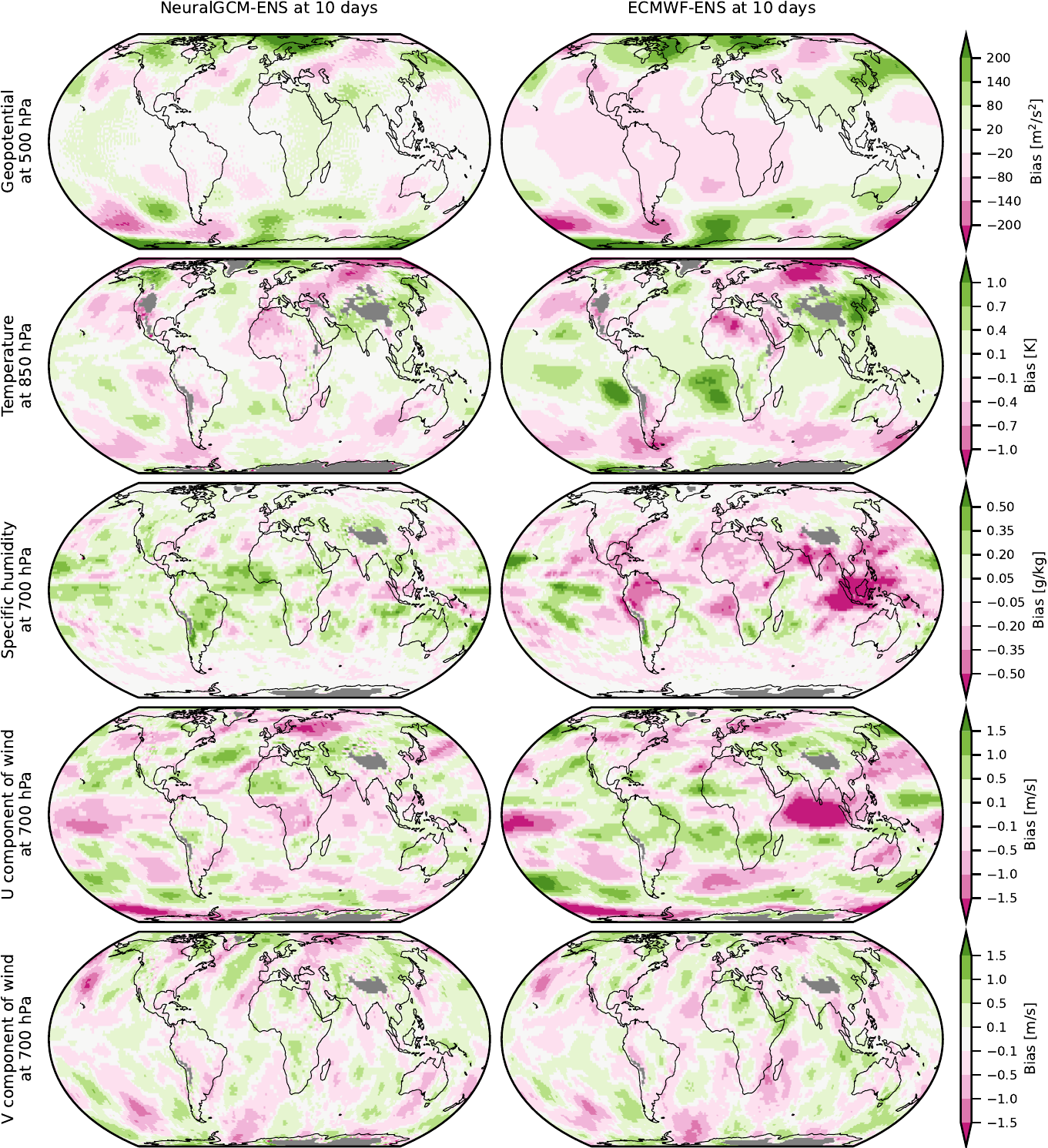}}}
\end{center}
\caption{Maps of average bias for NeuralGCM-ENS and ECMWF-ENS for all forecasts in 2020.}
\label{apx:fig:bias-map}
\end{figure*}

\begin{figure*}
\begin{center}
\makebox[\textwidth]{\colorbox{white}{\includegraphics[width=0.7\paperwidth]{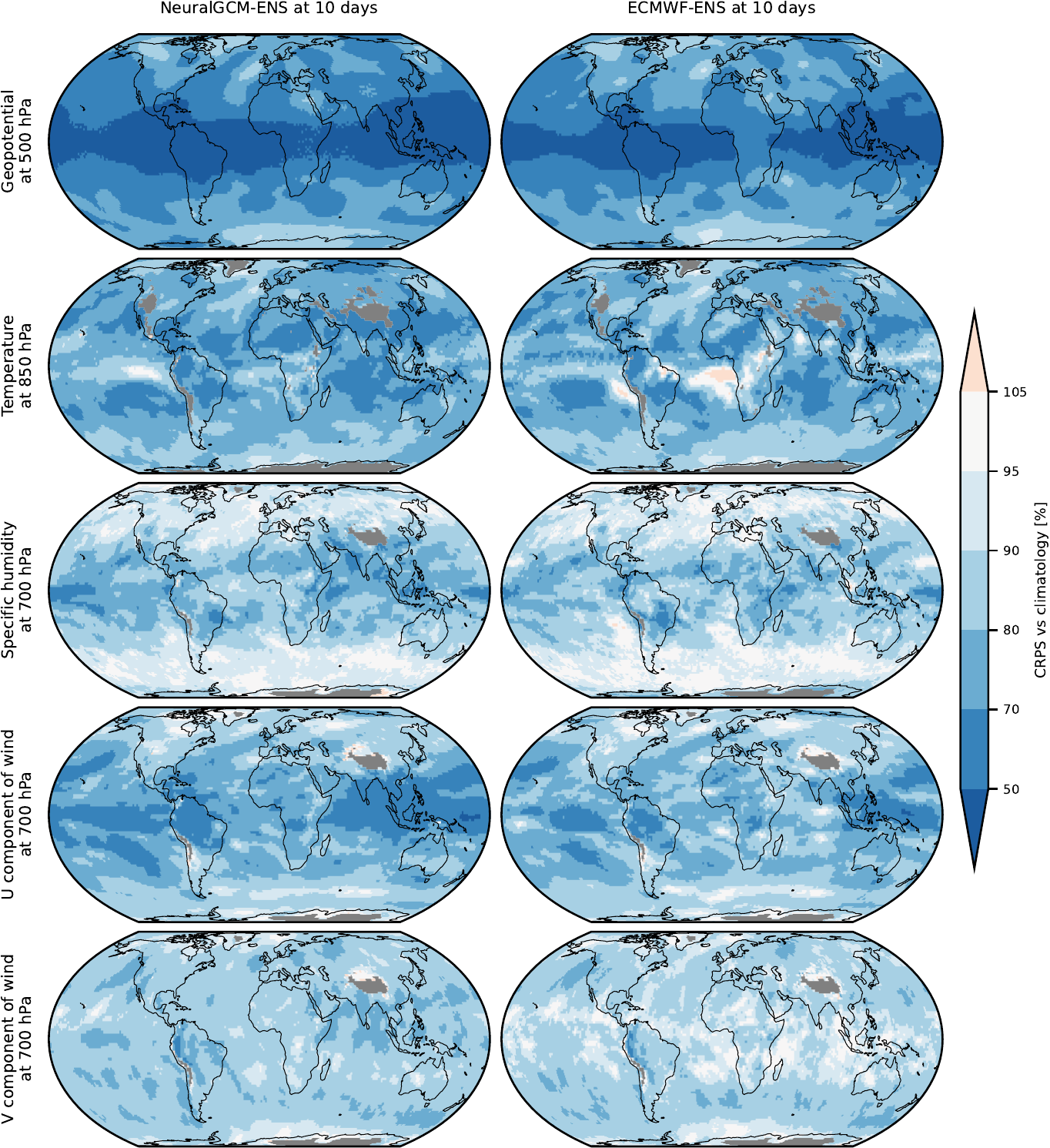}}}
\end{center}
\caption{Maps of average CRPS for NeuralGCM-ENS and ECMWF-ENS relative to a probabilistic climatology sampled from the years 1990-2019 for all forecasts in 2020.}
\label{apx:fig:crps-map}
\end{figure*}

\begin{figure*}
\begin{center}
\makebox[\textwidth]{\colorbox{white}{\includegraphics[width=0.7\paperwidth]{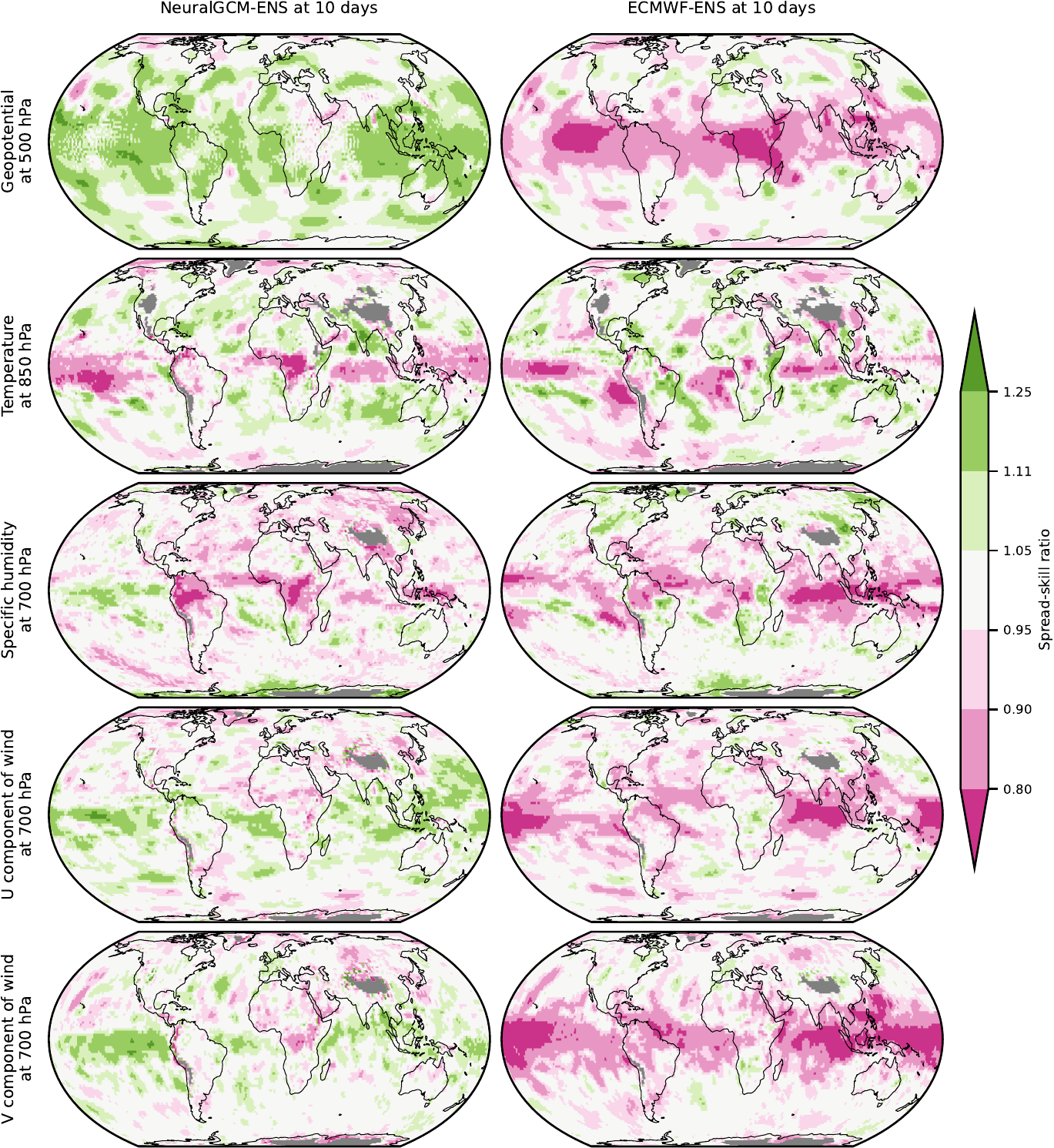}}}
\end{center}
\caption{Maps of spread-skill ratio for NeuralGCM-ENS and ECMWF-ENS relative to a probabilistic climatology sampled from the years 1990-2019 for all forecasts in 2020.}
\label{apx:fig:ssr-map}
\end{figure*}

Figures~\ref{apx:fig:rmse-map}, \ref{apx:fig:bias-map}, \ref{apx:fig:crps-map} and \ref{apx:fig:ssr-map} compare spatial metrics for 10-day forecasts from NeuralGCM-ENS and ECMWF-ENS across all core atmospheric variables on WeatherBench2 pressure levels.
Overall, NeuralGCM-ENS and ECMWF-ENS have similar spatial patterns of skill relative to climatology as measured by RMSE and CRPS.
Spatial biases for Neural-GCM are better overall than for ECMWF-ENS, and are also noticeably more evenly distributed.
Comparing spread-skill ratio shows that unlike ECMWF-ENS, NeuralGCM-ENS is not consistently under-dispersed in the tropics.

\subsection{Derived variables and spectra}\label{apx:subsec:derived_variables_and_consistency}

To better understand the consistency of ML weather forecasts, we calculate a variety of the variables that can be derived from geopotential, temperature, horizontal wind velocity and specific humidity:

\begin{enumerate}
    \item \textbf{Lapse rate} is given by
    \begin{align*}
        \frac{\partial T}{\partial z} = \left( \frac{\partial T}{\partial p} \right) \left( \frac{1}{g} \frac{\partial \Phi}{\partial p} \right)^{-1},
    \end{align*}
    where $T$ is temperature, $p$ is pressure, $g$ is the gravitional constant and $\Phi$ is geopotential.
    \item \textbf{Wind speed} is given by
    \begin{align*}
        \sqrt{u^2 + v^2} 
    \end{align*}
    where $u$ and $v$ denote zonal and meridional wind velocity.
    \item \textbf{Divergence} is given by
    \begin{align*}
        \nabla_p \cdot \mathbf{u} = \partial_x u + \partial_y v,
    \end{align*}
    where $\partial_x$ and $\partial_y$ denote the partial derivatives in the zonal and meridional directions.
    \item \textbf{Vorticity} is given by
    \begin{align*}
        \mathbf{\hat k} \cdot (\nabla_p \times \mathbf{u}) = \partial_x v - \partial_y u.
    \end{align*}
    \item \textbf{Vertical velocity} at pressure level $p_0$ under the assumption of hydrostatic balance $\partial p/\partial z = -\rho g$ is given by~\cite{durran2010numerical}
    \begin{align*}
        - \int_0^{p_0} (\nabla_p \cdot \mathbf{u})\,dp
        =
        - \int_0^{p_0}  [\partial_x u + \partial_y v]\,dp.
    \end{align*}
    \item \textbf{Eddy kinetic energy} is given by
    \begin{align*}
        \frac{1}{2} \left[ (u - \bar u)^2 + (v - \bar v)^2 \right],
    \end{align*}
    where $\bar u$ and $\bar v$ denote the longitudinal mean of zonal and meridional wind velocity, respectively.
    \item \textbf{Geostrophic wind} is a horizontal vector given by
    \begin{align*}
        \frac{1}{2 \Omega \sin \phi} \mathbf{\hat k} \times \nabla_p \Phi
        = \frac{1}{2 \Omega \sin \phi} \left[ -\partial_y \Phi, \partial_x \Phi \right],
    \end{align*}
    where $\Omega$ is the rotational speed of the Earth, $\phi$ is latitude in radians, $\nabla_p$ is the gradient on constant pressure levels and $\Phi$ is geopotential~\cite{bonavita2023limitations}.
    \item \textbf{Ageostrophic wind} is the horizontal wind vector minus geostrophic wind.
    \item \textbf{Total column moisture} quantities are given by a vertical integral over pressure levels,
    \begin{align*}
        \frac{1}{g} \int q\,dp ,
    \end{align*}
    where $g$ is the gravitational constant and $q$ is the desired moisture species (humidity/vapor, cloud liquid or cloud ice).
    \item \textbf{Integrated vapor transport} is given by
    \begin{align*}
        \frac{1}{g} \sqrt{ \left( \int_{p_0}^{p_1}  q u \,dp\right)^2 + \left(\int_{p_0}^{p_1} q v \,dp\right)^2},
    \end{align*}
    where $q$ is specific humidity, $u$ is zonal wind, $v$ is meridional wind, $p_0 = \SI{300}{hPa}$ is the minimum pressure and $p_1 = \SI{1000}{hPa}$ is the maximum pressure~\cite{lam2022graphcast}.
    \item \textbf{Relative humidity} is given by
    \begin{align*}
        \frac{q/(1 - q)}{\epsilon e_s / (p - e_s)}
    \end{align*}
    where $\epsilon = \num{0.622}$ and
    $e_s$ the saturation vapor pressure in hPa is
    \begin{align*}
        e_s = 6.112 \exp[ 17.67 (T - 273.15) / (T - 29.65)],
    \end{align*}
    where $T$ is the temperature in Kelvin, which is the formula implemented by MetPy~\citeSI{metpy}.
\end{enumerate}

All derivatives are calculated using second-order finite differences, and vertical integrals are calculated using trapezoidal integration.
Code for calculating these variables has been added to the WeatherBench2 codebase~\cite{rasp2023weatherbench}.

\begin{figure*}
\thisfloatpagestyle{empty}
\begin{center}
\makebox[\textwidth]{\colorbox{white}{\includegraphics[width=0.85\paperwidth]{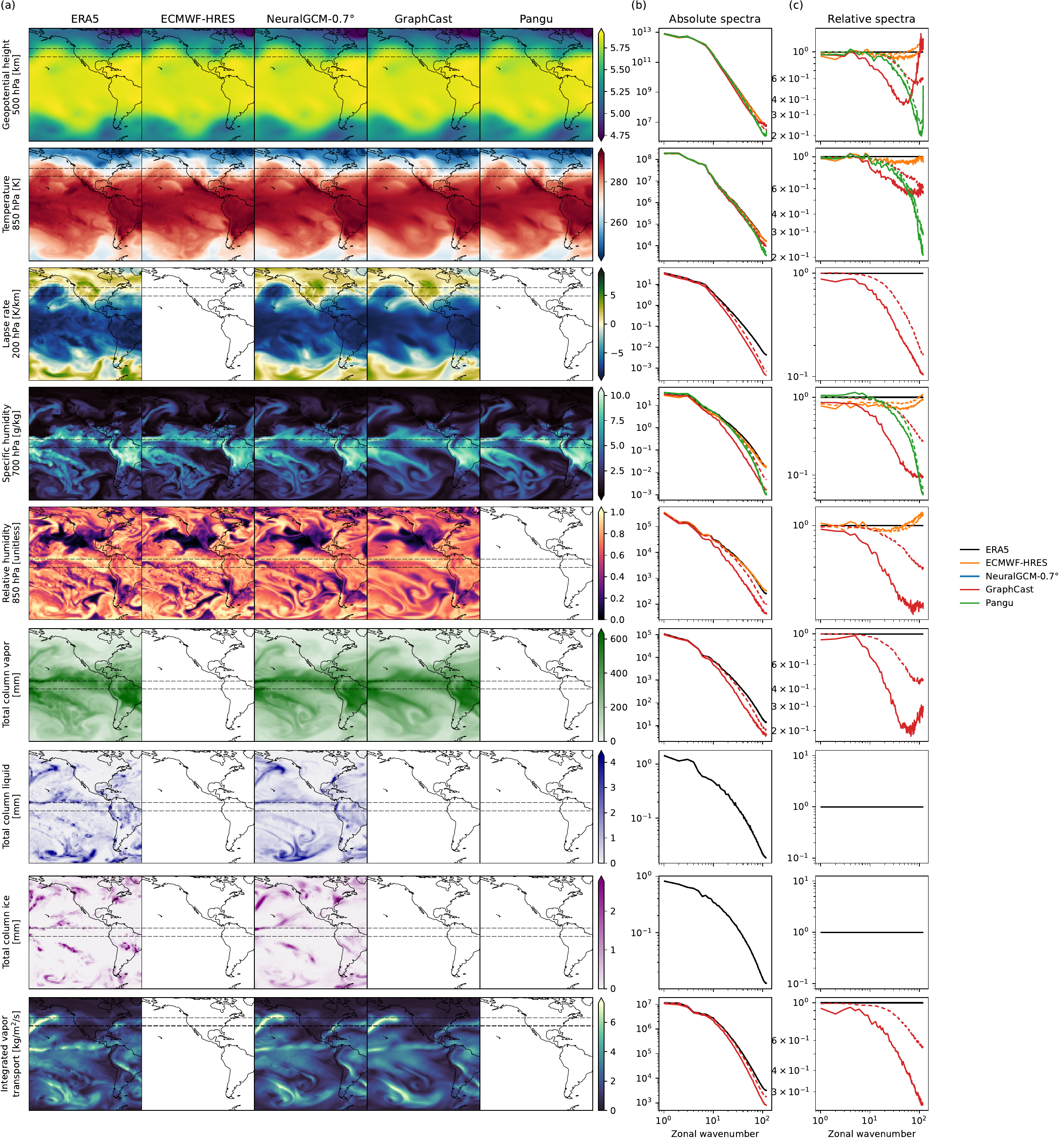}}}
\end{center}
\caption{Forecasts and power spectra for forecasts of thermodynamic variables.
(a) Example 7-day forecasts initialized at 2020-01-01T00.
(b) Absolute spectral density for 1-day (dashed) and 7-day (solid) forecasts of the indicated field, averaged over all forecasts initialized in 2020 and either over the tropics ($[-15^\circ, 15^\circ]$ North) or the extratropics ($[25^\circ, 55^\circ]$ North) as indicated by the area between dashed lines in panel (a).
(c) Normalized spectral density, given by the absolute spectral density divided by the spectral density of ERA5.
}\label{fig:physical_consistency_thermo}
\end{figure*}

\begin{figure*}
\begin{center}
\makebox[\textwidth]{\colorbox{white}{\includegraphics[width=0.85\paperwidth]{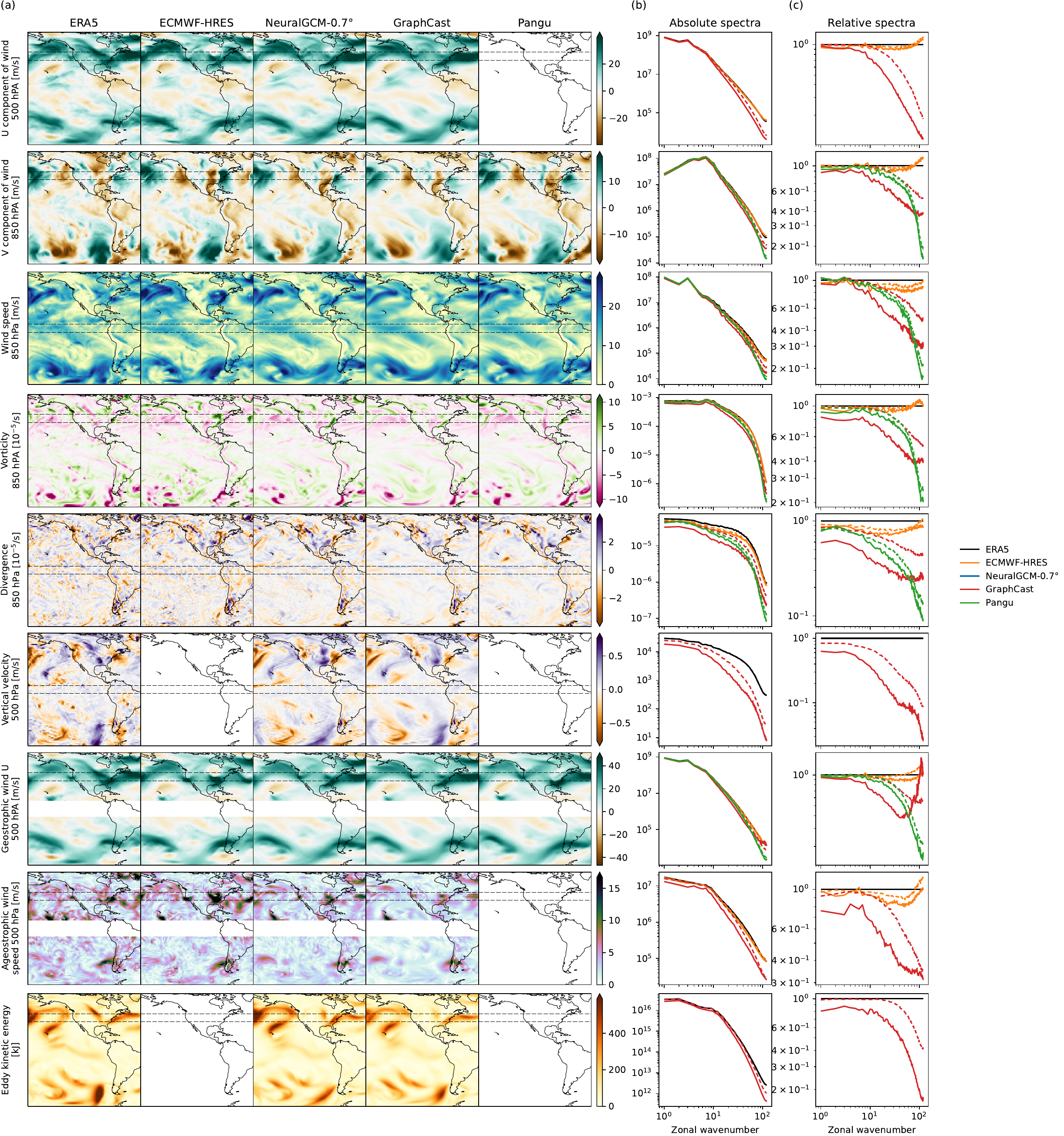}}}
\end{center}
\caption{Forecasts and power spectra for forecasts of wind variables, like in Fig.~\ref{fig:physical_consistency_thermo}. Note that vertical velocity is recalculated for GraphCast from horizontal wind, rather than using the vertical velocity prediction directly output by GraphCast.
}\label{fig:physical_consistency_wind}
\end{figure*}

Fig.~\ref{fig:physical_consistency_thermo} and Fig.~\ref{fig:physical_consistency_wind} plots a variety of example 7-day forecast fields and power spectra averaged over many weather forecasts, based archived weather forecasts from WeatherBench2.
To consistently compare predictions from models with different resolutions, all calculations are performed after conservative regridding to a 1.5$^{\circ}$ equiangular grid on 37 pressure levels.
Variables are excluded if calculating them requires fields not archived as part of WeatherBench2 for a particular forecast model, including vertical integrals or derivatives if not all pressure levels are available.
We omit NeuralGCM-ENS because conservative regridding from a 1.4$^\circ$ Gaussian to a 1.5$^\circ$ equiangular grid introduces aliasing artifacts that are particularly evident for derived variables.

\begin{figure*}
\begin{center}
\makebox[\textwidth]{\colorbox{white}{\includegraphics[width=0.85\paperwidth]{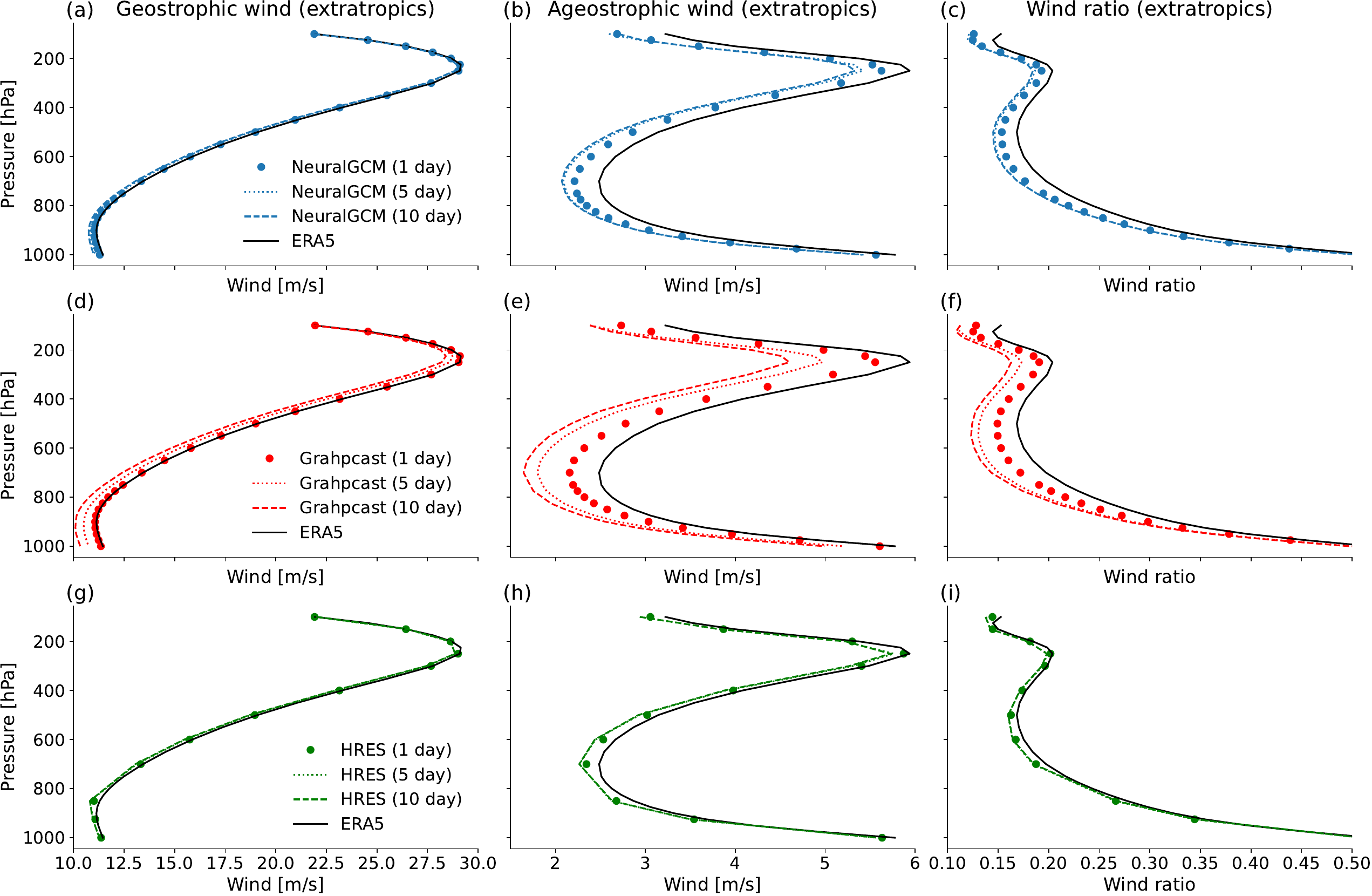}}}
\end{center}
\caption{Geostrophic balance in NeuralGCM, GraphCast and ECMWF-HRES. Vertical profiles of the extratropical intensity (averaged between latitude 30$^{\circ}$-70$^{\circ}$ in both hemispheres) and over initial conditions of (a,d,g) geostrophic wind, (b,e,h) ageostrophic wind and (c,f,i) the ratio of the intensity of ageostrophic wind over geostrophic wind for ERA5 (black continuous line in all panels), (a,b,c) NeuralGCM-$0.7^{\circ}$, (d,e,f) GraphCast and (g,h,i) ECMWF-HRES at lead times of 1 day, 5 days and 10 days.
}\label{sifig:geostrophic_wind_vertical}
\end{figure*}

\subsection{Visualization of ensemble weather forecasts}
\label{apx:subsec:ensemble_visualization}

\begin{figure*}
\begin{center}
\makebox[\textwidth]{\colorbox{white}{\includegraphics[width=0.85\paperwidth]{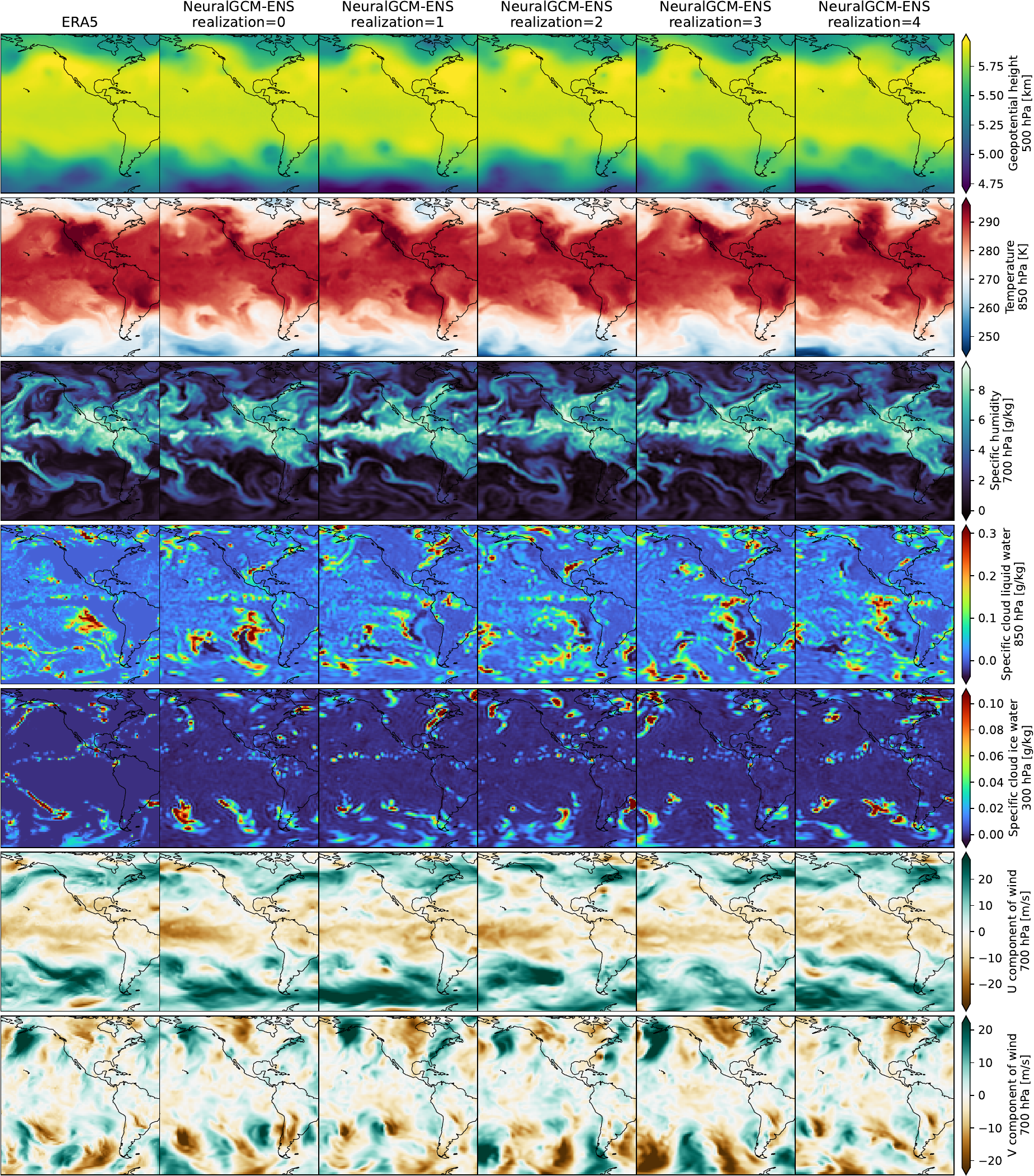}}}
\end{center}
\caption{Maps of five ensemble realizations of +15 day forecast fields from NeuralGCM-ENS initialized at 2020-08-22T12z, compared to ERA5.
}\label{sifig:ens-examples-maps}
\end{figure*}

\begin{figure*}
\begin{center}
\makebox[\textwidth]{\colorbox{white}{\includegraphics[width=0.85\paperwidth]{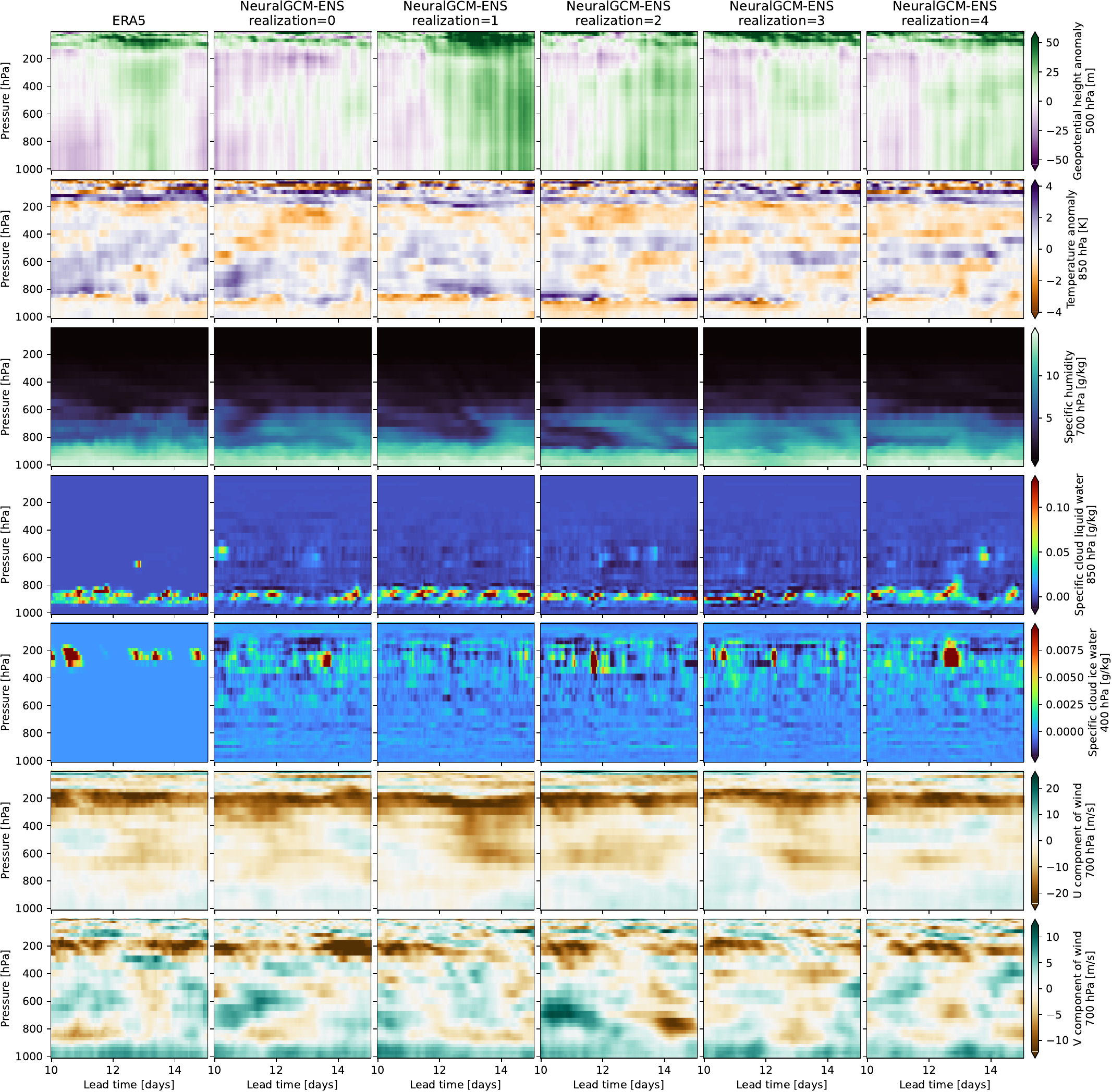}}}
\end{center}
\caption{Vertical profiles of five ensemble realizations of 10-15 day forecast fields at 0$^\circ$ N 0$^\circ$ W over the tropical Pacific ocean, from NeuralGCM-ENS initialized at 2020-08-22T12z, compared to ERA5.
}\label{sifig:ens-examples-tropics-vertical-profile}
\end{figure*}

\begin{figure*}
\begin{center}
\makebox[\textwidth]{\colorbox{white}{\includegraphics[width=0.85\paperwidth]{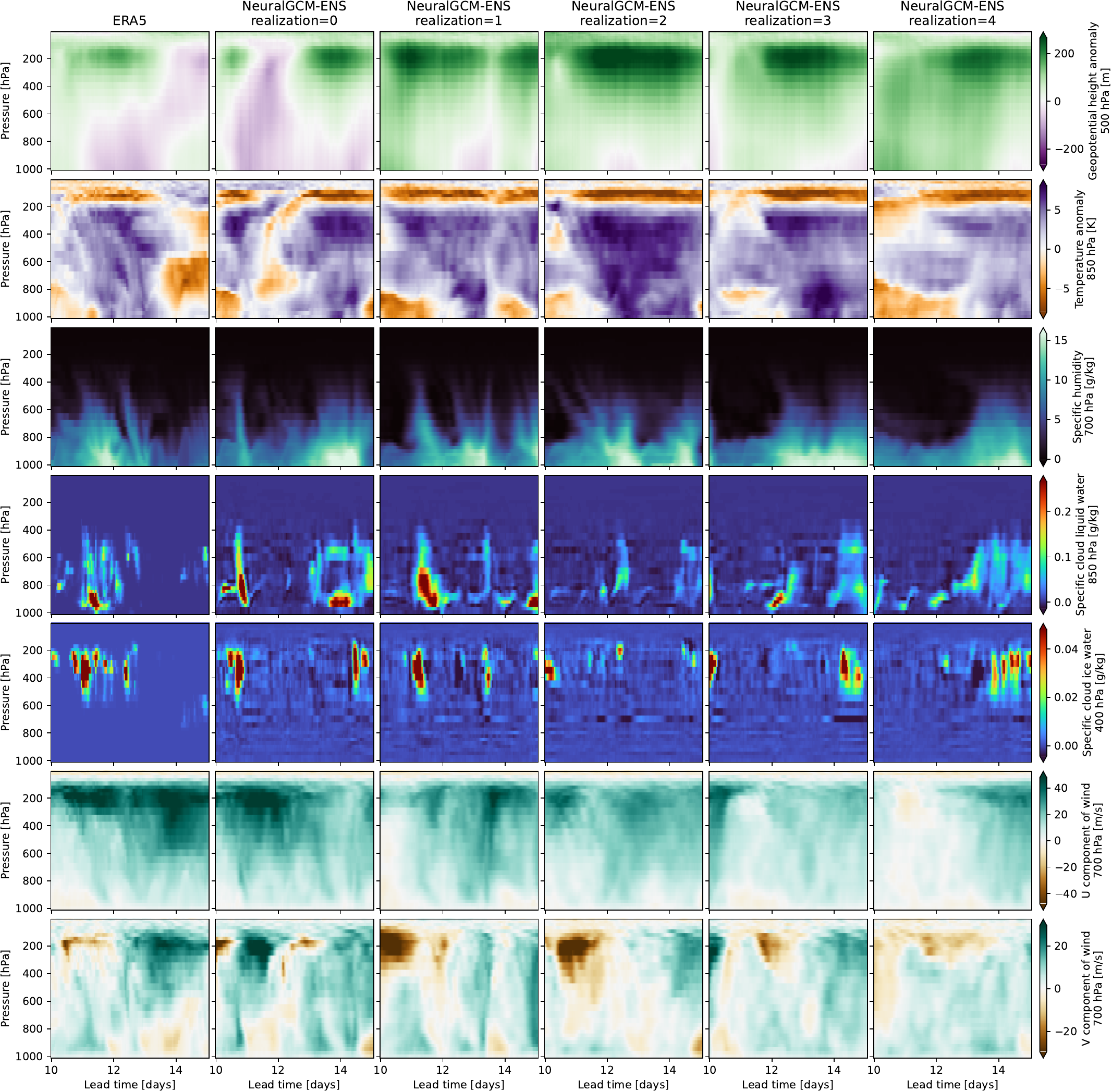}}}
\end{center}
\caption{Like Fig.~\ref{sifig:ens-examples-tropics-vertical-profile}, but at 42.3$^\circ$ N 71.1$^\circ$ W over Boston, MA, USA.
}\label{sifig:ens-examples-boston-vertical-profile}
\end{figure*}

To verify that NeuralGCM-ENS produces reasonable looking forecasts of all variables, we plot maps of forecast fields in Fig.~\ref{sifig:ens-examples-maps} and vertical profiles in Fig.~\ref{sifig:ens-examples-tropics-vertical-profile} and Fig.~\ref{sifig:ens-examples-boston-vertical-profile}.
Here we compare to ERA5 after conservative horizontal regridding to the native resolution of NeuralGCM-ENS.
The forecasts look qualitatively similar to ERA5, except the cloud variables (specific cloud liquid water content and specific cloud ice water content) occasionally take on small in magnitude negative values, which is not physically possible.
Fixing this issue represents an improvement opportunity for improving future versions of NeuralGCM.

\subsection{Evaluation of lower resolution models}
In Fig.~\ref{apx:fig:rmse-all-levels},\ref{apx:fig:rmsb-all-levels} we compare RMSE skill and root mean squared bias of NeuralGCM models at $0.7^{\circ}$, $1.4^{\circ}$ and $2.8^{\circ}$ resolutions, evaluated on the year $2020$. Similar to the main text we normalize the results against ECMWF-ENS for easier comparison. Across all atmospheric variables we find that increasing the level of detail improves RMSE. Note that the skill of NeuralGCM-$2.8^{\circ}$ is evaluated on $5.6^{\circ}$ and then rescaled by the ratio of ECMWF-HRES errors at $1.5^\circ$ and $5.6^\circ$ resolutions to estimate model performance without upsampling coarse predictions.

\begin{figure*}
\begin{center}
\makebox[\textwidth]{\colorbox{white}{\includegraphics[width=0.85\paperwidth]{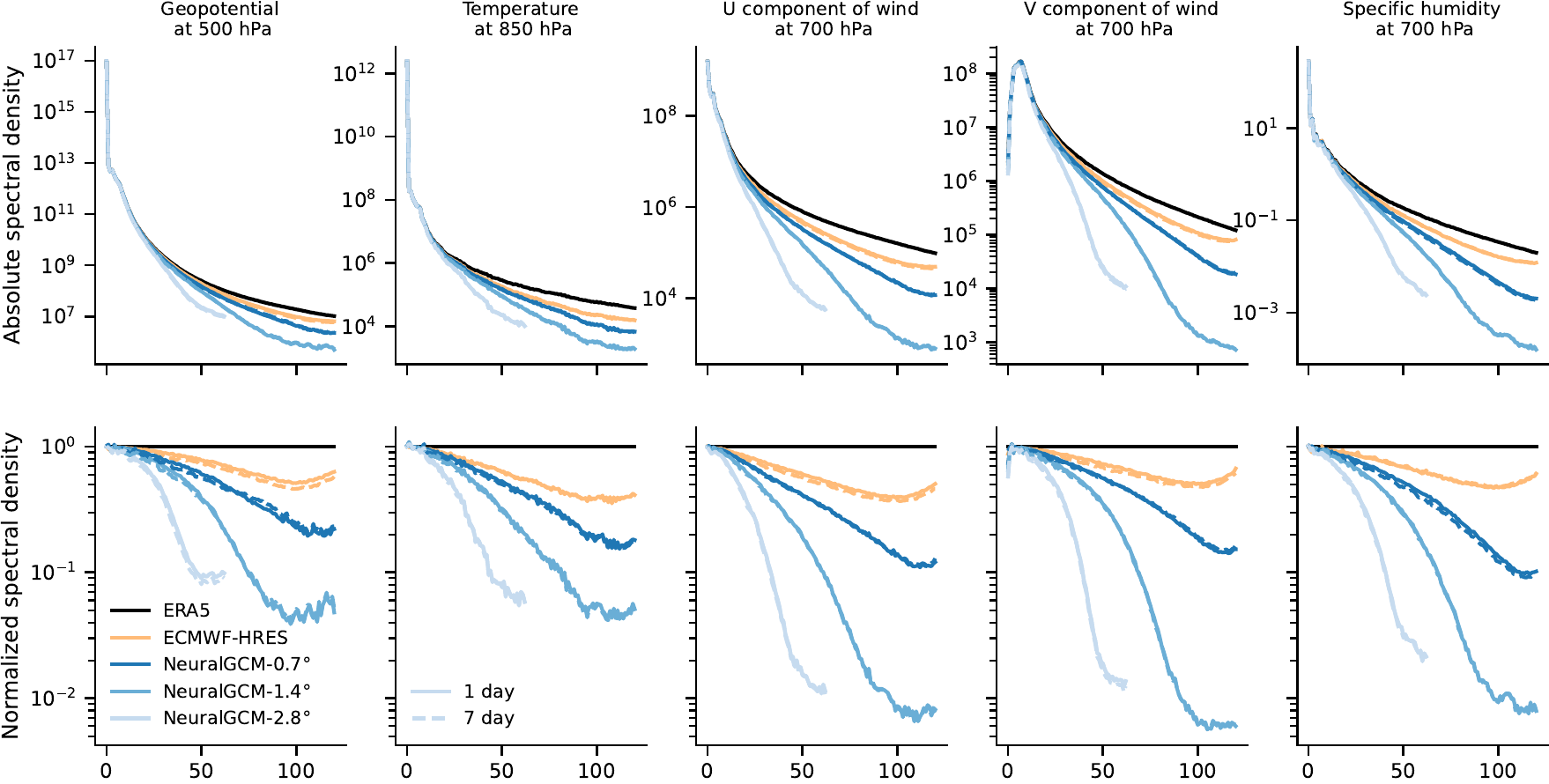}}}
\end{center}
\caption{Comparison of NeuralGCM models at different resolutions on zonal power spectra averaged over latitudes $15^{\circ}$ to $55^{\circ}$ for core atmospheric variables. (a) Absolute spectral density. (b) Spectral density normalized against ERA5.}\label{sifig:multiple_resolutions_spectra}
\end{figure*}

Fig.~\ref{sifig:multiple_resolutions_spectra} compares power spectra of core atmospheric variables for NeuralGCM models of different resolution at $1$ and $7$ days into the forecast.

\subsection{Diagnosing precipitation minus evaporation}\label{apx:sec:diagnose_precipitation}
To diagnose precipitation minus evaporation rate ($P-E$) we calculate:
\begin{equation}\label{eq:p_minus_e}
    P-E = \frac{1}{g}\int_{0}^{1} \sum_{\rm{i}} \left(\frac{dq}{dt}\right)_{\rm{i}}^{\rm{NN}} p_{\rm{s}} d\sigma
\end{equation}
where $p_{\rm{s}}$ is the surface pressure, and $\sum_{\rm{i}} (\frac{dq}{dt})_{\rm{i}}^{\rm{NN}}$ is the sum of the water species tendencies predicted by the neural network. To compare against reanalysis data we use the precipitation rate and evaporation rate from ERA5. 

\begin{figure*}
\centering
\includegraphics[width=1.0\textwidth]{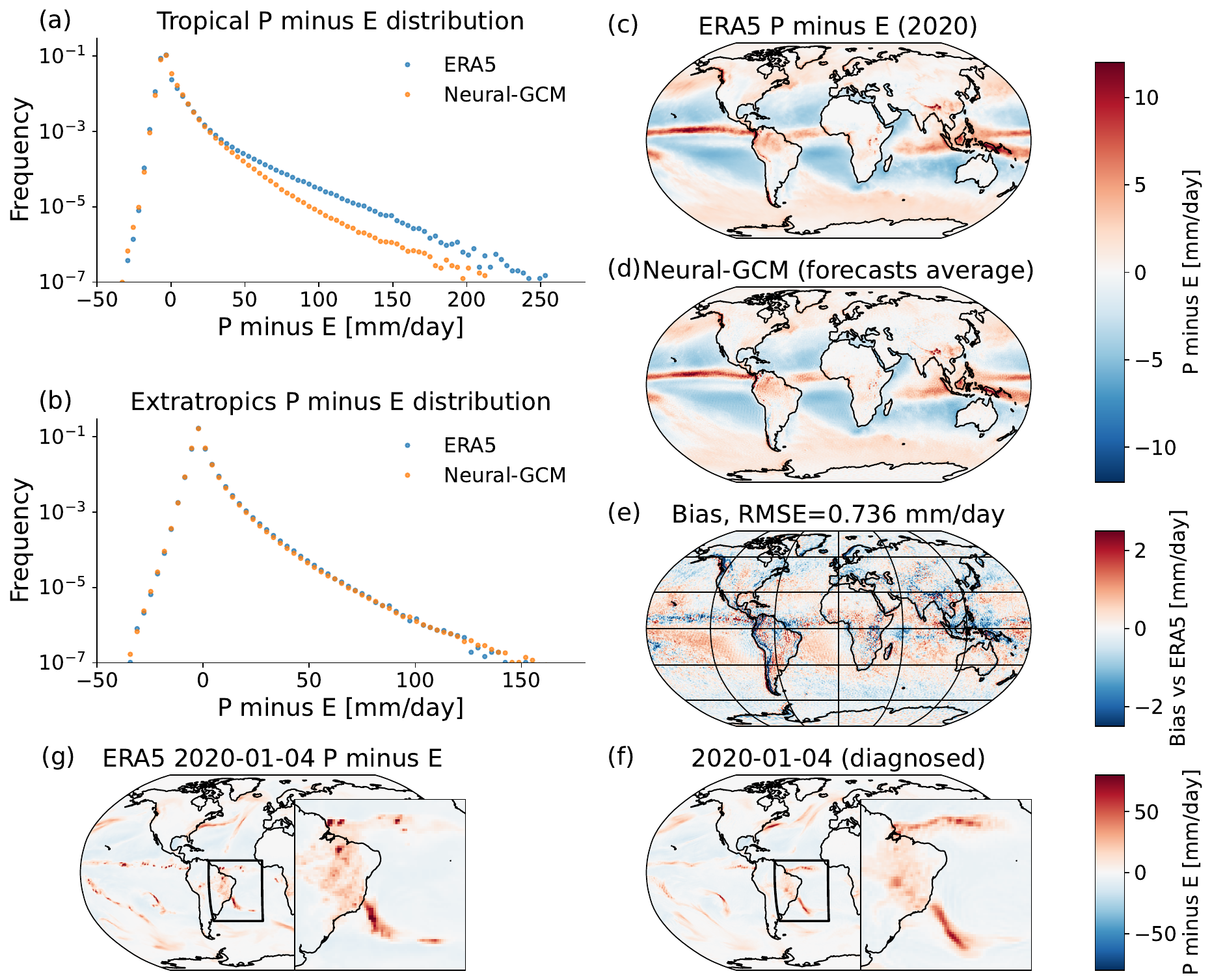}
\caption{Precipitation minus evaporation calculated from the third day of weather forecasts. (a) Tropical (latitudes $-20^{\circ}$ to $20^{\circ}$) Precipitation rate distribution, (b) Extratropical (latitudes $30^{\circ}$ to $70^{\circ}$ in both hemispheres), (c) mean precipitation minus evaporation for 2020 ERA5 and (d) NeuralGCM-$0.7^{\circ}$ (calculated from the third day of forecasts and averaged over initial conditions), (e) the bias between NeuralGCM-$0.7^{\circ}$ and ERA5, (f-g) Snapshot of daily precipitation for 01-04-2020 for (f) ERA5 and (g) NeuralGCM-$0.7^{\circ}$ (forecast initialized on 01-02-2020)}\label{sifig:P_minus_E_short_term}
\end{figure*}

Fig.~\ref{sifig:P_minus_E_short_term}a,b shows the daily distribution of precipitation minus evaporation rates from NeuralGCM-$0.7^{\circ}$ forecasts alongside ERA5. The distribution of precipitation minus evaporation rates has been normalized such that the area under the points is equal to one.
We observe that the precipitation rate distribution from NeuralGCM-$0.7^{\circ}$ aligns closely with the ERA5 distribution in the extratropics. However, NeuralGCM-$0.7^{\circ}$ tends to underestimate extreme events in the tropics.
For weather forecasting, we note that the average forecast (averaged across all initial conditions for the third day of prediction) aligns well with the spatial distribution seen in ERA5 (Fig.~\ref{sifig:P_minus_E_short_term}c,d).
To underscore the differences between ERA5's $P-E$ and NeuralGCM-$0.7^{\circ}$'s diagnosed $P-E$, we present snapshots of daily $P-E$. While there are overarching similarities between these snapshots, it's evident that in the tropics, NeuralGCM-$0.7^{\circ}$ tends to moderate extreme events (Fig.~\ref{sifig:P_minus_E_short_term}e,f).

\begin{figure*}
\centering
\includegraphics[width=1.0\textwidth]{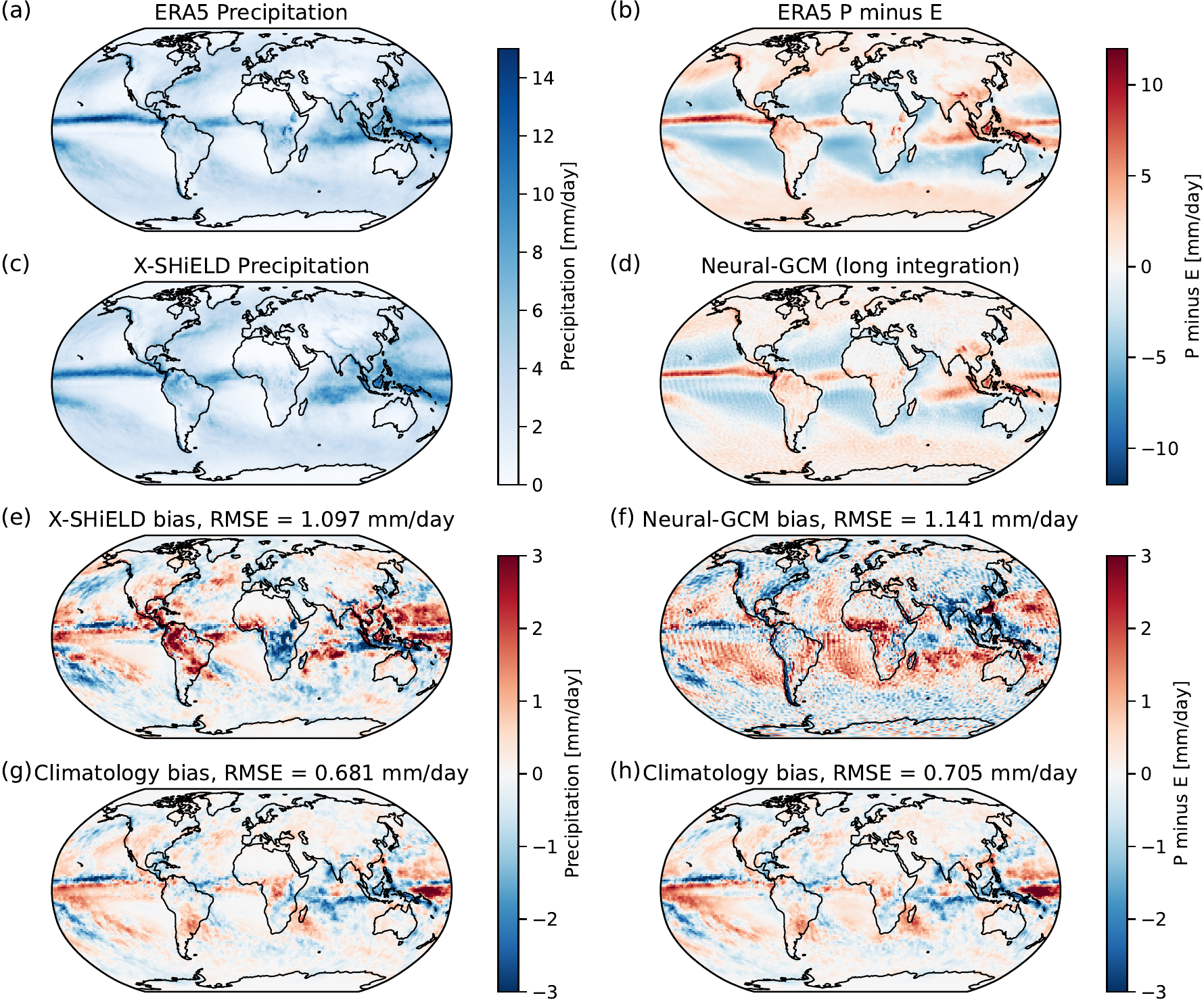}
\caption{Indirect comparison between precipitation bias in X-SHiELD and precipitation minus evaporation bias in NeuralGCM-$1.4^{\circ}$. Mean precipitation calculated between 01-19-2020 to 01-17-2021 for (a) ERA5 (b) X-SHiELD and the biases in (c) X-SHiELD and (d) climatology (ERA5 data averaged over 1990-2019). Mean precipitation minus evaporation calculated between 01-19-2020 to 01-17-2021 for (a) ERA5 (b) NeuralGCM-$1.4^{\circ}$ (initialized in October 18th 2019)  and the biases in (c) NeuralGCM-$1.4^{\circ}$ and (d) climatology (data averaged over 1990-2019). }\label{sifig:precip_climate_2020}
\end{figure*}

We note that the calculation of $P-E$ assumes that the dynamical core is responsible for all horizontal motions; thus, $P-E$ could be diagnosed using Eq.~\ref{eq:p_minus_e}. However, the physics module might also learn to correct errors originating from inaccuracies in the dynamical core (e.g., as a result of calculating advective tendencies on a coarse grid), which may introduce errors into the calculation of $P-E$. Given that $P-E$ calculated from NeuralGCM-$0.7^{\circ}$ appears generally consistent with ERA5 in the weather forecasting scenario (Fig.\ref{sifig:P_minus_E_short_term}), this suggests that any error is likely small. However, at lower resolutions, the error could be larger. For example, Fig~\ref{sifig:precip_climate_2020} displays the climatology of $P-E$ calculated with NeuralGCM-$1.4^{\circ}$, which shows clear artifacts, although it remains unclear whether these artifacts stem from the aforementioned error in calculation or another issue.

In this work, we only diagnosed precipitation minus evaporation from NeuralGCM. However, in future work, we plan to develop a scheme to reformulate NeuralGCM to predict precipitation and evaporation separately. This could be achieved, for example, by using conventional parameterization to estimate evaporation (and then calculating precipitation by adding it to P-E). Another approach could involve training a neural network (NN) specifically to predict evaporation, potentially optimized to predict the same fluxes as those in ERA5 data.

\subsection{Ablation tests}\label{apx:sec:ablation_tests}

\subsubsection{Different loss functions}\label{apx:subsec:ablation_loss}
To train deterministic NeuralGCM models we used a combination of 3 different loss functions (sections~\ref{apx:subsubsec:filtered_mse_loss}, \ref{apx:subsubsec:spectral_loss},\ref{apx:subsubsec:bias_loss}). To investigate the effect of utilizing these loss terms, we trained NeuralGCM-$2.8^\circ$ models while omitting certain loss terms and  compared the skill and spectrum of models employing various loss terms. Specifically, we trained models: (a) without the bias loss (section~\ref{apx:subsubsec:bias_loss}), (b) without applying filtering to the MSE loss (section~\ref{apx:subsubsec:filtered_mse_loss}), and (c) excluding spectral loss, bias loss, and filtering on the MSE, resulting in a simple MSE loss. 

\begin{figure*}
\centering
\includegraphics[width=0.7\textwidth]{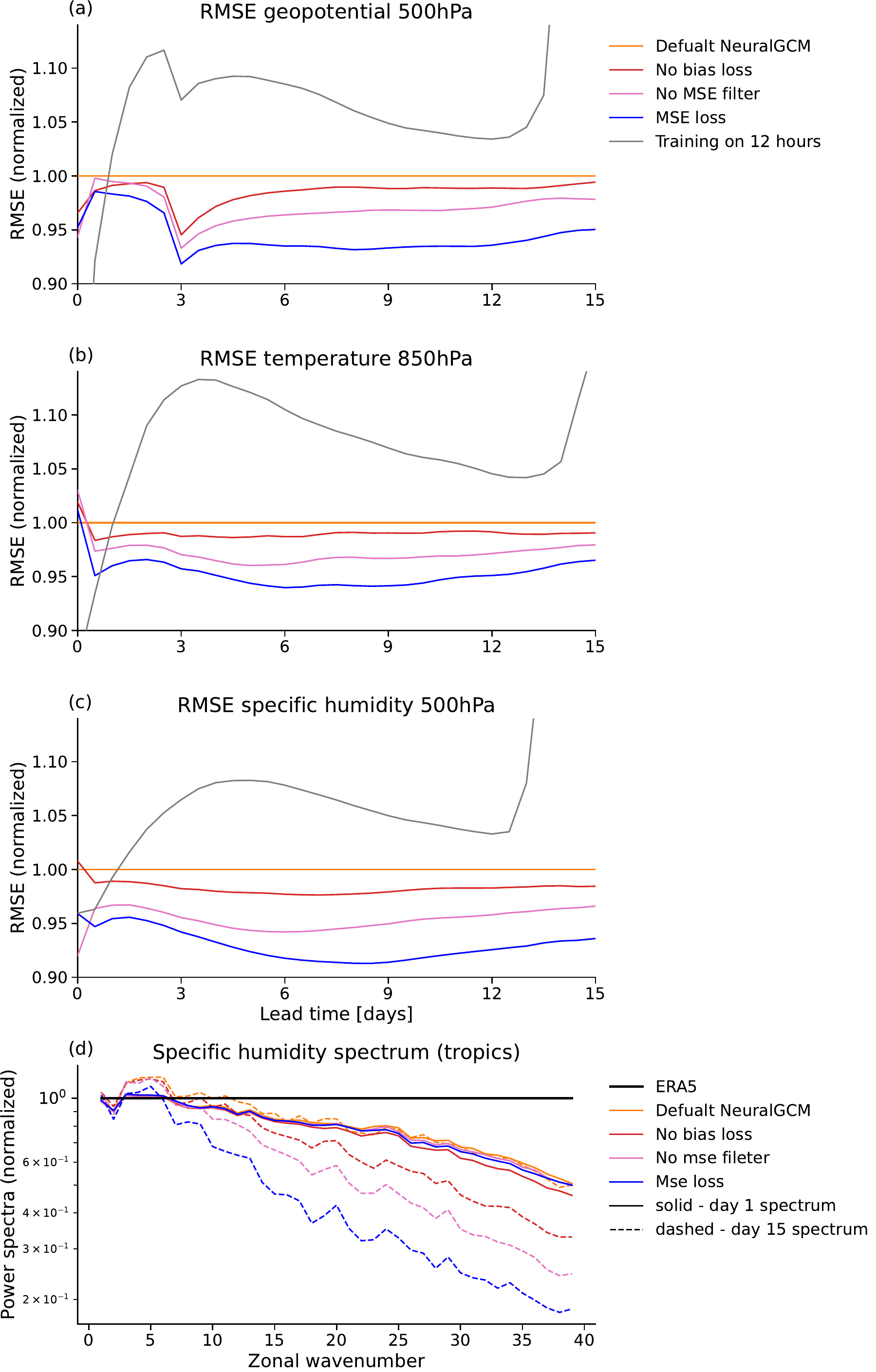}
\caption{
Accuracy and spectrum analysis for models using different loss terms and models trained on different rollout lengths. (a-c) RMSE normalized by the RMSE of the default NeuralGCM-$2.8^{\circ}$ model, as a function of prediction lead time for simulations utilizing various loss functions: the default NeuralGCM-$2.8^{\circ}$
  loss function (sum of filtered-MSE, spectral and bias losses; orange), a model trained without the bias loss term (red), a model trained without the filter on the MSE loss (pink), and a model trained with simple MSE loss function (without the spectral loss, without bias loss, and without the filter on the MSE; blue) and a model trained only up to 12 hour rollouts (grey), for (a) geopotential at 500 hPa, (b) temperature at 850 hPa, and (c) specific humidity at 700 hPa. Panel (d) depicts the zonal spectrum of specific humidity at 850 hPa in the tropics (defined from 15S-15N), normalized by ERA5 spectrum for the aforementioned models (except for the model model that is trained up to 12 hours as the spectrum diverges at day 15) and also includes the ERA5 spectrum (black). Solid lines show  1-day predictions and dashed lines show 15-day predictions. The zonal spectrum is truncated at wavenumber 40 to facilitate a clearer comparison between the models.
}\label{sifig:ablation_loss}
\end{figure*}

\begin{figure*}
\centering
\includegraphics[width=1.0\textwidth]{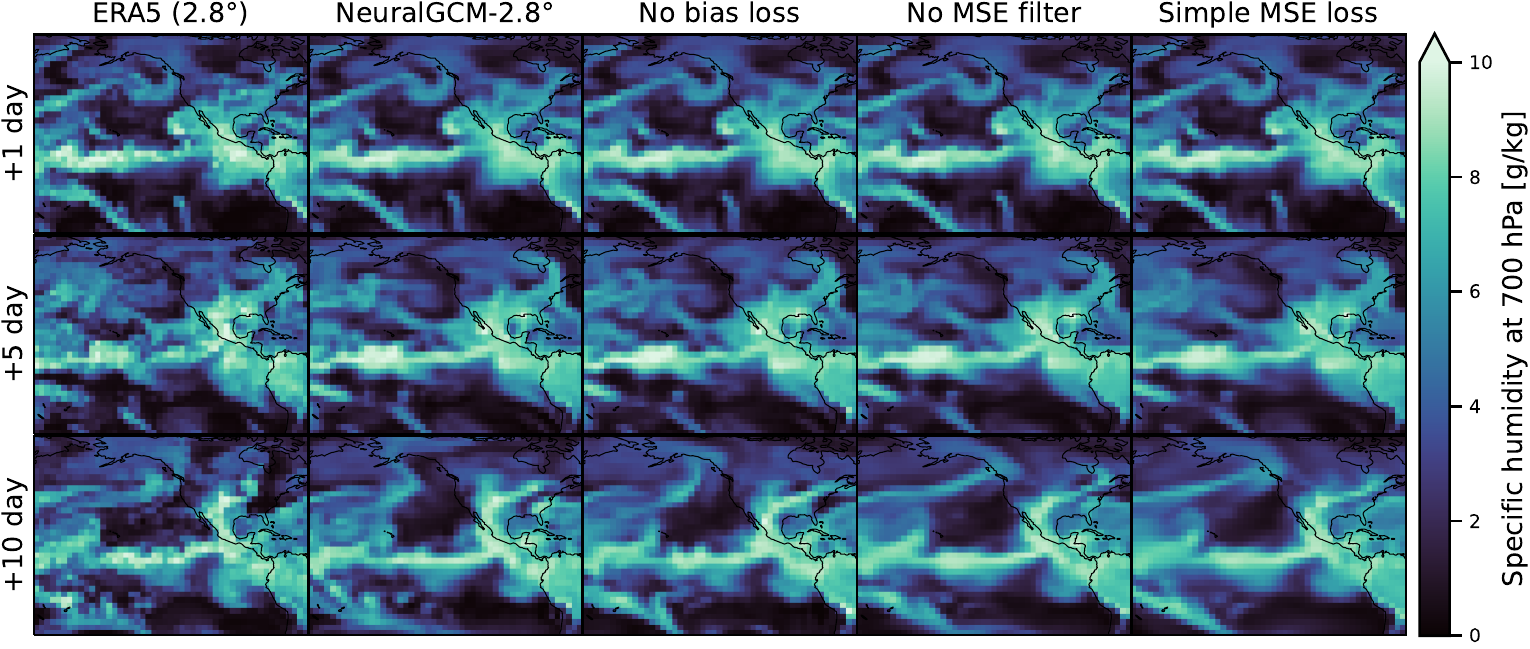}
\caption{
Case study on the impact of different loss functions on forecast snapshots. The plot illustrates the specific humidity at 700 hPa for 1-day, 5-day, and 10-day forecasts from each model across latitudes 30S-70N and longitudes 162-300 for ERA5, NeuralGCM-$2.8^\circ$ as well as NeuralGCM-$2.8^\circ$ variants trained without bias loss, without an MSE filter, and with a simple MSE loss (excluding spectral loss, bias loss, and MSE filter).
All models are at $2.8^\circ$ resolution (with ERA5 data regridded to match this resolution).
All forecasts are initialized at 2021-08-
25T00z.}\label{sifig:ablation_case_study}
\end{figure*}

We find that removing these loss terms, or removing the MSE loss filter, enhances the RMSE score  (Figs.~\ref{sifig:ablation_loss}a-c ).
For example, adopting solely the MSE loss - a common approach in ML algorithms for weather forecasting - enabled us to enhance our skill by several percent.
Namely, the RMSE of NeuralGCM deterministic models could be improved by using an MSE loss instead of the losses we used in the paper. 
However, removing these loss terms leads to a degradation of the simulation spectrum over time (Fig.~\ref{sifig:ablation_loss}d, Fig.~\ref{sifig:ablation_case_study}). Such a degradation of the spectrum over time would likely harm the ability to conduct climate simulations, and would make the NeuralGCM forecasts less useful as they would not resemble realistic atmospheric states.
The spectral degradation observed in the model employing MSE loss suggests that the primary cause of such degradation in previous ML-based models \cite{bi2023accurate,lam2022graphcast} could potentially be addressed by introducing new loss terms. However, these modifications might adversely affect the models' accuracy.

\subsubsection{Training on shorter rollouts}
To investigate the benefits of training on longer rollouts (i.e., optimizing for extended prediction periods), we trained a NeuralGCM-$2.8^\circ$ model exclusively on predictions up to 12 hours, while keeping all other model parameters identical to the default NeuralGCM-$2.8^\circ$ model. We find that the model trained on shorter rollouts not only exhibit deteriorated performance in terms of RMSE but, more critically, tend to be significantly less stable (Fig.~\ref{sifig:ablation_loss}a-c). For example, the sharp increase in RMSE after 13 days is associated with model instabilities (Fig.~\ref{sifig:ablation_loss}a-c).

Additionally, we attempted to train a model on 1-hour rollouts (which is a single physics timestep for NeuralGCM-$2.8^\circ$) but found that the trained model becomes unstable after very few days (not shown).

\subsubsection{Learning curve - pure ML vs hybrid}
\begin{figure*}
\centering
\includegraphics[width=0.7\textwidth]{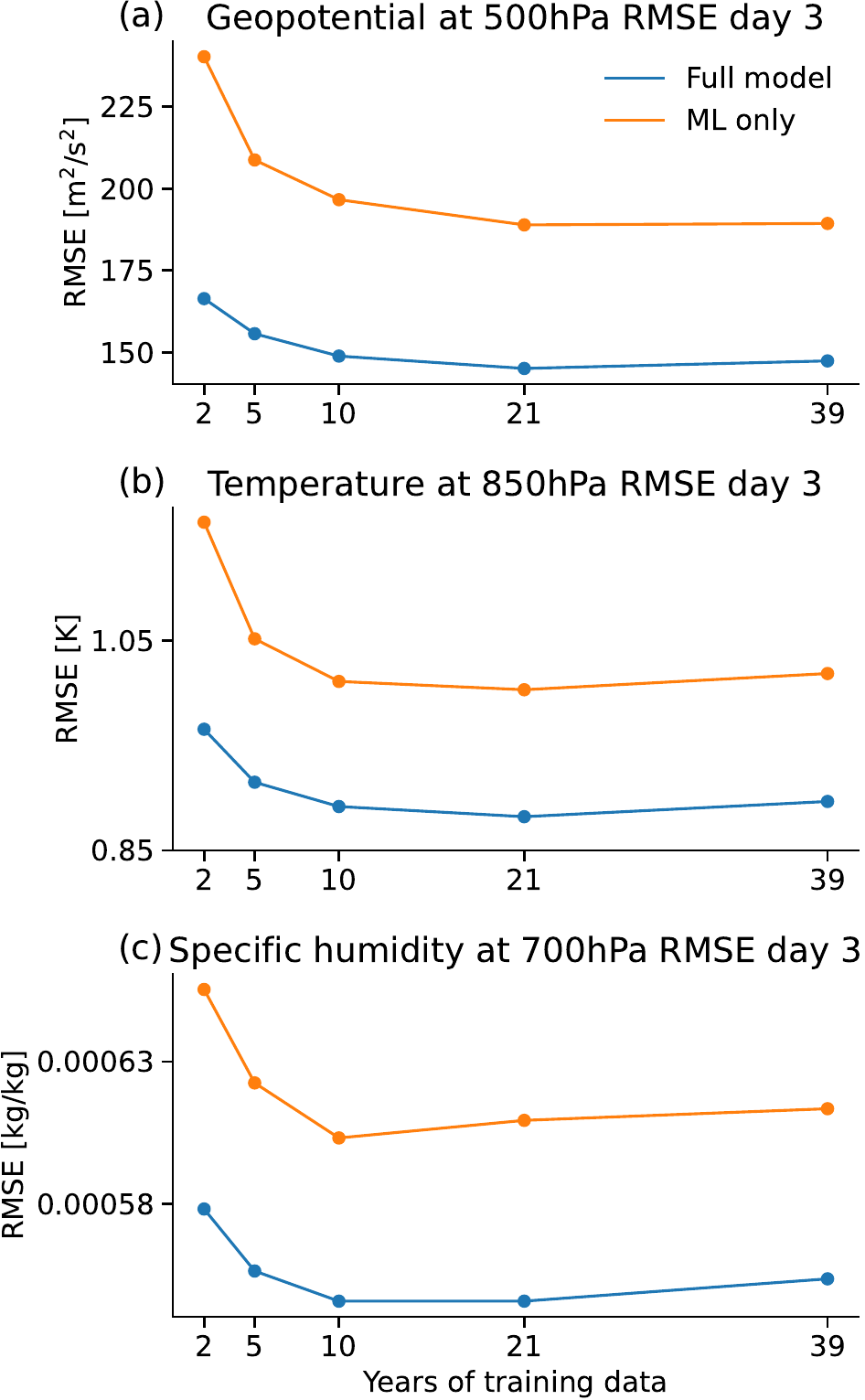}
\caption{Learning curves comparing the hybrid NeuralGCM model with the ML-only variant of NeuralGCM. 
RMSE as a function of the number of training years (with all models trained on the years leading up to 2017 and evaluated on data from 2018, utilizing varying lengths of training data) for both NeuralGCM-$2.8^{\circ}$
  (blue) and its ML-only counterpart, NeuralGCM-ML-only-$2.8^{\circ}$
  (orange), which is similar to NeuralGCM-$2.8^{\circ}$
  but without the dynamical core. The RMSE values are presented for forecasts at day 3 for (a) geopotential at 500hPa, (b) temperature at 850hPa, and (c) specific humidity at 700 hPa.}\label{sifig:Learning_curve}
\end{figure*}

To estimate the volume of training data which is beneficial for NeuralGCM  we trained several NeuralGCM-$2.8^{\circ}$ models using varying amounts of data (all models were trained on the most recent years up to 2017, but each utilized a different number of years for training). We find that for weather forecasting  NeuralGCM-$2.8^{\circ}$ does not benefit from incorporating more than the last 21 years of data (Fig.~\ref{sifig:Learning_curve}). This finding might be related to variations in the ERA5 dataset, in which recent years uses assimilated data sources which are more similar to the year on which we tested these models (2018). In contrast older years likely incorporate fewer or different assimilated data sources relative to the year on which we tested the model.

Next, we wanted to assess whether the NeuralGCM requires a different quantity of training data compared to a purely ML-based approach. To this end, we trained multiple NeuralGCM-ML-only-$2.8^{\circ}$ models, which are similar to NeuralGCM-$2.8^{\circ}$ but lack a dynamical core, making them solely ML-based. We find that incorporating the dynamical core enhances performance compared to ML-only model (Fig.~\ref{sifig:Learning_curve}), as anticipated, given that the single-column parameterization structure we employ limits the ability to accurately learn horizontal advective tendencies. Interestingly, we find that the inclusion of the dynamical core does not reduce the amount of necessary training data (both NeuralGCM-$2.8^{\circ}$ and  NeuralGCM-ML-only-$2.8^{\circ}$ have a similar learning curve - Fig.~\ref{sifig:Learning_curve}). We hypothesize that the single-column approach might inherently reduce the training data needed, suggesting that such a structure could potentially require less data than state-of-the-art ML models for weather forecasting. However, this idea remains speculative, as direct tests of this hypothesis have not yet been conducted. 

\subsection{Instability case study}\label{apx:sec:instability_case}
\begin{figure*}
\centering
\includegraphics[width=1.0\textwidth]{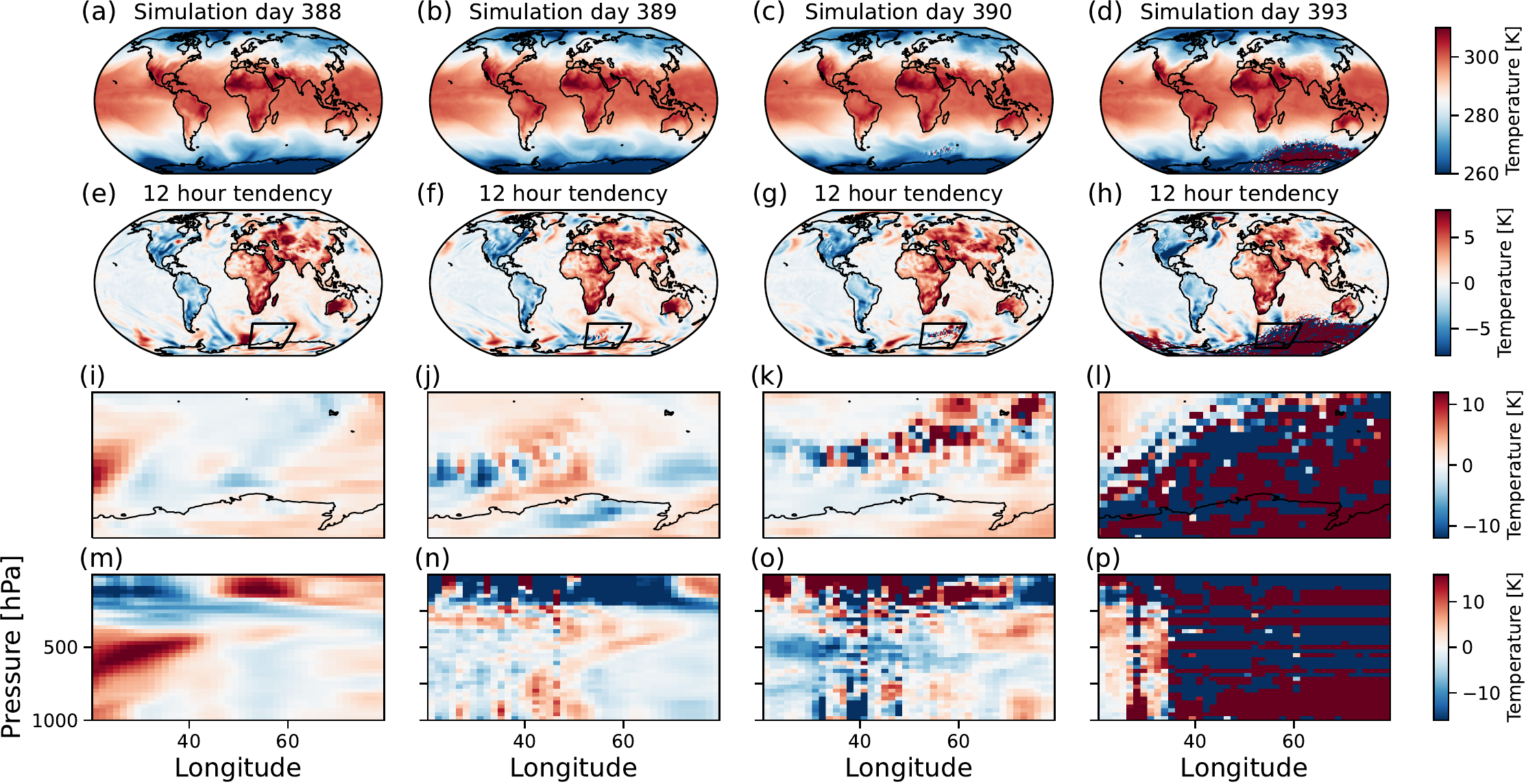}
\caption{Case study on instability in the NeuralGCM-$1.4^{\circ}$ model. (a-d) The temperature, (e-h) 12 hour temperature change at 1000hPa (i-l) zoom into longitudes 20 to 80 and latitudes 75S to 45S and (m-p) the vertical structure of 12 hour temperature change at latitude 60S, for different simulation lead times: day 388 (left column), day 389 (second column), day 390 (third column) and  day 392 (right column). The simulation was initialized at 09-18-2019.}
\label{sifig:instability_case_study_tl127}
\end{figure*}

\begin{figure*}
\centering
\includegraphics[width=1.0\textwidth]{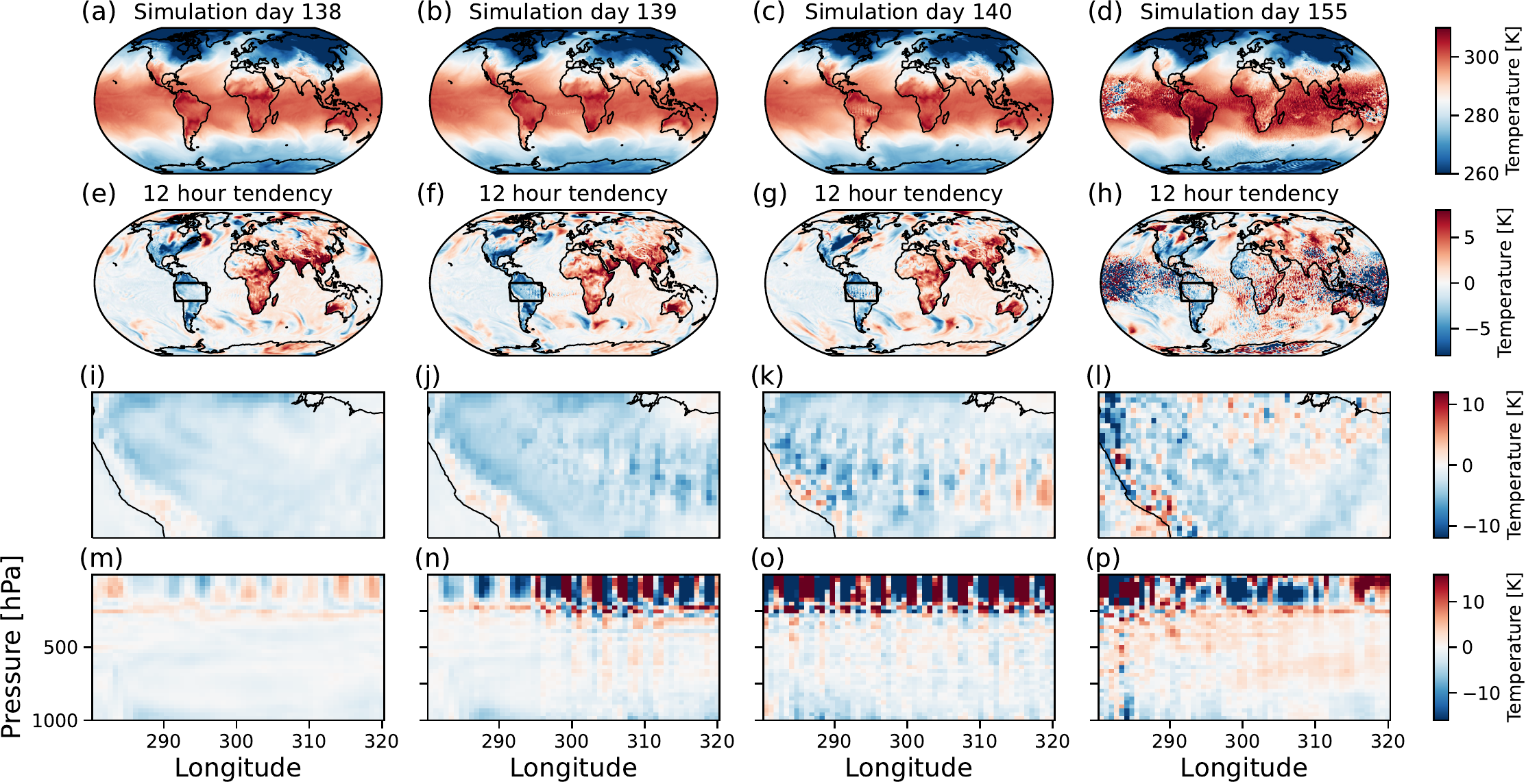}
\caption{Case study on instability in the NeuralGCM-$0.7^{\circ}$ model. (a-d) The temperature, (e-h) 12 hour temperature change at 1000hPa (i-l) zoom into longitudes 280 to 320 and latitudes 20S to the equator and (m-p) the vertical structure of 12 hour temperature change at latitude 10.1S, for different simulation lead times: day 138 (left column), day 139 (second column), day 140 (third column) and  day 155 (right column). The simulation was initialized at 08-29-2019.}
\label{sifig:instability_case_study_tl255}
\end{figure*}

Figs.~\ref{sifig:instability_case_study_tl127} and \ref{sifig:instability_case_study_tl255} show examples of an instabilities that were found in NeuralGCM-$1.4^{\circ}$ and NeuralGCM-$0.7^{\circ}$ simulations. The figures illustrates the progression of two instability case-studies over time. 

Fig.~\ref{sifig:instability_case_study_tl127} presents an example of instability observed in a NeuralGCM-$1.4^{\circ}$ simulation. Specifically, it illustrates that on day 388, the temperature field in the simulation appears realistic. However, by day 389, the 12-hour temperature tendencies at 1000 hPa (Fig.~\ref{sifig:instability_case_study_tl127}j) exhibit an unrealistic strong local tendencies at the South of the Indian Ocean. This unrealistic signal continues to intensify over time, and by day 392, the instability has significantly intensified and propagated throughout the Indian ocean (Fig.~\ref{sifig:instability_case_study_tl127}d).

Fig.~\ref{sifig:instability_case_study_tl255} presents an example of instability observed in a NeuralGCM-$0.7^{\circ}$ simulation. Specifically, it illustrates that on day 138, the temperature field in the simulation appears realistic. However, by day 139, the 12-hour temperature tendencies in the stratosphere (Fig.~\ref{sifig:instability_case_study_tl255}n) and at 1000 hPa (Fig.~\ref{sifig:instability_case_study_tl255}j) exhibit an unrealistic strong high wavenumber component. This unrealistic signal continues to intensify over time, and by day 155, the instability has significantly intensified and propagated throughout the tropics (Fig.~\ref{sifig:instability_case_study_tl255}d).

\clearpage
\section{Additional climate evaluations}\label{apx:sec:additional_climate_evaluation}

\subsection{Seasonal cycle}\label{apx:sec:seasonal cycle}

To assess the skill of our model for simulating seasonal cycles, we conduct a comprehensive comparison between NeuralGCM-$1.4^{\circ}$ resolution and ERA5.
We ran $2$-year simulations with 37 different initial conditions spaced at 10 days for the year 2019. Out of these 37 initial conditions, 35 successfully completed full two years without encountering model instability.
The comparison is done for the year 2020, focusing on several key aspects. 
We begin by examining the global mean temperature at 850 hPa and find that NeuralGCM-$1.4^{\circ}$ closely resembles ERA5, both in terms of mean temperature and variability for all stable initial conditions (Fig.~\ref{fig:climatology}a).

\begin{figure*}
\centering
\includegraphics[width=1.0\textwidth]{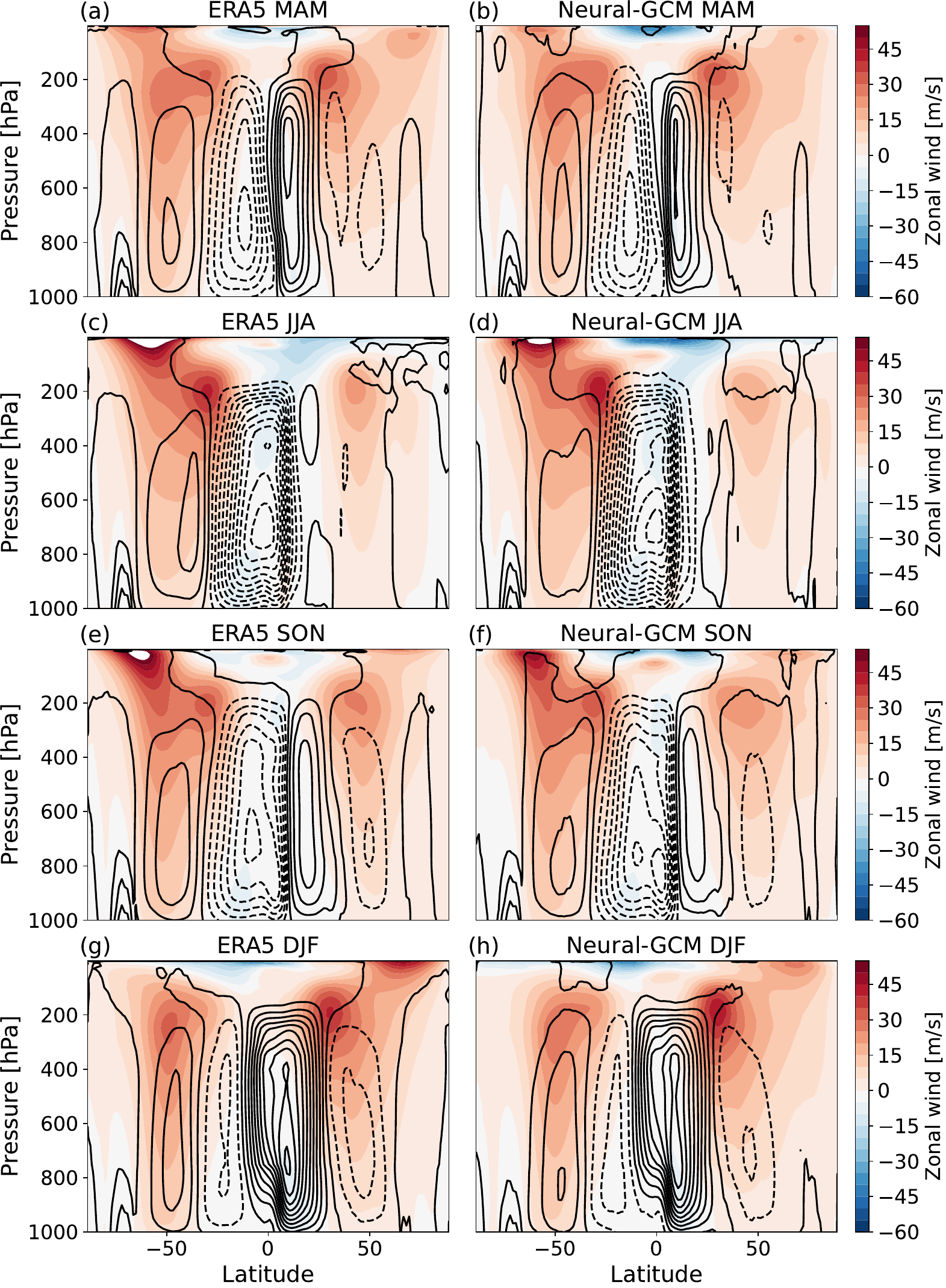}
\caption{The Hadley cell circulation and zonal wind for different seasons in NeuralGCM and ERA5. The mass streamfunction (contours) and zonal-mean zonal wind (colors) as a function of pressure and latitude averaged over different seasons during 2020 for (a,c,e,f) ERA5 and (b,d,f,h) NeuralGCM-$1.4^{\circ}$ simulation initialized in October 18th 2019. Solid contours indicate positive values of mass streamfunction and dashed contours indicate negative values and contour intervals are \SI{2e10}{kg/s}.}
\label{sifig:Hadley_cell_u_wind}
\end{figure*}

\begin{figure*}
\centering
\includegraphics[width=1.0\textwidth]{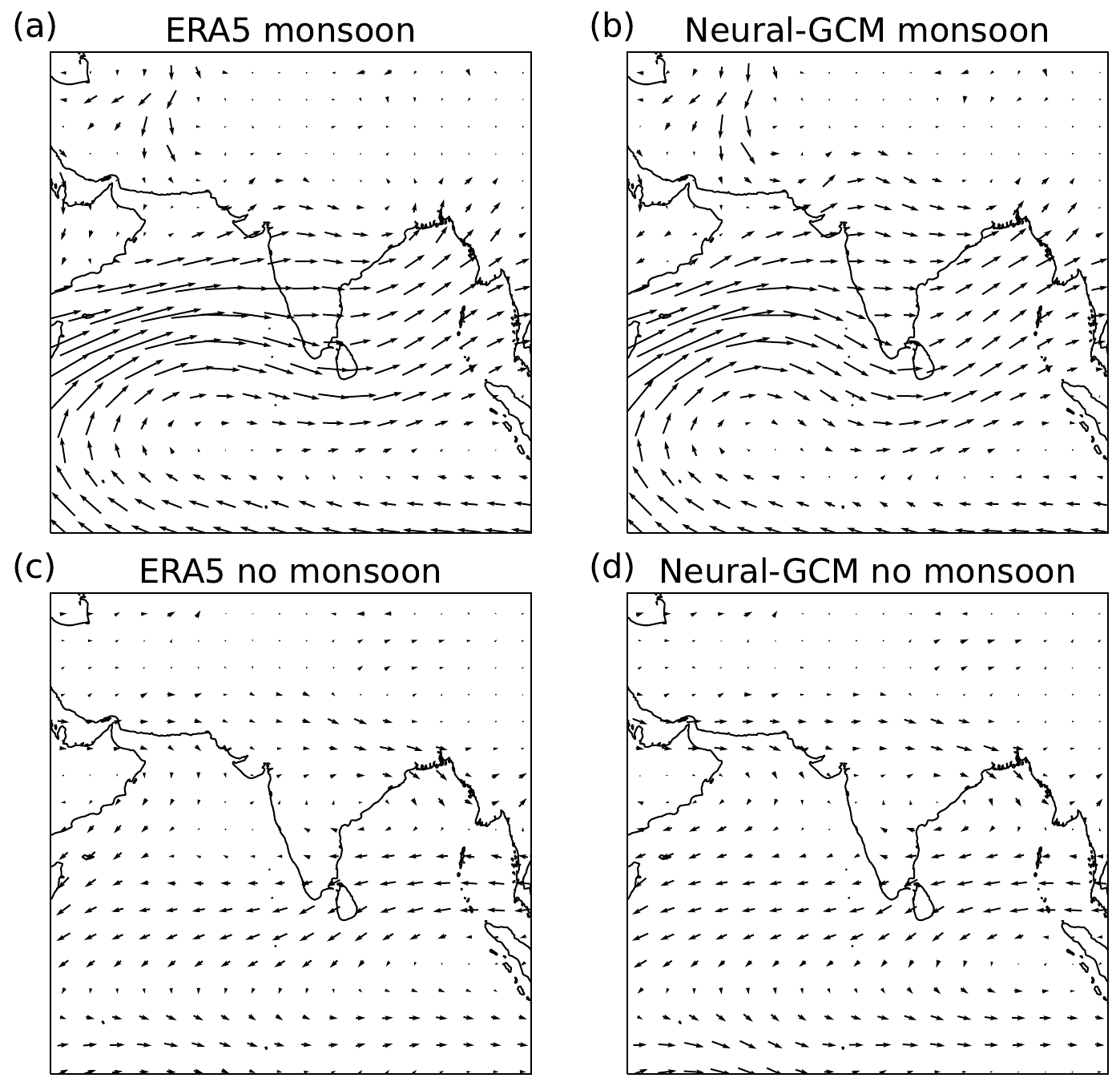}
\caption{Winds during Monsoon and no Monsoon months in NeuralGCM and ERA5. Quiver plot of the time mean wind at 850 hPa for the Indian Monsoon months (defined here as June 15th to Septemper 15th; panels a and b) and no Monsoon months (defined here as January 1st to April 1st; panels b and d) for 2020 for (a,c) ERA5 (b,d) NeuralGCM-$1.4^{\circ}$ simulation (initialized in October 18th 2019)}\label{sifig:monsoon_2020}
\end{figure*}

To example the tropical circulation we analyze the Hadley cell seasonal cycle and its amplitude. The Hadley cell circulation is characterized by computing the mass streamfunction $\psi(\phi, p)$, which quantifies the mass transport between latitudes and altitudes. The mass streamfunction is
\begin{equation}
\psi(\phi, p) = \frac{2 \pi cos(\phi)}{g}\int_p^{p_{\rm{s}}} \bar{v} dp,
\end{equation}
where $\phi$ is the latitude, p is the pressure, g is the acceleration due to gravity, $p_{\rm{s}}$ is the surface pressure and $\bar{v}$ is the zonal mean meridional velocity.
We find that  NeuralGCM-$1.4^{\circ}$ is able to capture both the seasonal cycle and the amplitude of the Hadley cell circulation (Fig.~\ref{sifig:Hadley_cell_u_wind}). We also demonstrate in Fig.~\ref{sifig:monsoon_2020} that NeuralGCM-$1.4^{\circ}$ can accurately capture the spatial structure of the wind during the Indian monsoon and non-monsoon months.

\begin{figure*}
\centering
\includegraphics[width=1.0\textwidth]{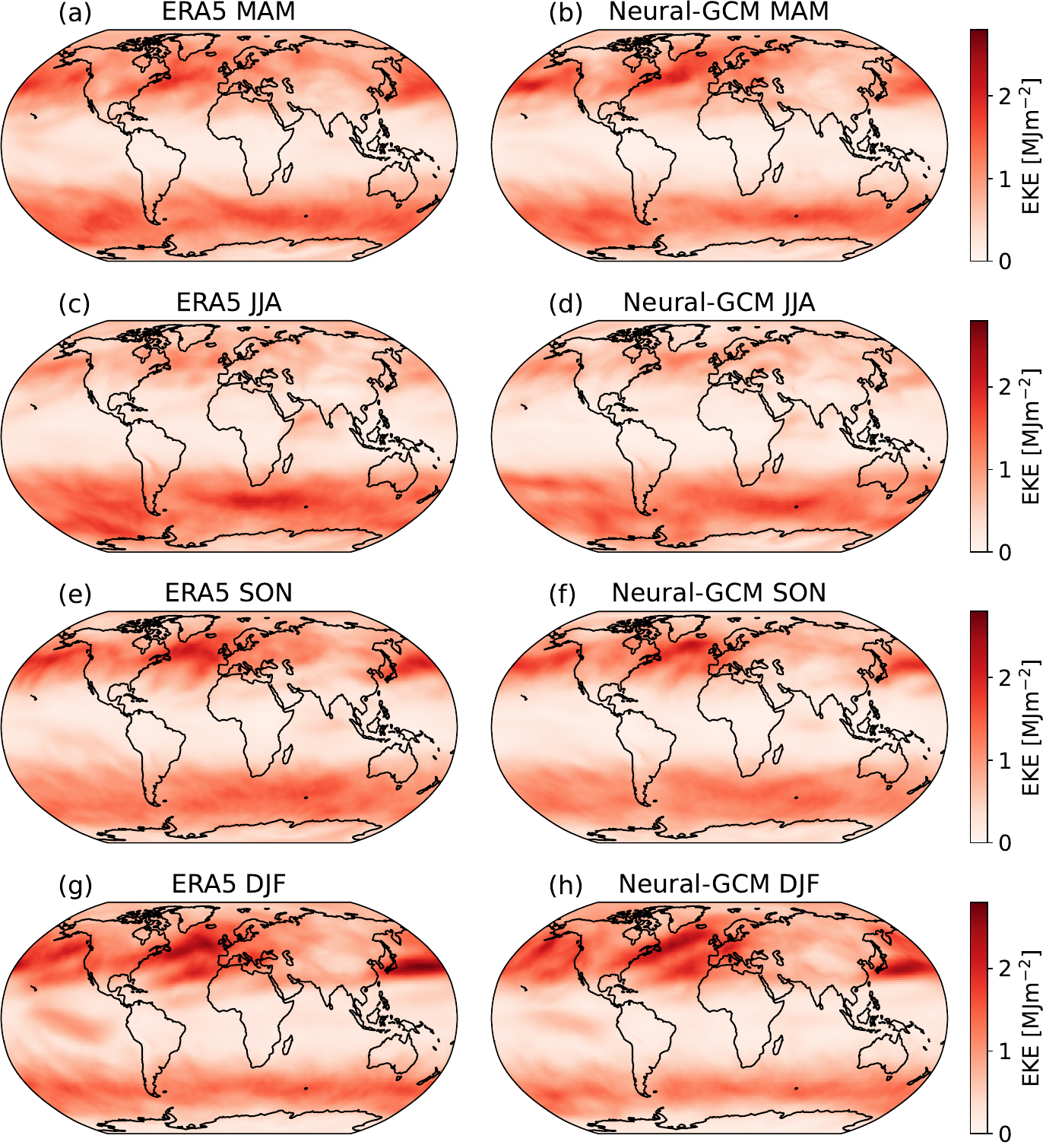}
\caption{Eddy kinetic energy in NeuralGCM and ERA5. Vertically integrated Eddy Kinetic Energy (EKE) as a function of longitude and latitude averaged over different seasons during 2020 for (a,c,e,f) ERA5 and (b,d,f,h) NeuralGCM-$1.4^{\circ}$ simulation (initialized in October 18th 2019).}\label{sifig:EKE_2020}
\end{figure*}

To characterize the extratropical circulation, we plot zonal-mean zonal wind in different seasons and find that NeuralGCM-$1.4^{\circ}$ has a very similar structure compared to ERA5 (Fig.~\ref{sifig:Hadley_cell_u_wind}) with the exception that above 30hPa there are noticeable differences as NeuralGCM-$1.4^{\circ}$ does not optimize its predictions for these levels (since the upper-most level exist in sigma coordinates is $\approx 0.03$). 
We also consider the extratropical storm tracks and we compare the seasonal cycle of the eddy kinetic energy (EKE). EKE is computed as:
\begin{equation}
\rm{EKE} = \int_{150hPa}^{1000hPa} \frac{1}{2g}(u'^2 + v'^2) dp,
\end{equation}
where the prime symbol denotes deviations from the instantaneous zonal mean (i.e., $v' = v - \bar{v}$, where $\bar{v}$ represents the zonal mean of the variable), and g is the gravitational acceleration. 
We find that NeuralGCM-$1.4^{\circ}$ successfully captures the seasonal cycle of EKE, displaying the correct seasonal and spatial structure of EKE (Fig.~\ref{sifig:EKE_2020}).

\begin{figure*}
\centering
\includegraphics[width=1.0\textwidth]{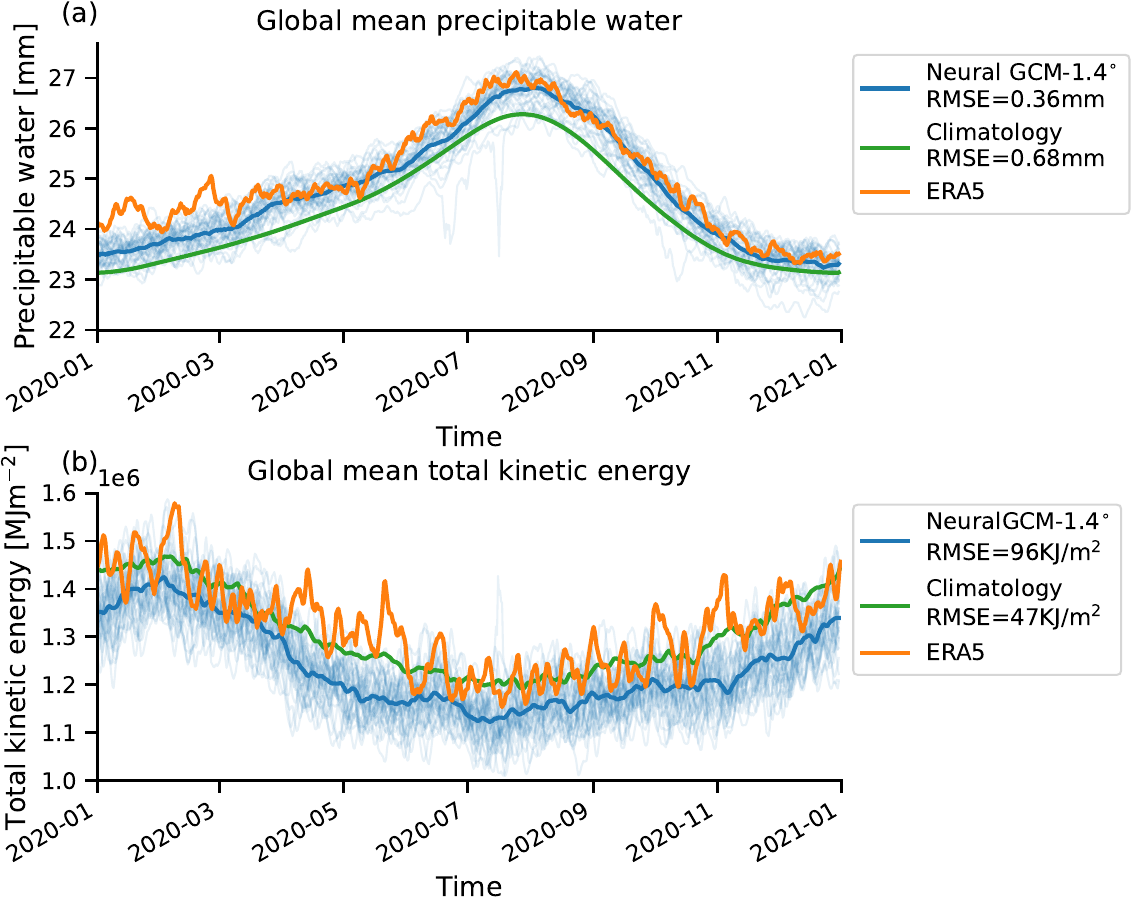}
\caption{Yearly cycle of precipitable water and total kinetic energy. Global mean (a) precipitable water and (b) total kinetic energy for 2020 ERA5 for 2020 (orange), climatology (defined as the averaged temperature between 1990-2019; green), and for NeuralGCM-$1.4^{\circ}$ for 2020 for simulations initialized every 10 days during 2019 (thick blue represents the ensemble mean, and thin blue lines indicate different initial conditions).}\label{sifig:TKE_2020}
\end{figure*}

We also illustrate the global annual cycle of atmospheric water and total kinetic energy ($\rm{TKE} = \int_{150hPa}^{1000hPa} \frac{1}{2g}(u^2 + v^2) dp$), demonstrating that there is no visible drift in these quantities and that we achieve a realistic annual cycle (\ref{sifig:TKE_2020}). This is in contrast to a recent attempt at hybrid modelling found that the atmosphere tend to dry out rapidly \cite{kwa2023machine}. The ensemble mean of NeuralGCM-$1.4^{\circ}$'s precipitable water accurately captures the magnitude seen in ERA5 during 2020, yielding a smaller RMSE than the climatological value. However, the ensemble mean TKE of NeuralGCM-$1.4^{\circ}$ exhibits a slight bias, trending lower than the ERA5 values.

\begin{figure*}
\centering
\includegraphics[width=1.0\textwidth]{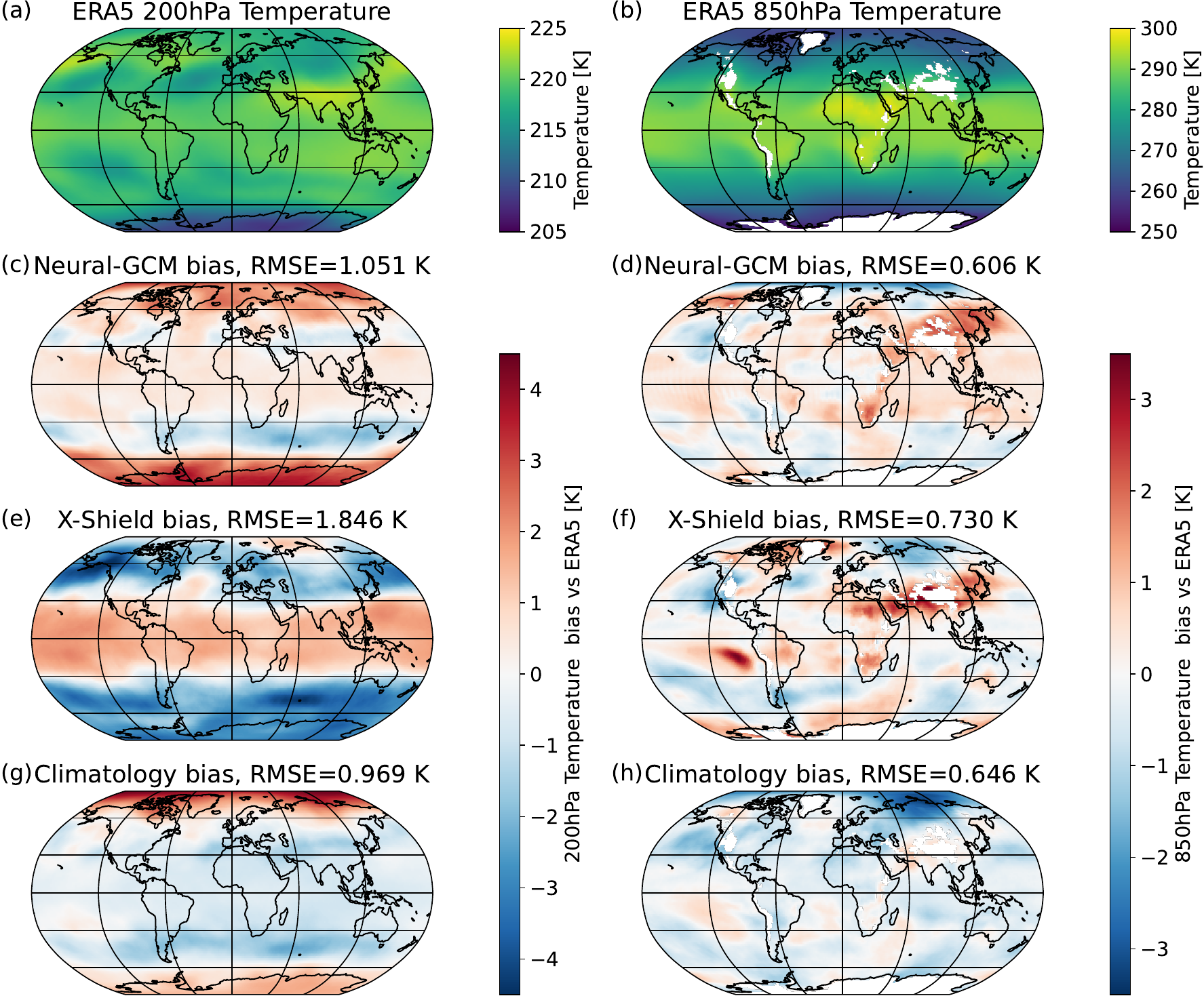}
\caption{Yearly temperature bias for NeuralGCM and X-SHiELD. Mean temperature between 01/19/2020 to 01/17/2020 for (a) ERA5 at 200hPa and (b) 850hPa. (c,d) the bias in the temperature for NeuralGCM-$1.4^{\circ}$, (e,f) the bias in X-SHiELD and (g,h) the bias in climatology (calculated from 1990-2019). NeuralGCM-$1.4^{\circ}$ was initialized in 18th of October (similar to X-SHiELD)}\label{sifig:temperature_bias_200_850_2020}
\end{figure*}

\begin{figure*}
\centering
\includegraphics[width=0.8\textwidth]{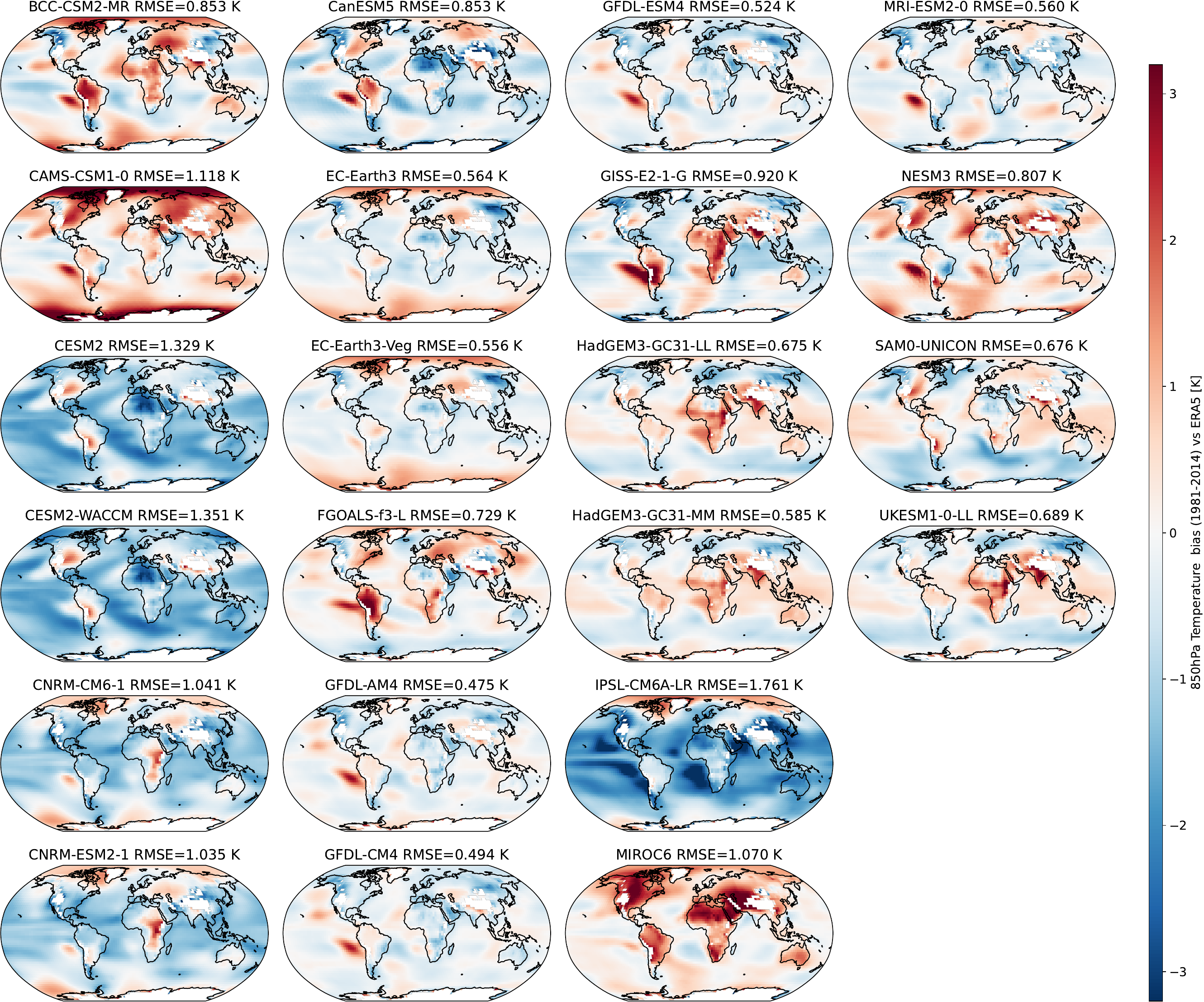}
\caption{850hPa temperature bias averaged over 1981-2014 in 22 models from CMIP6 AMIP simulations. \label{sifig:AMIP_1981_2014_bias_CESM_ens}}
\end{figure*}

\begin{figure*}
\centering
\includegraphics[width=0.8\textwidth]{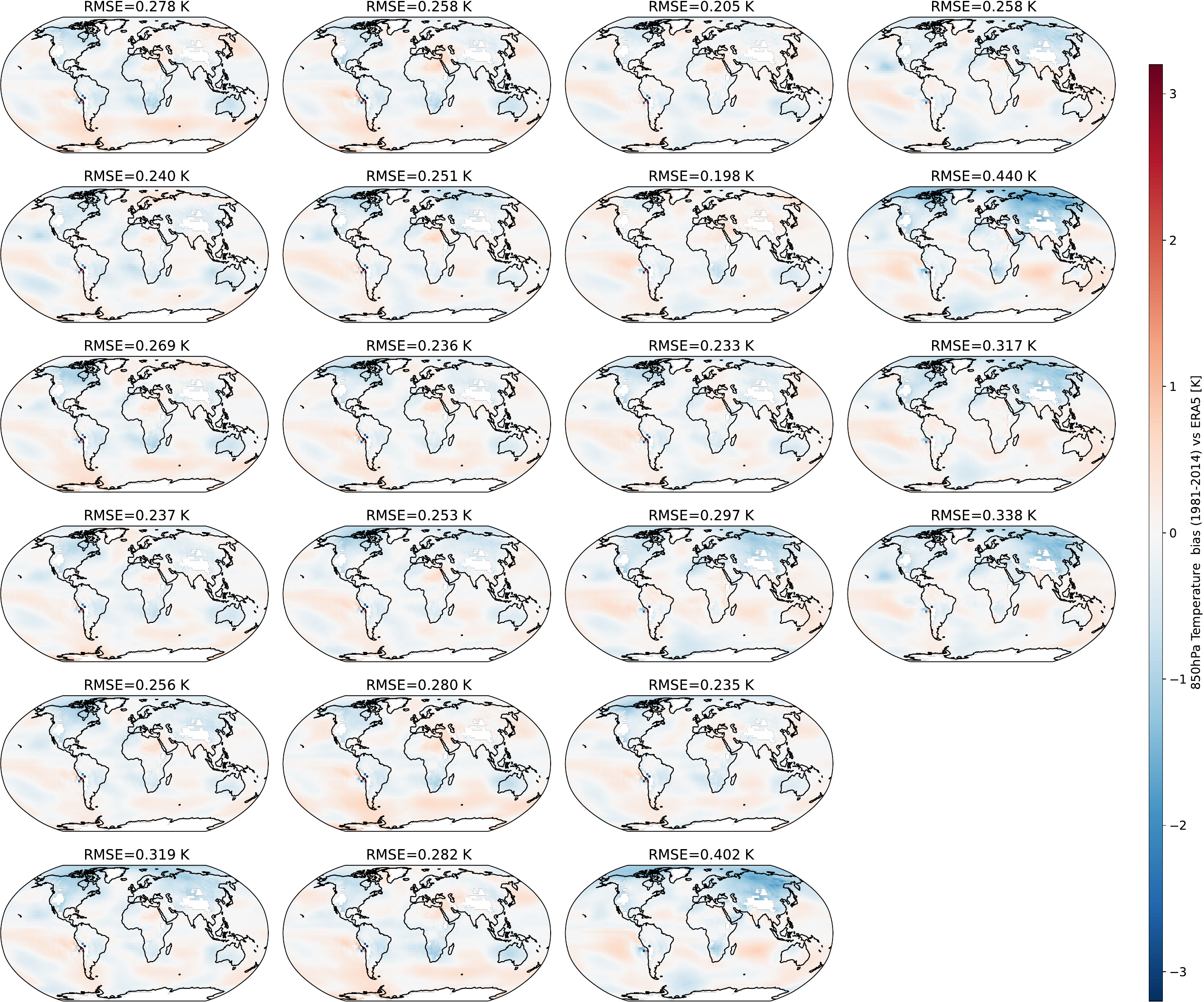}
\caption{850hPa temperature bias averaged over 1981-2014 in 22 NeuralGCM-$2.8^{\circ}$ simulations. \label{sifig:AMIP_1981_2014_bias_ngcm}}
\end{figure*}

\subsection{Tropical cyclone tracking}\label{apx:sec:TC_tracking}
To assess NeuralGCM's capability to simulate Tropical Cyclones (TCs), we use the TempestExtremes tracker \cite{ullrich2021tempestextremes} to identify TC tracks. To compare NeuralGCM TC results to ERA5 we first detect TCs in ERA5 native resolution (0.25$^{\circ}$) using the default configuration of TempestExtremes. 

The default configuration of TempestExtremes, used for the native ERA5 resolution (0.25$^{\circ}$), utilizes sea level pressure (SLP) as the feature-tracking variable. Candidate TCs are initially identified based on SLP minima, and a closed contour criterion is applied, demanding an SLP increase of at least 2 hPa within a $5.5^{\circ}$ Great Circle Distance (GCD) from the candidate point. Additionally, the difference between geopotential on the 300 and 500 hPa surfaces must decrease by $58.8$ $\rm{m^2s^{-2}}$ within a $6.5^{\circ}$ GCD from the candidate geopotential, taken as the maximum geopotential height difference within 1$^{\circ}$ GCD of the candidate location. These candidates are then linked over time to form TC paths. These paths have a maximum allowable distance of 8$^{\circ}$ between consecutive candidates, feature a maximum allowable gap size of $24$ hours (representing time periods with no TC identification) and have a minimum trajectory length of $54$ hours. For at least $10$ time steps, the underlying orography has to be less than 150m, the storm formation is constrained within latitudes of $-50^{\circ}$ to $50^{\circ}$, and the 10m wind magnitude has to exceed $10$ $\rm{m/s}$. Tracking is conducted at 6-hour output intervals for all resolutions.

One challenge with TempestExtremes is its tendency to yield significantly different numbers of Tropical Cyclones (TCs) when applied to data at varying resolutions \citeSI{roberts2020impact}. To apply the tracker to NeuralGCM-$1.4^{\circ}$ and make meaningful comparisons with ERA5 data, we specify a set of tracking parameters that yield nearly identical TC tracks when ERA5 data is regridded to a resolution of $1.4^{\circ}$.

\begin{figure*}
\centering
\includegraphics[width=0.75\textwidth]{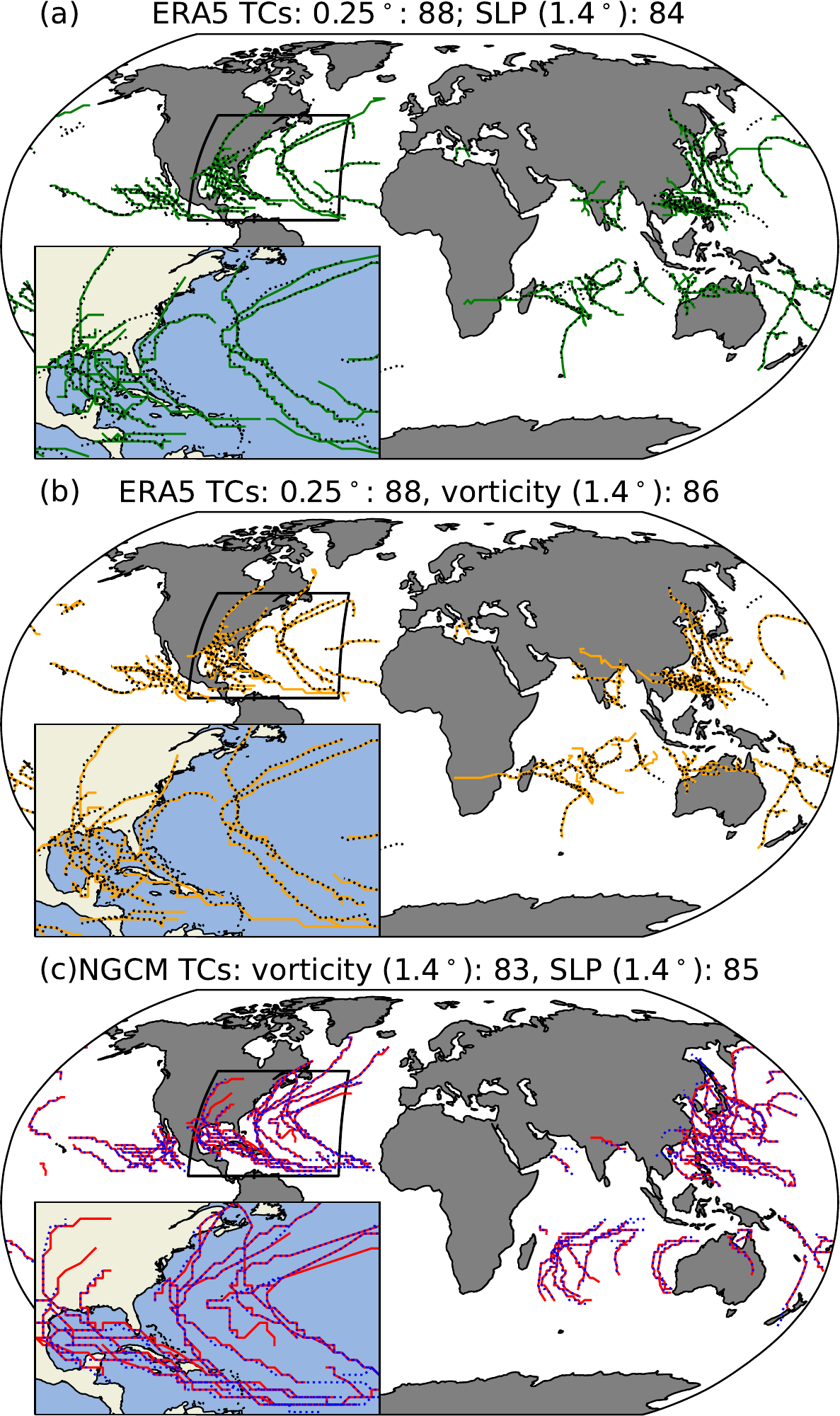}
\caption{Tropical cyclone tracks for different tracking criteria. Tropical Cyclone (TC) tracks identified from (a,b) ERA5 and (c) NeuralGCM-$1.4^{\circ}$ using different criteria and resolutions. (a) TC tracks from ERA5 native resolution ($0.25{^\circ}$) using sea level pressure criterion (SLP; dotted black) and ERA5 at $1.4{^\circ}$ resolution using SLP modified criterion (green). (b) TC tracks from ERA5 native resolution ($0.25{^\circ}$) using SLP criterion (dotted black) and ERA5 at $1.4{^\circ}$ resolution using vorticity criterion (orange).
(c) TC tracks from NeuralGCM-$1.4^{\circ}$ using SLP modified criterion (dotted blue) and vorticity criterion (red). The TC tracking was applied to all simulations between 01-19-2020 to 01-17-2021, which were the dates that were available for the X-SHiELD model. NeuralGCM-$1.4^{\circ}$ initialized on 10-18-2019 (similar to X-SHiELD). Insets show a zoom in into the Northern Atlantic region.}\label{sifig:TCs_xsheild_dates}
\end{figure*}

We fine-tuned these parameters so that TempestExtremes detected nearly identical TC tracks (as shown in Fig.~\ref{sifig:TCs_xsheild_dates}a) and a similar number of TCs, with 88 in the native resolution (using the default parameters for tracking) and 84 in the $1.4^{\circ}$ resolution (using the new set of parameters for tracking). The main differences is that we reduced the gradient requirements, necessitating an SLP increase of at least 0.6 hPa within a $5.5^{\circ}$ GCD radius of the candidate node and the 
difference between geopotential on the 300 and 500 hPa surfaces must decrease by $25.8$ $\rm{m^2s^{-2}}$ within a $6.5^{\circ}$ GCD from the candidate geopotential, taken as the maximum geopotential height difference within 1$^{\circ}$ GCD of the candidate location. It's worth noting that, for the $1.4^{\circ}$ tracking, we do not use the 10m wind criterion since this data was not available for the NeuralGCM simulation.

Next, we compare these results with TCs simulated by the X-SHiELD model (Fig.~\ref{fig:climatology}) when regridded to $1.4^{\circ}$, allowing for a fair comparison with NeuralGCM-$1.4^{\circ}$ at the same resolution. Since the X-SHiELD dataset lacks SLP information, we utilize vorticity for tracking TCs. In a manner similar to what is described in the previous paragraph, we identify a set of vorticity parameters that, when used with the TempestExtremes tracker on ERA5 data regridded to $1.4^{\circ}$ resolution, yield results that closely matched TCs in native ERA5 data tracked using SLP as the criterion. We successfully fine-tune the vorticity parameters to ensure that TempestExtremes detects nearly identical TC tracks (as shown in Fig.~\ref{sifig:TCs_xsheild_dates}b) when using the vorticity criterion compared to when using the tracker with native ERA5 resolution and the SLP criterion. This yields a similar number of TCs, with 88 in the native resolution and 86 when using the vorticity criterion in the $1.4^{\circ}$ resolution.
For vorticity tracking, we require a vorticity increase/decrease of at least $0.00006$ ${s^{-1}}$ within a $5.5^{\circ}$ GCD radius of the candidate location. Additionally, we impose the requirement for a decrease in geopotential height difference between the 300 hPa and 500 hPa of $25.8$ $\rm{m^2s^{-2}}$ within a $6.5^{\circ}$ GCD from the candidate geopotential. (see Table~\ref{apx:table:TC_parameters}).

We apply these new SLP and vorticity parameters to detect TCs in the NeuralGCM-$1.4^{\circ}$ simulation. We find that tracking TCs with both sets of parameters produces similar TC tracks (Fig.~\ref{sifig:TCs_xsheild_dates}c). These TC tracks also closely resemble TC tracks identified in ERA5 data in terms of number of TCs, as well as their locations and shapes (Fig.~\ref{sifig:TCs_xsheild_dates}).

We also employ the tracker to derive TC tracks from both ECMWF-ENS and NeuralGCM-ENS, each at a resolution of 1.4$^{\circ}$. However, since ECMWF-ENS lacks data for geopotential at 300 hPa, which is necessary for the conditions we used above, we track TCs for the ensemble using the sea-level pressure (SLP) criterion in conjunction with the vorticity criterion, as previously described for simulations at the 1.4$^{\circ}$ resolution.

\begin{table}[]
\begin{tabular}{|l|l|l|}
\hline{}
& Criterion 1& Criterion 2\\ \hline
\multicolumn{1}{|l|}{\begin{tabular}[c]{@{}l@{}}Sea level pressure \\ tracking ($0.25{^\circ}$)\end{tabular}} & \begin{tabular}[c]{@{}l@{}}SLP change \\ of 200Pa over 5.5 GCD\end{tabular}                   & \begin{tabular}[c]{@{}l@{}}Difference between geopotential surfaces \\ at 300 and 500 hPa decrease\\ by at least -58.8$m^2/s^2$ over 6.5 GCD\end{tabular} \\ \hline
\multicolumn{1}{|l|}{\begin{tabular}[c]{@{}l@{}}Sea level pressure \\ tracking ($1.4{^\circ}$)\end{tabular}}  & \begin{tabular}[c]{@{}l@{}}SLP change \\ of 60Pa over 5.5 GCD\end{tabular}                    & \begin{tabular}[c]{@{}l@{}}Difference between geopotential surfaces \\ at 300 and 500 hPa decrease\\ by at least -25.8$m^2/s^2$ over 6.5 GCD\end{tabular} \\ \hline
\multicolumn{1}{|l|}{\begin{tabular}[c]{@{}l@{}}Vorticity \\ tracking ($1.4{^\circ}$)\end{tabular}}         & \begin{tabular}[c]{@{}l@{}}Vorticity at 850hPa change \\ of $\pm0.00006 s^{-1}$  over 5.5GCD\end{tabular} & \begin{tabular}[c]{@{}l@{}}Difference between geopotential surfaces \\ at 300 and 500 hPa decrease\\ by at least -25.8$m^2/s^2$ over 6.5 GCD\end{tabular} \\ \hline
\multicolumn{1}{|l|}{\begin{tabular}[c]{@{}l@{}}SLP ensemble\\ tracking ($1.4{^\circ}$)\end{tabular}}         & \begin{tabular}[c]{@{}l@{}}SLP change \\ of 60Pa over 5.5 GCD\end{tabular} & \begin{tabular}[c]{@{}l@{}}Vorticity at 850hPa change \\ of $\pm0.00006 s^{-1}$  over 5.5GCD\end{tabular} \\ \hline
\end{tabular}
\caption{\label{apx:table:TC_parameters} The parameters used for TC tracking when using different model resolutions. }
\end{table}

\begin{figure*}
\centering
\includegraphics[width=\textwidth]{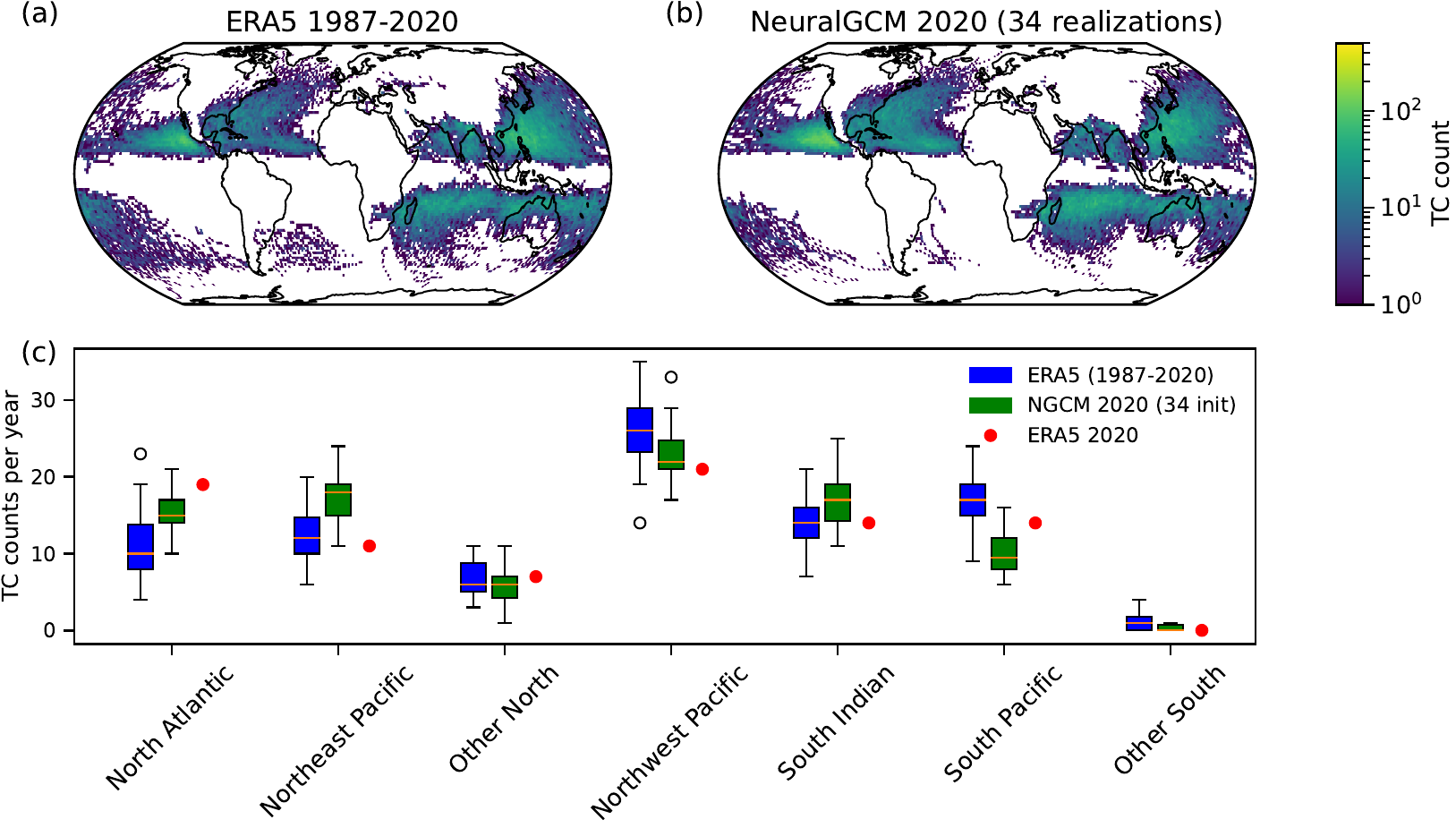}
\caption{Tropical Cyclone densities and annual regional counts. (a) Tropical Cyclone (TC) density from ERA5 data spanning 1987-2020. (b) TC density from NeuralGCM-$1.4^{\circ}$ for 2020, generated using 34 different initial conditions all initialized in 2019. (c) Box plot depicting the annual number of TCs across different regions, based on ERA5 data (1987-2020), NeuralGCM-$1.4^{\circ}$ for 2020 (34 initial conditions), and red markers show ERA5 for 2020. In the box plots, the red line represents the median; the box delineates the first to third quartiles; the whiskers extend to 1.5 times the interquartile range (Q1 - 1.5IQR and Q3 + 1.5IQR), and outliers are shown as individual dots. Each year is defined from January 19th to January 17th of the following year, aligning with data availability from XSHiELD. For NeuralGCM simulations, the 3 initial conditions starting in January 2019 exclude data for January 17th, 2021, as these runs spanned only two years.}\label{sifig:TCs_34_years}
\end{figure*}

\begin{figure*}
\centering
\includegraphics[width=\textwidth]{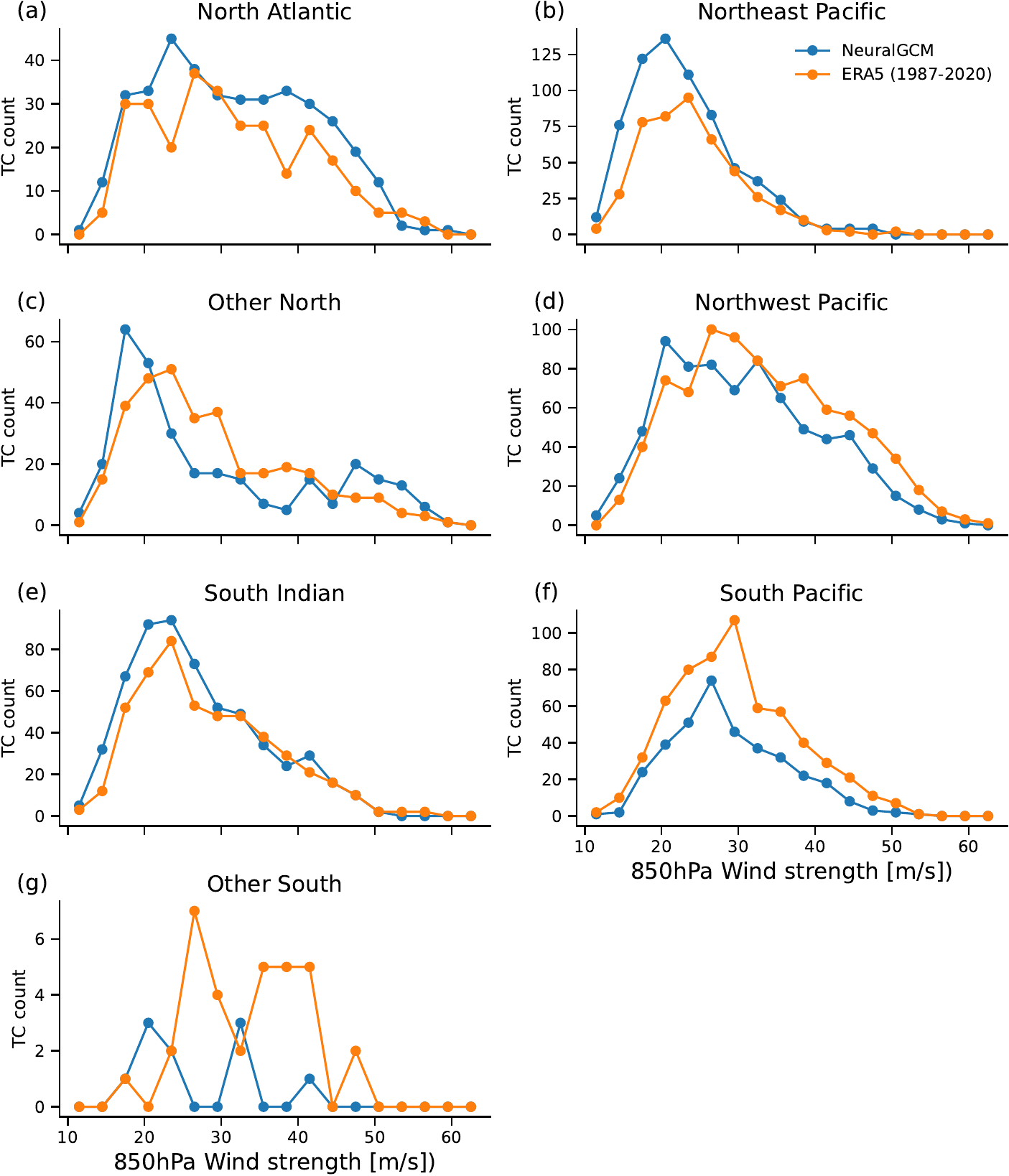}
\caption{Tropical Cyclone maximum wind distribution in NeuralGCM vs. ERA5. Number of Tropical Cyclones (TCs) as a function of maximum wind speed at 850hPa across different regions, based on ERA5 data (1987-2020; in orange), and NeuralGCM-$1.4^{\circ}$ for 2020 (34 initial conditions; in blue). Each year is defined from January 19th to January 17th of the following year, aligning with data availability from XSHiELD. For NeuralGCM simulations, the 3 initial conditions starting in January 2019 exclude data for January 17th, 2021, as these runs spanned only two years.}
\label{sifig:TCs_wind_34_years}
\end{figure*}

\subsection{CMIP6 models used in AMIP runs}\label{apx:sec:AMIP_models_used}
We analyzed data from 22 CMIP6 models that were configured to use prescribed sea surface temperatures (SST), known as AMIP runs, taken from Google's Public Dataset program stored on Google Cloud Storage. Among these, for the following 17 models — BCC-CSM2-MR, CAMS-CSM1-0, CESM2, CESM2-WACCM, CanESM5, EC-Earth3, EC-Earth3-Veg, FGOALS-f3-L, GFDL-AM4, GFDL-CM4, GFDL-ESM4, GISS-E2-1-G, IPSL-CM6A-LR, MIROC6, MRI-ESM2-0, NESM3, and SAM0-UNICON — we utilized the ``r1i1p1f1'' variant identifier to which we had access. However, for the remaining five models, we resorted to using alternative variant identifiers: ``r1i1p1f2'' for CNRM-CM6-1 and CNRM-ESM2-1, ``r2i1p1f3'' for HadGEM3-GC31-LL, ``r1i1p1f3'' for HadGEM3-GC31-MM, and ``r1i1p1f2'' for UKESM1-0-LL.
Additionally, although data from the E3SM model was available from Google's Public Dataset program, we chose not to include it in our analysis due to some temperature artifacts near Antarctica.

\subsection{Generalizing to unseen data}\label{apx:sec:extrapolation}

One motivation for incorporating strong physics priors into NeuralGCM is to improve performance on out-of-sample data, such as that caused by climate change or systematic changes in Earth observing systems. Physically consistent weather models should remain accurate even under different climates or with weather conditions outside their training data.
First, we study the ability of NeuralGCM models to generalize weather forecasts to warmer years. Later, we investigate its ability to simulate warmer climates.

\subsubsection{Weather forecasting in warmer years}\label{apx:subsec:weather_extrapolation}

\begin{figure*}
\centering
\includegraphics[width=0.65\textwidth]{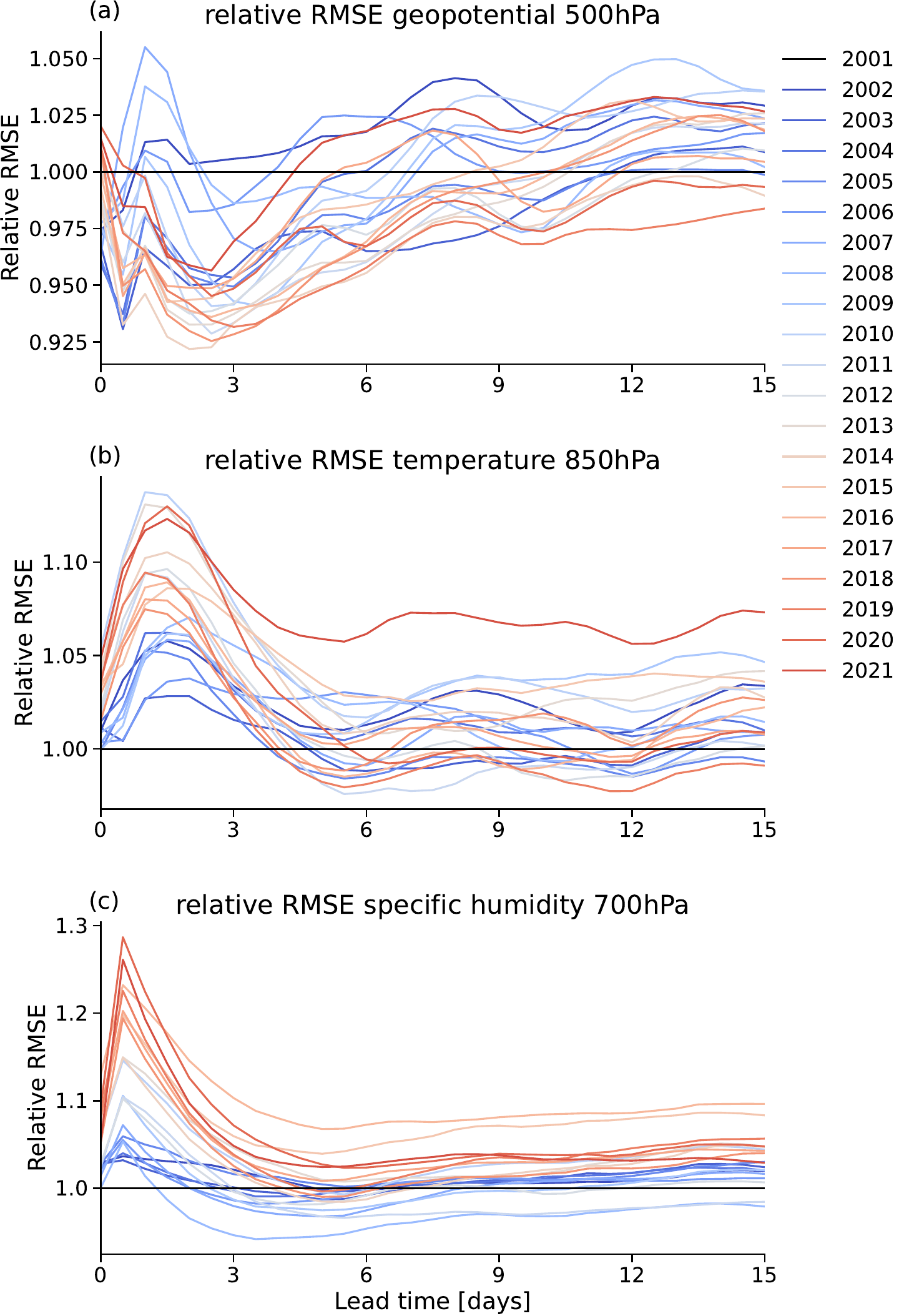}
\caption{
Extrapolation of mid-range weather forecast skill over 21 years. Relative Root Mean Square Error (RMSE), normalized to the year 2001, for a NeuralGCM-$2.8^{\circ}$ model trained on 1979-2000 ERA5 data. Results are shown for (a) 500 hPa geopotential height, (b) 850 hPa temperature, and (c) 700 hPa specific humidity as a function of forecast lead time.}
\label{sifig:extrapolation_weather_21_years}
\end{figure*}

To assess the extrapolation capabilities of NeuralGCM in the context of weather predictions, we trained a NeuralGCM-$2.8^{\circ}$
model using ERA5 data exclusively from 1979 to 2000. We then tested its performance across 21 future years. Figure~\ref{sifig:extrapolation_weather_21_years} demonstrates the model's skill in mid-range weather prediction for various years and variables. For geopotential, forecast skill remains remarkably consistent even when extrapolating 20 years into the future. While there is a noticeable decline in the 4-day forecast accuracy for temperature and specific humidity, skill remains stable for temperature and exhibits only a slight deterioration for specific humidity in longer-range predictions.

Next, we wanted to compare the ability of NeuralGCM-$0.7^{\circ}$ to generalize to future years against that of GraphCast. 
Because the final version of NeuralGCM-$0.7^{\circ}$ was trained on data through 2019, to produce additional data for comparisons we also consider a developmental version NeuralGCM-$0.7^{\circ}$-2017 only trained through 2017 data.
This model uses the same architecture described previously with three major differences:

\begin{enumerate}
\item It does not include a surface embedding or surface forcing features (sea surface temperature, sea ice concentration).
\item It does not include the memory feature.
\item Its loss is scaled differently in time, like $(1 + (\tau/24))^{-1}$ instead of $(1 + (\tau/40)^4)^{-1/2}$.
\end{enumerate}

\begin{figure*}
\begin{center}
\makebox[\textwidth]{\colorbox{white}{\includegraphics[width=0.85\paperwidth]{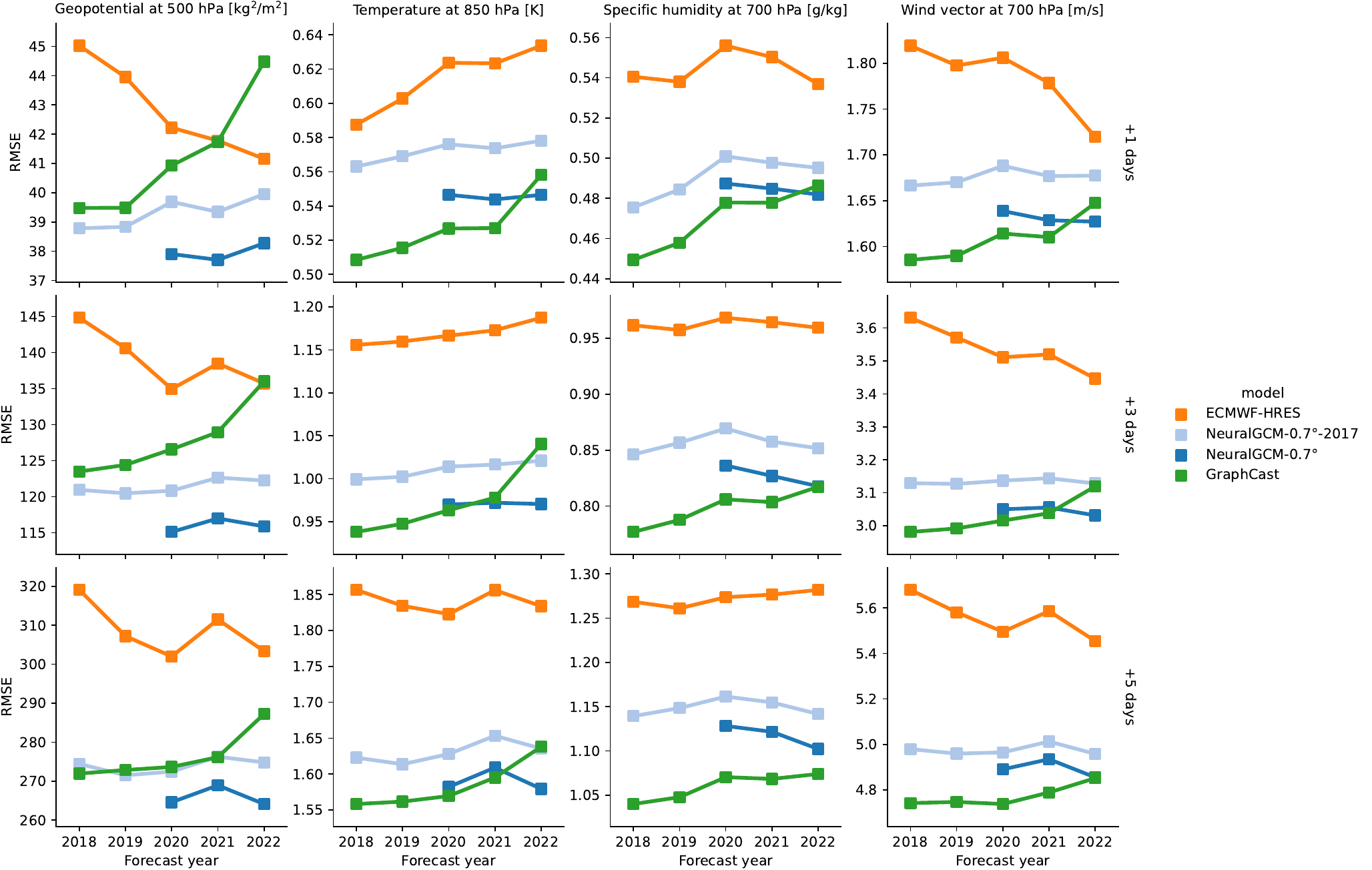}}}
\end{center}
\caption{RMSE for 1-, 3- and 5-day forecasts, starting in different forecast outside the training datasets for NeuralGCM and GraphCast.
RMSE is averaged over forecasts initialized every 12 hours at midnight and noon UTC in the indicated year.
}\label{sifig:5years}
\end{figure*}

In Fig.~\ref{sifig:5years} we compare accuracy trends of GraphCast, NeuralGCM-$0.7^{\circ}$, NeuralGCM-$0.7^{\circ}$-2017 and ECMWF-HRES evaluated on ERA5 data from years $2018$-$2022$.
In this experiment GraphCast and NeuralGCM-$0.7^{\circ}$-2017 variants were trained on data until $2017$.
ECMWF-HRES and NeuralGCM models display little variation over evaluation years, while GraphCast accuracy degrades by a few percent when evaluated on 2022 forecasts ($5$ years more recent than the training domain), consistent with the degraded performance noted in the GraphCast paper.
Forecasts of geopotential and at short lead times have the largest increases in error.

\subsubsection{Extrapolation to warmer climates}\label{apx:subsec:climate_extrapolation}
All NeuralGCM models have been trained solely on ERA5 data. As with all machine learning models, we do not expect the NeuralGCM models to extrapolate beyond their training regime. However, we acknowledge that extrapolation might be reasonable in cases where the input/output distributions do not substantially change compared to the training data. Therefore, we aimed to test whether NeuralGCM can still provide reasonable results for climates that are moderately to substantially warmer.

To assess the capability of NeuralGCM to simulate warmer climates, we conducted AMIP-like simulations with warmer SST (including +1K, +2K, and +4K SST) over a duration of 34 years using NeuralGCM-$2.8^\circ$. As a baseline for comparison, we used AMIP and AMIP +4K SST runs of the CESM model (CESM AMIP +4K was available until 2013). Out of 8 different runs with initial conditions spaced every 50 days during the year 1980, 6 (for AMIP +1K), 6 (for AMIP +2K), and 4 (for AMIP +4K) simulations were stable for the entire 34-year period. Since we compare to a single realization of CESM, we have chosen to use simulations that were initialized on April 10th, as simulations started on this date remained stable for 34 years across all SST warming levels. 

\begin{figure*}
\centering
\includegraphics[width=\textwidth]{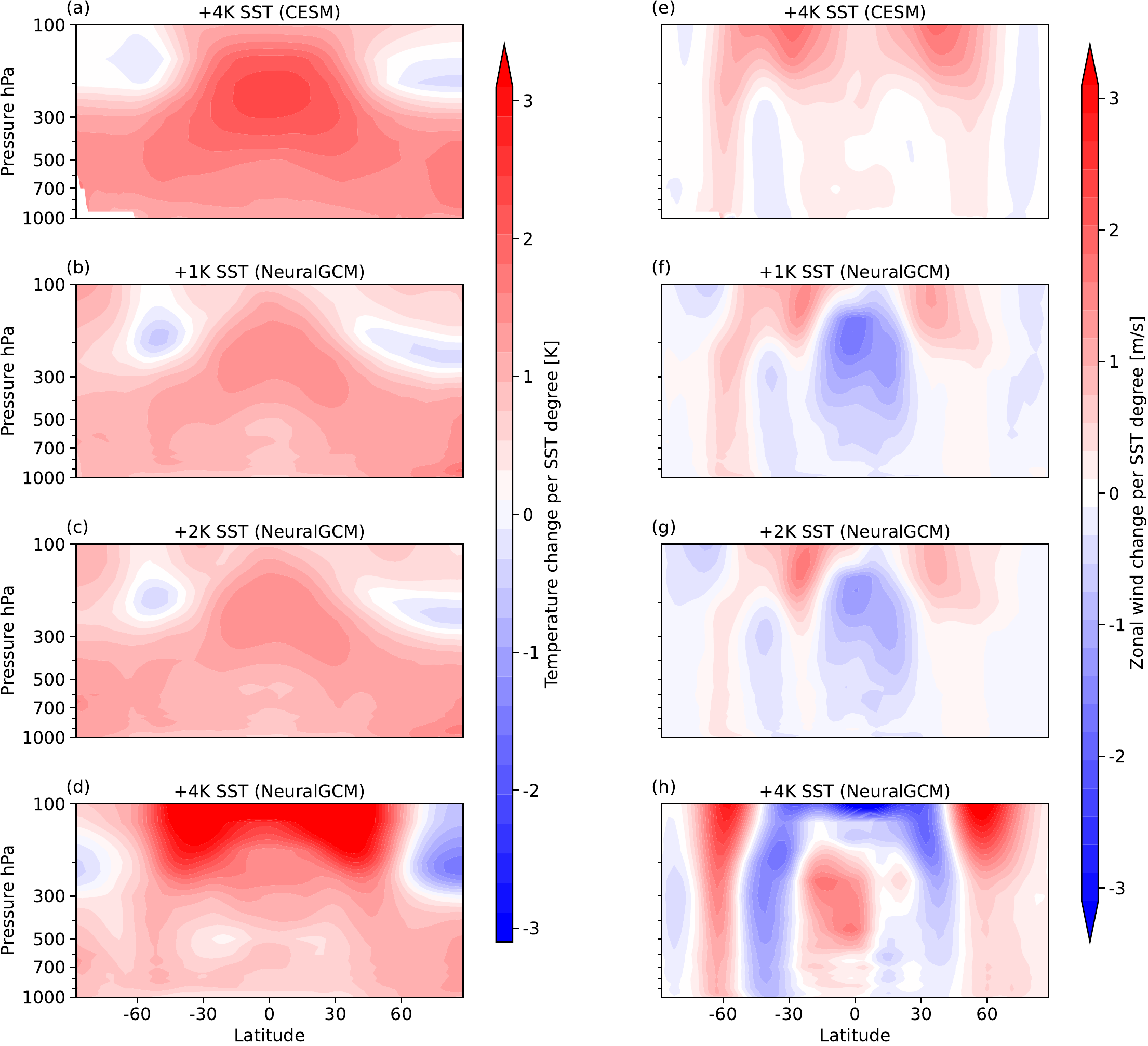}
\caption{Response of temperature and zonal wind to  sea surface temperature (SST) warming. Left panels depict the zonal mean temperature response per degree of SST increase as a function of latitude and pressure, calculated as the difference between AMIP runs with increased SST to AMIP simulations averaged over 1981-2013: (a) CESM with a +4K SST warming, (b-d) NeuralGCM with +1K, +2K, and +4K SST warming, respectively. Right panels illustrate the corresponding zonal mean zonal wind response per SST degree increase for (e) CESM with a +4K SST enhancement and (f-h) NeuralGCM with +1K, +2K, and +4K SST warming, respectively. All NeuralGCM models used initial condition from April 10th 1980.}
\label{sifig:extrapolation_temperature_wind}
\end{figure*}

In Fig.~\ref{sifig:extrapolation_temperature_wind}, we test the response of NeuralGCM model runs to SST warming. We find that the zonal mean temperature structure and the zonal mean zonal wind response in NeuralGCM for +1K and +2K SST warming runs are broadly consistent with the CESM response. Specifically, robust features of climate warming, such as upper tropospheric warming, polar amplification, and Southern Hemisphere polar jet shift \citeSI{vallis2015response}, are present in all simulations. However, there are also some differences in the responses, for example, the zonal mean zonal wind changes in the tropics have opposite signs in CESM and NeuralGCM.
We find that the NeuralGCM +4K SST warming simulation shows a very different response compared to NeuralGCM simulations with +1K and +2K SST warming. This likely indicates that when using NeuralGCM in a climate driven by +4K SST warming, the results provided by NeuralGCM are unrealistic.

\begin{figure*}
\centering
\includegraphics[width=0.82\textwidth]{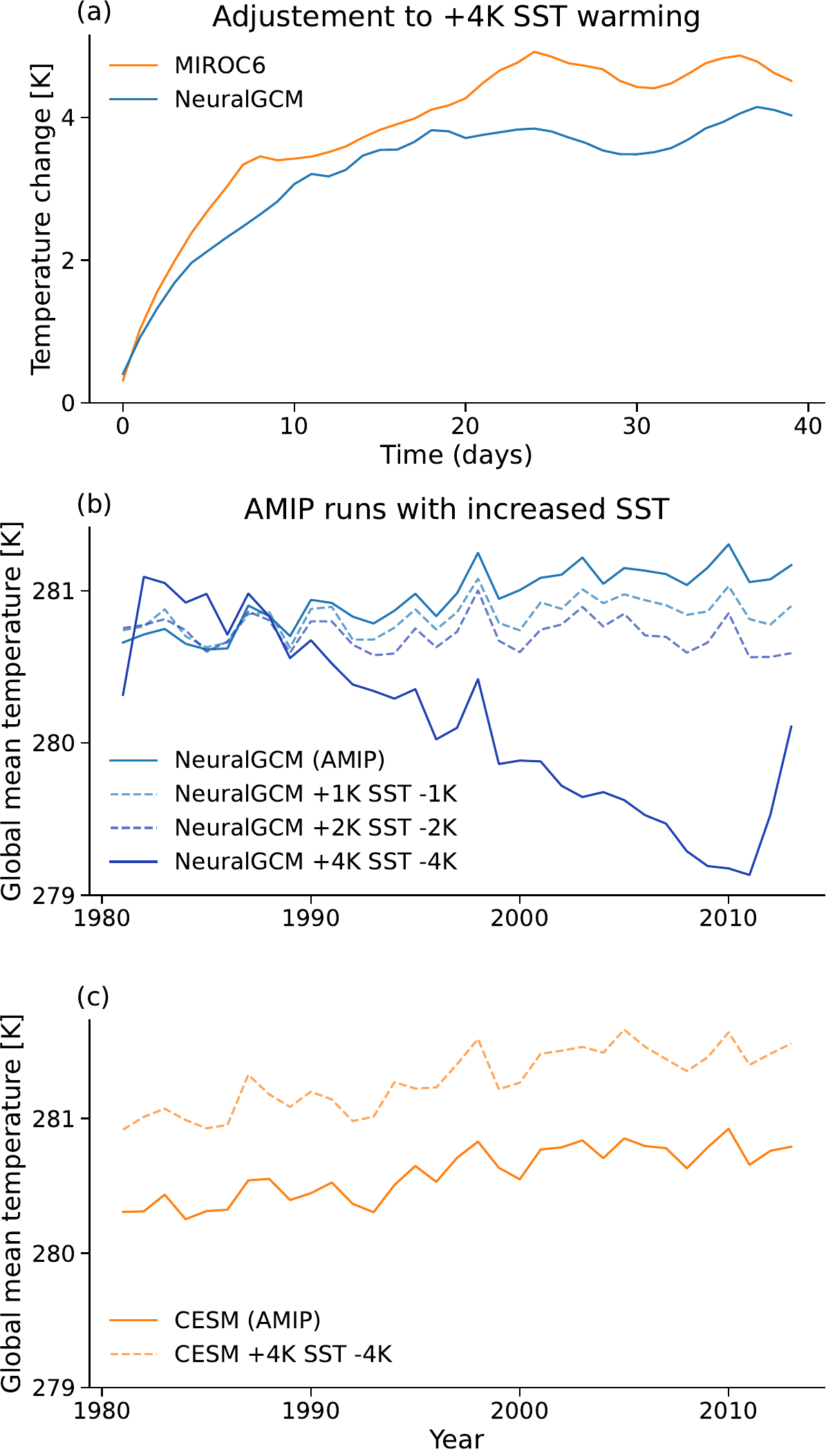}
\caption{Temporal dynamics of temperature changes in AMIP simulations with warmed sea surface temperature (SST). Panel (a) shows the 40-day 850hPa global temperature response to a +4K SST increase in MIROC6 (defined as the difference between AMIP +4K and the standard AMIP run; represented in orange) and NeuralGCM (in blue). Panel (b) presents the global mean temperature at 850hPa across various NeuralGCM AMIP scenarios, including standard runs and those with +1K, +2K, and +4K SST enhancements, adjusting the global mean by subtracting the corresponding SST warming. Panel (c) displays the global mean temperature at 850hPa for both the CESM AMIP standard run and the CESM AMIP scenario with a +4K SST increase, applying the same SST adjustment methodology. Panels b and c cover the period from 1981 to 2013, with 2013 being the final year for which CESM AMIP +4K SST data were available. All NeuralGCM simulations commenced from the initial condition on April 10th, 1980.}
\label{sifig:extrapolation_temperature_treds_AMIP_SST_warming}
\end{figure*}

Fig.~\ref{sifig:extrapolation_temperature_treds_AMIP_SST_warming}a demonstrates that the mean global temperature at 850hPa adjusts in response to an increase in SST by +4K on the same timescale in both NeuralGCM and MIROC6. We use MIROC6 for comparison because the daily temperature data for both AMIP and AMIP +4K SST scenarios were available from Google Cloud Storage for MIROC6. This implies that the short-term response of NeuralGCM is similar to that of a physics-based model.

In Fig.~\ref{sifig:extrapolation_temperature_treds_AMIP_SST_warming}b and c, we present the global mean temperature at 850hPa for NeuralGCM and CESM simulations. The most notable observation is that, unlike CESM simulations, NeuralGCM simulations with warmer SST do not exhibit the same trend as NeuralGCM with the AMIP configuration. This suggests that NeuralGCM, when simulating a warmer climate, experiences a climate drift. The climate drift is particularly evident for the +4K SST warming (a scenario we have already noted NeuralGCM does not realistically extrapolate to), but it is also observable in the +1K and +2K SST warming scenarios, albeit to a lesser extent. For example, after 34 years of model integration, the temperature difference between the AMIP run and AMIP + 1K SST -1K is approximately 0.25K. We note that the drift we find in NeuralGCM AMIP +1K and +2K SST warming is substantially lower than the warming itself. Therefore, we believe it is still appropriate to compare the response to warming in Figure~\ref{sifig:extrapolation_temperature_treds_AMIP_SST_warming}.

In the future, NeuralGCM could improve its ability to generalize to different climates, even those it has not trained on, by using a mix of strategies. While climate-invariant methods \citeSI{beucler2024climate} can help, training NeuralGCM on data obtained from simulations (e.g., of warmer climates) and incorporating physically-based models (e.g., radiation schemes) will likely be necessary.

\end{appendices}


\begin{thebibliography}{10}
\expandafter\ifx\csname url\endcsname\relax
    \def\url#1{\burl{#1}}\fi
\expandafter\ifx\csname urlprefix\endcsname\relax\def\urlprefix{URL }\fi
\providecommand{\bibinfo}[2]{#2}
\providecommand{\eprint}[2][]{\url{#2}}
\providecommand{\doi}[1]{\url{https://doi.org/#1}}
\bibcommenthead

\bibitem{Bauer2015}
\bibinfo{author}{Bauer, P.}, \bibinfo{author}{Thorpe, A.} \& \bibinfo{author}{Brunet, G.}
\newblock \bibinfo{title}{The quiet revolution of numerical weather prediction}.
\newblock \emph{\bibinfo{journal}{Nature}} \textbf{\bibinfo{volume}{525}}, \bibinfo{pages}{47--55} (\bibinfo{year}{2015}).
\newblock \urlprefix\url{http://dx.doi.org/10.1038/nature14956}.

\bibitem{Balaji2022-kp}
\bibinfo{author}{Balaji, V.} \emph{et~al.}
\newblock \bibinfo{title}{Are general circulation models obsolete?}
\newblock \emph{\bibinfo{journal}{Proc. Natl. Acad. Sci. U. S. A.}} \textbf{\bibinfo{volume}{119}}, \bibinfo{pages}{e2202075119} (\bibinfo{year}{2022}).
\newblock \urlprefix\url{http://dx.doi.org/10.1073/pnas.2202075119}.

\bibitem{hourdin2017_tuning}
\bibinfo{author}{Hourdin, F.} \emph{et~al.}
\newblock \bibinfo{title}{The art and science of climate model tuning}.
\newblock \emph{\bibinfo{journal}{Bull. Am. Meteorol. Soc.}} \textbf{\bibinfo{volume}{98}}, \bibinfo{pages}{589--602} (\bibinfo{year}{2017}).
\newblock \urlprefix\url{https://doi.org/10.1175/BAMS-D-15-00135.1}.

\bibitem{bony2005marine}
\bibinfo{author}{Bony, S.} \& \bibinfo{author}{Dufresne, J.-L.}
\newblock \bibinfo{title}{Marine boundary layer clouds at the heart of tropical cloud feedback uncertainties in climate models}.
\newblock \emph{\bibinfo{journal}{Geophysical Research Letters}} \textbf{\bibinfo{volume}{32}} (\bibinfo{year}{2005}).

\bibitem{webb2013origins}
\bibinfo{author}{Webb, M.~J.}, \bibinfo{author}{Lambert, F.~H.} \& \bibinfo{author}{Gregory, J.~M.}
\newblock \bibinfo{title}{Origins of differences in climate sensitivity, forcing and feedback in climate models}.
\newblock \emph{\bibinfo{journal}{Climate Dynamics}} \textbf{\bibinfo{volume}{40}}, \bibinfo{pages}{677--707} (\bibinfo{year}{2013}).

\bibitem{sherwood2014spread}
\bibinfo{author}{Sherwood, S.~C.}, \bibinfo{author}{Bony, S.} \& \bibinfo{author}{Dufresne, J.-L.}
\newblock \bibinfo{title}{Spread in model climate sensitivity traced to atmospheric convective mixing}.
\newblock \emph{\bibinfo{journal}{Nature}} \textbf{\bibinfo{volume}{505}}, \bibinfo{pages}{37--42} (\bibinfo{year}{2014}).

\bibitem{PalmerStevens2019}
\bibinfo{author}{Palmer, T.} \& \bibinfo{author}{Stevens, B.}
\newblock \bibinfo{title}{The scientific challenge of understanding and estimating climate change}.
\newblock \emph{\bibinfo{journal}{Proc. Natl. Acad. Sci. U. S. A.}} \textbf{\bibinfo{volume}{116}}, \bibinfo{pages}{24390--24395} (\bibinfo{year}{2019}).
\newblock \urlprefix\url{http://dx.doi.org/10.1073/pnas.1906691116}.

\bibitem{fischer2013robust}
\bibinfo{author}{Fischer, E.~M.}, \bibinfo{author}{Beyerle, U.} \& \bibinfo{author}{Knutti, R.}
\newblock \bibinfo{title}{Robust spatially aggregated projections of climate extremes}.
\newblock \emph{\bibinfo{journal}{Nature Climate Change}} \textbf{\bibinfo{volume}{3}}, \bibinfo{pages}{1033--1038} (\bibinfo{year}{2013}).

\bibitem{field2012managing}
\bibinfo{author}{Field, C.~B.}
\newblock \emph{\bibinfo{title}{Managing the risks of extreme events and disasters to advance climate change adaptation: special report of the intergovernmental panel on climate change}}  (\bibinfo{publisher}{Cambridge University Press}, \bibinfo{address}{Cambridge, UK}, \bibinfo{year}{2012}).

\bibitem{rasp2023weatherbench}
\bibinfo{author}{Rasp, S.} \emph{et~al.}
\newblock \bibinfo{title}{{WeatherBench 2: A benchmark for the next generation of data-driven global weather models}} (\bibinfo{year}{2023}).
\newblock \bibinfo{note}{Preprint at \url{http://arxiv.org/abs/2308.15560}}.

\bibitem{keisler2022forecasting}
\bibinfo{author}{Keisler, R.}
\newblock \bibinfo{title}{Forecasting global weather with graph neural networks}.
\newblock \emph{\bibinfo{journal}{arXiv preprint arXiv:2202.07575}}  (\bibinfo{year}{2022}).

\bibitem{bi2023accurate}
\bibinfo{author}{Bi, K.} \emph{et~al.}
\newblock \bibinfo{title}{Accurate medium-range global weather forecasting with 3d neural networks}.
\newblock \emph{\bibinfo{journal}{Nature}} \textbf{\bibinfo{volume}{619}}, \bibinfo{pages}{533--538} (\bibinfo{year}{2023}).

\bibitem{lam2022graphcast}
\bibinfo{author}{Lam, R.} \emph{et~al.}
\newblock \bibinfo{title}{Learning skillful medium-range global weather forecasting}.
\newblock \emph{\bibinfo{journal}{Science}} \textbf{\bibinfo{volume}{382}}, \bibinfo{pages}{1416--1421} (\bibinfo{year}{2023}).
\newblock \urlprefix\url{https://www.science.org/doi/abs/10.1126/science.adi2336}.

\bibitem{hersbach2020era5}
\bibinfo{author}{Hersbach, H.} \emph{et~al.}
\newblock \bibinfo{title}{The {ERA5} global reanalysis}.
\newblock \emph{\bibinfo{journal}{Quarterly Journal of the Royal Meteorological Society}} \textbf{\bibinfo{volume}{146}}, \bibinfo{pages}{1999--2049} (\bibinfo{year}{2020}).

\bibitem{Zhou2019-next-gen-GFS}
\bibinfo{author}{Zhou, L.} \emph{et~al.}
\newblock \bibinfo{title}{Toward convective-scale prediction within the next generation global prediction system}.
\newblock \emph{\bibinfo{journal}{Bull. Am. Meteorol. Soc.}} \textbf{\bibinfo{volume}{100}}, \bibinfo{pages}{1225--1243} (\bibinfo{year}{2019}).
\newblock \urlprefix\url{https://journals.ametsoc.org/view/journals/bams/100/7/bams-d-17-0246.1.xml}.

\bibitem{bonavita2023limitations}
\bibinfo{author}{Bonavita, M.}
\newblock \bibinfo{title}{On the limitations of data-driven weather forecasting models}.
\newblock \emph{\bibinfo{journal}{arXiv preprint arXiv:2309.08473}}  (\bibinfo{year}{2023}).

\bibitem{weyn2020improving}
\bibinfo{author}{Weyn, J.~A.}, \bibinfo{author}{Durran, D.~R.} \& \bibinfo{author}{Caruana, R.}
\newblock \bibinfo{title}{Improving data-driven global weather prediction using deep convolutional neural networks on a cubed sphere}.
\newblock \emph{\bibinfo{journal}{Journal of Advances in Modeling Earth Systems}} \textbf{\bibinfo{volume}{12}}, \bibinfo{pages}{e2020MS002109} (\bibinfo{year}{2020}).

\bibitem{watt2023ace}
\bibinfo{author}{Watt-Meyer, O.} \emph{et~al.}
\newblock \bibinfo{title}{{ACE}: A fast, skillful learned global atmospheric model for climate prediction}.
\newblock \emph{\bibinfo{journal}{arXiv preprint arXiv:2310.02074}}  (\bibinfo{year}{2023}).

\bibitem{Bretherton2023-ym}
\bibinfo{author}{Bretherton, C.~S.}
\newblock \bibinfo{title}{Old dog, new trick: Reservoir computing advances machine learning for climate modeling}.
\newblock \emph{\bibinfo{journal}{Geophysical Research Letters}} \textbf{\bibinfo{volume}{50}}, \bibinfo{pages}{e2023GL104174} (\bibinfo{year}{2023}).

\bibitem{Reichstein2019review}
\bibinfo{author}{Reichstein, M.} \emph{et~al.}
\newblock \bibinfo{title}{Deep learning and process understanding for data-driven earth system science}.
\newblock \emph{\bibinfo{journal}{Nature}} \textbf{\bibinfo{volume}{566}}, \bibinfo{pages}{195--204} (\bibinfo{year}{2019}).
\newblock \urlprefix\url{https://doi.org/10.1038/s41586-019-0912-1}.

\bibitem{brenowitz2019spatially}
\bibinfo{author}{Brenowitz, N.~D.} \& \bibinfo{author}{Bretherton, C.~S.}
\newblock \bibinfo{title}{Spatially extended tests of a neural network parametrization trained by coarse-graining}.
\newblock \emph{\bibinfo{journal}{Journal of Advances in Modeling Earth Systems}} \textbf{\bibinfo{volume}{11}}, \bibinfo{pages}{2728--2744} (\bibinfo{year}{2019}).

\bibitem{rasp2018deep}
\bibinfo{author}{Rasp, S.}, \bibinfo{author}{Pritchard, M.~S.} \& \bibinfo{author}{Gentine, P.}
\newblock \bibinfo{title}{Deep learning to represent subgrid processes in climate models}.
\newblock \emph{\bibinfo{journal}{Proceedings of the National Academy of Sciences}} \textbf{\bibinfo{volume}{115}}, \bibinfo{pages}{9684--9689} (\bibinfo{year}{2018}).

\bibitem{yuval2020stable}
\bibinfo{author}{Yuval, J.} \& \bibinfo{author}{O’Gorman, P.~A.}
\newblock \bibinfo{title}{Stable machine-learning parameterization of subgrid processes for climate modeling at a range of resolutions}.
\newblock \emph{\bibinfo{journal}{Nature communications}} \textbf{\bibinfo{volume}{11}}, \bibinfo{pages}{3295} (\bibinfo{year}{2020}).

\bibitem{kwa2023machine}
\bibinfo{author}{Kwa, A.} \emph{et~al.}
\newblock \bibinfo{title}{Machine-learned climate model corrections from a global storm-resolving model: Performance across the annual cycle}.
\newblock \emph{\bibinfo{journal}{Journal of Advances in Modeling Earth Systems}} \textbf{\bibinfo{volume}{15}}, \bibinfo{pages}{e2022MS003400} (\bibinfo{year}{2023}).

\bibitem{arcomano2023hybrid}
\bibinfo{author}{Arcomano, T.}, \bibinfo{author}{Szunyogh, I.}, \bibinfo{author}{Wikner, A.}, \bibinfo{author}{Hunt, B.~R.} \& \bibinfo{author}{Ott, E.}
\newblock \bibinfo{title}{A hybrid atmospheric model incorporating machine learning can capture dynamical processes not captured by its physics-based component}.
\newblock \emph{\bibinfo{journal}{Geophysical Research Letters}} \textbf{\bibinfo{volume}{50}}, \bibinfo{pages}{e2022GL102649} (\bibinfo{year}{2023}).

\bibitem{han2023ensemble}
\bibinfo{author}{Han, Y.}, \bibinfo{author}{Zhang, G.~J.} \& \bibinfo{author}{Wang, Y.}
\newblock \bibinfo{title}{An ensemble of neural networks for moist physics processes, its generalizability and stable integration}.
\newblock \emph{\bibinfo{journal}{Journal of Advances in Modeling Earth Systems}} \textbf{\bibinfo{volume}{15}}, \bibinfo{pages}{e2022MS003508} (\bibinfo{year}{2023}).

\bibitem{Gelbrecht2023differentiable}
\bibinfo{author}{Gelbrecht, M.}, \bibinfo{author}{White, A.}, \bibinfo{author}{Bathiany, S.} \& \bibinfo{author}{Boers, N.}
\newblock \bibinfo{title}{Differentiable programming for earth system modeling}.
\newblock \emph{\bibinfo{journal}{Geoscientific Model Development}} \textbf{\bibinfo{volume}{16}}, \bibinfo{pages}{3123--3135} (\bibinfo{year}{2023}).
\newblock \urlprefix\url{https://gmd.copernicus.org/articles/16/3123/2023/}.

\bibitem{bradbury2018jax}
\bibinfo{author}{Bradbury, J.} \emph{et~al.}
\newblock \bibinfo{title}{{JAX}: composable transformations of {P}ython+{N}um{P}y programs} (\bibinfo{year}{2018}).
\newblock \urlprefix\url{http://github.com/google/jax}.

\bibitem{Bourke1974-spectral}
\bibinfo{author}{Bourke, W.}
\newblock \bibinfo{title}{{A Multi-Level Spectral Model. I. Formulation and Hemispheric Integrations}}.
\newblock \emph{\bibinfo{journal}{Mon. Weather Rev.}} \textbf{\bibinfo{volume}{102}}, \bibinfo{pages}{687--701} (\bibinfo{year}{1974}).
\newblock \urlprefix\url{https://journals.ametsoc.org/view/journals/mwre/102/10/1520-0493_1974_102_0687_amlsmi_2_0_co_2.xml}.

\bibitem{durran2010numerical}
\bibinfo{author}{Durran, D.~R.}
\newblock \emph{\bibinfo{title}{Numerical methods for fluid dynamics: With applications to geophysics}} \bibinfo{edition}{Second} edn, Vol.~\bibinfo{volume}{32} (\bibinfo{publisher}{Springer}, \bibinfo{address}{New York}, \bibinfo{year}{2010}).

\bibitem{wang2022non}
\bibinfo{author}{Wang, P.}, \bibinfo{author}{Yuval, J.} \& \bibinfo{author}{O’Gorman, P.~A.}
\newblock \bibinfo{title}{Non-local parameterization of atmospheric subgrid processes with neural networks}.
\newblock \emph{\bibinfo{journal}{Journal of Advances in Modeling Earth Systems}} \textbf{\bibinfo{volume}{14}}, \bibinfo{pages}{e2022MS002984} (\bibinfo{year}{2022}).

\bibitem{daley1981normal}
\bibinfo{author}{Daley, R.}
\newblock \bibinfo{title}{Normal mode initialization}.
\newblock \emph{\bibinfo{journal}{Reviews of Geophysics}} \textbf{\bibinfo{volume}{19}}, \bibinfo{pages}{450--468} (\bibinfo{year}{1981}).

\bibitem{whitaker2013implicit}
\bibinfo{author}{Whitaker, J.~S.} \& \bibinfo{author}{Kar, S.~K.}
\newblock \bibinfo{title}{Implicit--explicit runge--kutta methods for fast--slow wave problems}.
\newblock \emph{\bibinfo{journal}{Monthly weather review}} \textbf{\bibinfo{volume}{141}}, \bibinfo{pages}{3426--3434} (\bibinfo{year}{2013}).

\bibitem{gilleland2009intercomparison}
\bibinfo{author}{Gilleland, E.}, \bibinfo{author}{Ahijevych, D.}, \bibinfo{author}{Brown, B.~G.}, \bibinfo{author}{Casati, B.} \& \bibinfo{author}{Ebert, E.~E.}
\newblock \bibinfo{title}{Intercomparison of spatial forecast verification methods}.
\newblock \emph{\bibinfo{journal}{Weather and forecasting}} \textbf{\bibinfo{volume}{24}}, \bibinfo{pages}{1416--1430} (\bibinfo{year}{2009}).

\bibitem{Gneiting2007ProperScoring}
\bibinfo{author}{Gneiting, T.} \& \bibinfo{author}{Raftery, A.~E.}
\newblock \bibinfo{title}{{Strictly Proper Scoring Rules, Prediction, and Estimation}}.
\newblock \emph{\bibinfo{journal}{J. Am. Stat. Assoc.}} \textbf{\bibinfo{volume}{102}}, \bibinfo{pages}{359--378} (\bibinfo{year}{2007}).
\newblock \urlprefix\url{https://doi.org/10.1198/016214506000001437}.

\bibitem{rasp2018neural}
\bibinfo{author}{Rasp, S.} \& \bibinfo{author}{Lerch, S.}
\newblock \bibinfo{title}{Neural networks for postprocessing ensemble weather forecasts}.
\newblock \emph{\bibinfo{journal}{Monthly Weather Review}} \textbf{\bibinfo{volume}{146}}, \bibinfo{pages}{3885--3900} (\bibinfo{year}{2018}).

\bibitem{pacchiardi2021probabilistic}
\bibinfo{author}{Pacchiardi, L.}, \bibinfo{author}{Adewoyin, R.}, \bibinfo{author}{Dueben, P.} \& \bibinfo{author}{Dutta, R.}
\newblock \bibinfo{title}{Probabilistic forecasting with generative networks via scoring rule minimization}.
\newblock \emph{\bibinfo{journal}{arXiv preprint arXiv:2112.08217}}  (\bibinfo{year}{2021}).

\bibitem{Fortin2014-spread-skill}
\bibinfo{author}{Fortin, V.}, \bibinfo{author}{Abaza, M.}, \bibinfo{author}{Anctil, F.} \& \bibinfo{author}{Turcotte, R.}
\newblock \bibinfo{title}{Why should ensemble spread match the {RMSE} of the ensemble mean?}
\newblock \emph{\bibinfo{journal}{J. Hydrometeorol.}} \textbf{\bibinfo{volume}{15}}, \bibinfo{pages}{1708--1713} (\bibinfo{year}{2014}).
\newblock \urlprefix\url{http://journals.ametsoc.org/doi/10.1175/JHM-D-14-0008.1}.

\bibitem{holton2004introduction}
\bibinfo{author}{Holton, J.~R.}
\newblock \emph{\bibinfo{title}{An introduction to dynamic meteorology}} \bibinfo{edition}{Fifth} edn (\bibinfo{publisher}{Elsevier Academic Press}, \bibinfo{address}{Waltham, MA, USA}, \bibinfo{year}{2004}).

\bibitem{cheng2022impact}
\bibinfo{author}{Cheng, K.-Y.} \emph{et~al.}
\newblock \bibinfo{title}{Impact of warmer sea surface temperature on the global pattern of intense convection: insights from a global storm resolving model}.
\newblock \emph{\bibinfo{journal}{Geophysical Research Letters}} \textbf{\bibinfo{volume}{49}}, \bibinfo{pages}{e2022GL099796} (\bibinfo{year}{2022}).

\bibitem{stevens2019dyamond}
\bibinfo{author}{Stevens, B.} \emph{et~al.}
\newblock \bibinfo{title}{Dyamond: the dynamics of the atmospheric general circulation modeled on non-hydrostatic domains}.
\newblock \emph{\bibinfo{journal}{Progress in Earth and Planetary Science}} \textbf{\bibinfo{volume}{6}}, \bibinfo{pages}{1--17} (\bibinfo{year}{2019}).

\bibitem{ullrich2021tempestextremes}
\bibinfo{author}{Ullrich, P.~A.} \emph{et~al.}
\newblock \bibinfo{title}{Tempestextremes v2. 1: A community framework for feature detection, tracking, and analysis in large datasets}.
\newblock \emph{\bibinfo{journal}{Geoscientific Model Development}} \textbf{\bibinfo{volume}{14}}, \bibinfo{pages}{5023--5048} (\bibinfo{year}{2021}).

\bibitem{haimberger2008toward}
\bibinfo{author}{Haimberger, L.}, \bibinfo{author}{Tavolato, C.} \& \bibinfo{author}{Sperka, S.}
\newblock \bibinfo{title}{Toward elimination of the warm bias in historic radiosonde temperature records—some new results from a comprehensive intercomparison of upper-air data}.
\newblock \emph{\bibinfo{journal}{Journal of Climate}} \textbf{\bibinfo{volume}{21}}, \bibinfo{pages}{4587--4606} (\bibinfo{year}{2008}).

\bibitem{eyring2016overview}
\bibinfo{author}{Eyring, V.} \emph{et~al.}
\newblock \bibinfo{title}{Overview of the coupled model intercomparison project phase 6 (cmip6) experimental design and organization}.
\newblock \emph{\bibinfo{journal}{Geoscientific Model Development}} \textbf{\bibinfo{volume}{9}}, \bibinfo{pages}{1937--1958} (\bibinfo{year}{2016}).

\bibitem{mitchell2020vertical}
\bibinfo{author}{Mitchell, D.~M.}, \bibinfo{author}{Lo, Y.~E.}, \bibinfo{author}{Seviour, W.~J.}, \bibinfo{author}{Haimberger, L.} \& \bibinfo{author}{Polvani, L.~M.}
\newblock \bibinfo{title}{The vertical profile of recent tropical temperature trends: Persistent model biases in the context of internal variability}.
\newblock \emph{\bibinfo{journal}{Environmental Research Letters}} \textbf{\bibinfo{volume}{15}}, \bibinfo{pages}{1040b4} (\bibinfo{year}{2020}).

\bibitem{ruiz2013estimating}
\bibinfo{author}{Ruiz, J.~J.}, \bibinfo{author}{Pulido, M.} \& \bibinfo{author}{Miyoshi, T.}
\newblock \bibinfo{title}{Estimating model parameters with ensemble-based data assimilation: A review}.
\newblock \emph{\bibinfo{journal}{Journal of the Meteorological Society of Japan. Ser. II}} \textbf{\bibinfo{volume}{91}}, \bibinfo{pages}{79--99} (\bibinfo{year}{2013}).

\bibitem{schneider2017earth}
\bibinfo{author}{Schneider, T.}, \bibinfo{author}{Lan, S.}, \bibinfo{author}{Stuart, A.} \& \bibinfo{author}{Teixeira, J.}
    \newblock \bibinfo{title}{Earth system modeling 2.0: A blueprint for models that learn from observations and targeted high-resolution simulations}.
    \newblock \emph{\bibinfo{journal}{Geophysical Research Letters}} \textbf{\bibinfo{volume}{44}}, \bibinfo{pages}{12--396} (\bibinfo{year}{2017}).
    
    \bibitem{Schneider2024opinion}
    \bibinfo{author}{Schneider, T.}, \bibinfo{author}{Leung, L.~R.} \& \bibinfo{author}{Wills, R. C.~J.}
    \newblock \bibinfo{title}{Opinion: Optimizing climate models with process-knowledge, resolution, and {AI}}.
    \newblock \emph{\bibinfo{journal}{EGUsphere [preprint]}}  (\bibinfo{year}{2024}).
    \newblock \urlprefix\url{https://egusphere.copernicus.org/preprints/2024/egusphere-2024-20/}.
    
    \bibitem{sutskever2014sequence}
    \bibinfo{author}{Sutskever, I.}, \bibinfo{author}{Vinyals, O.} \& \bibinfo{author}{Le, Q.~V.}
    \newblock \bibinfo{title}{Sequence to sequence learning with neural networks}.
    \newblock \emph{\bibinfo{journal}{Advances in neural information processing systems}} \textbf{\bibinfo{volume}{27}} (\bibinfo{year}{2014}).
    
    \end{thebibliography}

\clearpage

\begin{thebibliography}{10}
\makeatletter
\addtocounter{\@listctr}{49}
\makeatother
\expandafter\ifx\csname url\endcsname\relax
\def\url#1{\burl{#1}}\fi
\expandafter\ifx\csname urlprefix\endcsname\relax\def\urlprefix{URL }\fi
\providecommand{\bibinfo}[2]{#2}
\providecommand{\eprint}[2][]{\url{#2}}
\providecommand{\doi}[1]{\url{https://doi.org/#1}}
\bibcommenthead

\bibitem{Jouppi2023-rw}
\bibinfo{author}{Jouppi, N.} \emph{et~al.}
\newblock \emph{\bibinfo{title}{Tpu v4: An optically reconfigurable supercomputer for machine learning with hardware support for embeddings}}, ISCA '23 (\bibinfo{publisher}{Association for Computing Machinery}, \bibinfo{address}{New York, NY, USA}, \bibinfo{year}{2023}).
\newblock \urlprefix\url{https://doi.org/10.1145/3579371.3589350}.

\bibitem{Henry2019-it}
\bibinfo{author}{Henry, G.}, \bibinfo{author}{Tang, P. T.~P.} \& \bibinfo{author}{Heinecke, A.}
\newblock \emph{\bibinfo{title}{Leveraging the bfloat16 artificial intelligence datatype for higher-precision computations}} (\bibinfo{publisher}{IEEE}, \bibinfo{address}{Kyoto, Japan}, \bibinfo{year}{2019}).
\newblock \urlprefix\url{https://ieeexplore.ieee.org/document/8877427/}.

\bibitem{Xu2021-fe}
\bibinfo{author}{Xu, Y.} \emph{et~al.}
\newblock \bibinfo{title}{{GSPMD}: General and scalable parallelization for {ML} computation graphs}  (\bibinfo{year}{2021}).
\newblock \urlprefix\url{http://arxiv.org/abs/2105.04663}.

\bibitem{shard_map}
\bibinfo{author}{Douglas, S.}, \bibinfo{author}{Vikram, S.}, \bibinfo{author}{Bradbury, J.}, \bibinfo{author}{Zhang, Q.} \& \bibinfo{author}{Johnson, M.}
\newblock \bibinfo{title}{shmap (shard\_map) for simple per-device code} (\bibinfo{year}{2023}).
\newblock \urlprefix\url{https://jax.readthedocs.io/en/latest/jep/14273-shard-map.html}.
\newblock \bibinfo{note}{Accessed: 2023-10-21}.

\bibitem{Palmer2009SPPT}
\bibinfo{author}{Palmer, T.~N.} \emph{et~al.}
\newblock \bibinfo{title}{{Stochastic parametrization and model uncertainty}} (\bibinfo{year}{2009}).
\newblock \urlprefix\url{https://www.ecmwf.int/sites/default/files/elibrary/2009/11577-stochastic-parametrization-and-model-uncertainty.pdf}.

\bibitem{battaglia2018relational}
\bibinfo{author}{Battaglia, P.~W.} \emph{et~al.}
\newblock \bibinfo{title}{Relational inductive biases, deep learning, and graph networks}.
\newblock \emph{\bibinfo{journal}{arXiv preprint arXiv:1806.01261}}  (\bibinfo{year}{2018}).

\bibitem{jablonowski2011pros}
\bibinfo{author}{Jablonowski, C.} \& \bibinfo{author}{Williamson, D.~L.}
\newblock \bibinfo{title}{The pros and cons of diffusion, filters and fixers in atmospheric general circulation models}.
\newblock \emph{\bibinfo{journal}{Numerical techniques for global atmospheric models}} \bibinfo{pages}{381--493} (\bibinfo{year}{2011}).

\bibitem{Jablonowski2011-qf}
\bibinfo{author}{Jablonowski, C.} \& \bibinfo{author}{Williamson, D.~L.}
\newblock \bibinfo{title}{ in \textit{The pros and cons of diffusion, filters and fixers in atmospheric general circulation models}} (eds \bibinfo{editor}{Lauritzen, P.}, \bibinfo{editor}{Jablonowski, C.}, \bibinfo{editor}{Taylor, M.} \& \bibinfo{editor}{Nair, R.}) \emph{\bibinfo{booktitle}{Numerical Techniques for Global Atmospheric Models}} \bibinfo{pages}{381--493} (\bibinfo{publisher}{Springer Berlin Heidelberg}, \bibinfo{address}{Berlin, Heidelberg}, \bibinfo{year}{2011}).
\newblock \urlprefix\url{https://doi.org/10.1007/978-3-642-11640-7_13}.

\bibitem{kingma2014adam}
\bibinfo{author}{Kingma, D.~P.} \& \bibinfo{author}{Ba, J.}
\newblock \bibinfo{title}{Adam: A method for stochastic optimization}.
\newblock \emph{\bibinfo{journal}{arXiv preprint arXiv:1412.6980}}  (\bibinfo{year}{2014}).

\bibitem{metpy}
\bibinfo{author}{May, R.~M.} \emph{et~al.}
\newblock \bibinfo{title}{Metpy: A meteorological python library for data analysis and visualization}.
\newblock \emph{\bibinfo{journal}{Bulletin of the American Meteorological Society}} \textbf{\bibinfo{volume}{103}}, \bibinfo{pages}{E2273 -- E2284} (\bibinfo{year}{2022}).
\newblock \urlprefix\url{https://journals.ametsoc.org/view/journals/bams/103/10/BAMS-D-21-0125.1.xml}.

\bibitem{roberts2020impact}
\bibinfo{author}{Roberts, M.~J.} \emph{et~al.}
\newblock \bibinfo{title}{Impact of model resolution on tropical cyclone simulation using the highresmip--primavera multimodel ensemble}.
\newblock \emph{\bibinfo{journal}{Journal of Climate}} \textbf{\bibinfo{volume}{33}}, \bibinfo{pages}{2557--2583} (\bibinfo{year}{2020}).

\bibitem{vallis2015response}
\bibinfo{author}{Vallis, G.~K.}, \bibinfo{author}{Zurita-Gotor, P.}, \bibinfo{author}{Cairns, C.} \& \bibinfo{author}{Kidston, J.}
\newblock \bibinfo{title}{Response of the large-scale structure of the atmosphere to global warming}.
\newblock \emph{\bibinfo{journal}{Quarterly Journal of the Royal Meteorological Society}} \textbf{\bibinfo{volume}{141}}, \bibinfo{pages}{1479--1501} (\bibinfo{year}{2015}).

\bibitem{beucler2024climate}
\bibinfo{author}{Beucler, T.} \emph{et~al.}
\newblock \bibinfo{title}{Climate-invariant machine learning}.
\newblock \emph{\bibinfo{journal}{Science Advances}} \textbf{\bibinfo{volume}{10}}, \bibinfo{pages}{eadj7250} (\bibinfo{year}{2024}).

\end{thebibliography}
\end{document}